\documentclass[twocolumn,trackchanges]{aastex631} 
\usepackage{CLASSY_Preamble}

\usepackage{amssymb}	
\usepackage{natbib}
\usepackage[figuresright]{rotating}
\usepackage{threeparttable}
\usepackage{subfigure}
\usepackage{soul}

\newcommand{\PI}{CLASSY~I}
\newcommand{\PII}{CLASSY~II}
\newcommand{\PIII}{CLASSY~III}

\newcommand{\lya}{Ly$\alpha$}

\newcommand{\lyaesc}{$f^{\mathrm{Ly\alpha}}_\mathrm{esc}$}

\newcommand{\vtrough}{$v^\mathrm{Ly\alpha}_\mathrm{trough}$}
\newcommand{\vsep}{$\Delta v_\mathrm{Ly\alpha}$}

\newcommand{\vsiii}{$v^\mathrm{outflow}_\mathrm{Si\ II}$}
\newcommand{\tlac}{\texttt{tlac}}
\defcitealias{Berg22}{\PI}
\defcitealias{James22}{\PII}
\defcitealias{Xu2022}{\PIII}

\begin{document}

\submitjournal{AASJournal ApJ}
\shortauthors{Hu et al.}
\shorttitle{CLASSY \lya\ Profiles}

\title{CLASSY VII \lya\ Profiles: The Structure and Kinematics of Neutral Gas and Implications for 
LyC Escape in Reionization-Era Analogs\footnote{
Based on observations made with the NASA/ESA Hubble Space Telescope,
obtained from the Data Archive at the Space Telescope Science Institute, which
is operated by the Association of Universities for Research in Astronomy, Inc.,
under NASA contract NAS 5-26555.}}

\author[0000-0003-3424-3230]{Weida Hu}
\affiliation{Department of Physics, University of California, Santa Barbara, Santa Barbara, CA 93106, USA}

\author[0000-0001-9189-7818]{Crystal L. Martin}
\affiliation{Department of Physics, University of California, Santa Barbara, Santa Barbara, CA 93106, USA}

\author[0000-0003-2491-060X]{Max Gronke}
\affiliation{Max-Planck Institute for Astrophysics, Karl-Schwarzschild-Str. 1, D-85741 Garching, Germany}

\author[0000-0002-5659-4974]{Simon Gazagnes}
\affiliation{Department of Astronomy, The University of Texas at Austin, 2515 Speedway, Stop C1400, Austin, TX 78712, USA}

\author[0000-0001-8587-218X]{Matthew Hayes}
\affiliation{Stockholm University, Department of Astronomy and Oskar Klein Centre for Cosmoparticle Physics, AlbaNova University Centre, SE-10691, Stockholm, Sweden}

\author[0000-0002-0302-2577]{John Chisholm}
\affiliation{Department of Astronomy, The University of Texas at Austin, 2515 Speedway, Stop C1400, Austin, TX 78712, USA}

\author[0000-0003-1127-7497]{Timothy Heckman}
\affiliation{Center for Astrophysical Sciences, Department of Physics \& Astronomy, Johns Hopkins University, Baltimore, MD 21218, USA}

\author[0000-0003-2589-762X]{Matilde Mingozzi}
\affiliation{Space Telescope Science Institute, 3700 San Martin Drive, Baltimore, MD 21218, USA}

\author[0000-0002-4430-8846]{Namrata Roy}
\affiliation{Center for Astrophysical Sciences, Department of Physics \& Astronomy, Johns Hopkins University, Baltimore, MD 21218, USA}

\author[0000-0002-9132-6561]{Peter Senchyna}
\affiliation{Carnegie Observatories, 813 Santa Barbara Street, Pasadena, CA 91101, USA}

\author[0000-0002-9217-7051]{Xinfeng Xu}
\affiliation{Center for Astrophysical Sciences, Department of Physics \& Astronomy, Johns Hopkins University, Baltimore, MD 21218, USA}

\author[0000-0002-4153-053X]{Danielle A. Berg}
\affiliation{Department of Astronomy, The University of Texas at Austin, 2515 Speedway, Stop C1400, Austin, TX 78712, USA}

\author[0000-0003-4372-2006]{Bethan L. James}
\affiliation{AURA for ESA, Space Telescope Science Institute, 3700 San Martin Drive, Baltimore, MD 21218, USA}

\author[0000-0001-6106-5172]{Daniel P. Stark}
\affiliation{Steward Observatory, The University of Arizona, 933 N Cherry Ave, Tucson, AZ, 85721, USA}

\author[0000-0002-2644-3518]{Karla Z. Arellano-C\'{o}rdova}
\affiliation{Department of Astronomy, The University of Texas at Austin, 2515 Speedway, Stop C1400, Austin, TX 78712, USA}

\author[0000-0002-6586-4446]{Alaina Henry}
\affiliation{Center for Astrophysical Sciences, Department of Physics \& Astronomy, Johns Hopkins University, Baltimore, MD 21218, USA}
\affiliation{Space Telescope Science Institute, 3700 San Martin Drive, Baltimore, MD 21218, USA}

\author[0000-0002-6790-5125]{Anne E. Jaskot}
\affiliation{Department of Astronomy, Williams College, Williams town, MA 01267, USA}

\author[0000-0002-5320-2568]{Nimisha Kumari}
\affiliation{AURA for ESA, Space Telescope Science Institute, 3700 San Martin Drive, Baltimore, MD 21218, USA}

\author[0000-0002-8809-4608]{Kaelee S. Parker}
\affiliation{Department of Astronomy, The University of Texas at Austin, 2515 Speedway, Stop C1400, Austin, TX 78712, USA}

\author[0000-0002-9136-8876]{Claudia Scarlata}
\affiliation{Minnesota Institute for Astrophysics, University of Minnesota, 116 Church Street SE, Minneapolis, MN 55455, USA}

\author[0000-0001-8289-3428]{Aida Wofford}
\affiliation{Instituto de Astronom\'ia, Universidad Nacional Aut\'onoma de M\'exico, Unidad Acad\'emica en Ensenada, Km 103 Carr. Tijuana-Ensenada, Ensenada 22860, Mexico}

\author[0000-0001-5758-1000]{Ricardo O. Amor\'in}
\affiliation{Instituto de Investigaci\'on Multidisciplinar en Ciencia y Tecnolog\'ia, Universidad de La Serena, Raul Bitr\'an 1305, La Serena 2204000, Chile}
\affiliation{Departamento de Astronomía, Universidad de La Serena, Av. Juan Cisternas 1200 Norte, La Serena 1720236, Chile}

\author{Naunet Leonhardes-Barboza}
\affiliation{Wellesley College, 106 Central Street, Wellesley, MA 02481, USA}

\author[0000-0003-4359-8797]{Jarle Brinchmann}
\affiliation{Instituto de Astrof\'isica e Ci\^encias do Espaço, Universidade do Porto, CAUP, Rua das Estrelas, PT4150-762 Porto, Portugal}

\author{Cody Carr}
\affiliation{Minnesota Institute for Astrophysics, University of Minnesota, 116 Church Street SE, Minneapolis, MN 55455, USA}

\begin{abstract}
Lyman-alpha line profiles are a powerful probe of ISM structure, outflow speed, and Lyman continuum escape fraction.
In this paper, we present the \lya\ line profiles of the COS Legacy Archive Spectroscopic SurveY, a sample rich in 
spectroscopic analogs of reionization-era galaxies. A large fraction of the spectra show a complex profile, consisting 
of a double-peaked \lya\ emission profile in the bottom of a damped, \lya\ absorption trough.  Such profiles reveal an 
inhomogeneous interstellar medium (ISM). We successfully fit the damped \lya\ absorption (DLA) and the \lya\ emission 
profiles separately, but with complementary covering factors, a surprising result because this approach requires no 
\lya\ exchange between high-$N_\mathrm{HI}$ and low-$N_\mathrm{HI}$ paths. The combined distribution of column densities 
is qualitatively similar to the bimodal distributions observed in numerical simulations. We find an inverse relation 
between \lya\ peak separation and the [O \textsc{iii}]/[O \textsc{ii}] flux ratio, confirming that the covering fraction 
of Lyman-continuum-thin sightlines increases as the \lya\ peak separation decreases.  We combine measurements of \lya\ 
peak separation and \lya\ red peak asymmetry in a diagnostic diagram which identifies six Lyman continuum leakers in 
the CLASSY sample. We find a strong correlation between the \lya\ trough velocity and the outflow velocity measured from 
interstellar absorption lines.  We argue that greater vignetting of the blueshifted \lya\ peak, relative to the 
redshifted peak, is the source of the well-known discrepancy between shell-model parameters and directly measured 
outflow properties. The CLASSY sample illustrates how scattering of \lya\ photons outside the spectroscopic  aperture 
reshapes \lya\ profiles as the distances to these compact starbursts span a large range.
\end{abstract} 


\section{Introduction} 
\label{sec:intro}

The Epoch of Reionization (EoR) marks a period in the history of the Universe when the emergence of galaxies
ionized most of the neutral hydrogen in the intergalactic medium (IGM). Observations suggest that the first
ionized pockets in the IGM grew around the largest overdensities of galaxies \citep[e.g.,][]{Hu2021,Hayes2023b,Tilvi2020}. The massive stars in those galaxies are likely the source of the ionizing photons, 
the Lyman continuum (LyC) at wavelengths $\lambda<912$ \AA\ \citep[e.g.,][]{Robertson2015}. 
How this ionizing radiation leaks out of the dense structures where early galaxies form, however, is not well
understood.  A small column density of neutral hydrogen, $N_\mathrm{HI} \approx 1.6 \times 10^{17}$~cm$^{-2}$, will absorb a 
LyC photon. Exactly how feedback from massive stars opens pathways for LyC escape \citep{Ma2020,Kakiichi2021} 
sets the timeline for cosmic reionization \citep[e.g.,][]{Mason2019,Finkelstein2019,Naidu2020}.
Direct observations of the escaping LyC photons are not possible during the EoR because of attenuation by
the IGM \citep[][and references therein]{Ouchi2020}, so indirect tracers LyC escape and outflows are needed.

Lyman-$\alpha$ is the most commonly detected emission line from high-redshift galaxies \citep{Partridge1967}. 
The channels through which \lya\ photons emerge from galaxies appear to be tightly related to the pathways 
of LyC escape \citep{Chisholm2018,Gazagnes2020} 
{because the origins of \lya\ photons, H \textsc{ii} regions, are illuminated by the LyC photons arising from central massive stars.}
Even low column densities of neutral hydrogen in these channels
scatter  \lya\ photons many times, altering their direction and frequency. Their random walk redistributes 
photons flux from the line core into the line wings, and this reshaping of the line profile imprints information 
about the outflow velocity, column density, and ISM structure on the emergent line profile 
\citep{Verhamme2006,Dijkstra2014,Verhamme2015,Chisholm2018,Gazagnes2020,Saldana-Lopez2022}.  

In the absence of absorption by dust, all the \lya\ photons eventually escape from the galaxy, and radiative 
transfer calculations demonstrate some general properties of the line profiles. For example,
analytic solutions for static slabs and spheres yield emerging \lya\ spectra with symmetric redshifted and 
blueshifted peaks \citep{Neufeld1990,Dijkstra2006}. Bulk motion requires Monte Carlo techniques, 
and these calculations demonstrate that outflowing gas produces an asymmetric profile which has a 
stronger redshifted component regardless of the outflow geometry and structure \citep{Verhamme2006,Verhamme2012}.

The most commonly applied radiative transfer model, the homogeneous shell model, assumes an expanding, 
spherical shell of neutral hydrogen \citep{Verhamme2006}. Over a wide range of outflow properties, the emergent 
line profile has a P Cygni shape characterized by a redshifted emission line with a broad red wing plus a blueshifted
absorption trough. For a very low H $\textsc{i}$ column density, some emission from the near side of the thin shell
is transmitted, producing blueshifted emission instead of absorption. Whereas a very high column density shell 
will trap a \lya\ photon until it is eventually absorbed by a dust grain, and the emergent line profile has become
that of a damped \lya\ absorber (DLA), a completely-dark absorption trough with very broad wings. The shell model 
does a good job of reproducing the diversity of commonly observed \lya\ profile shapes
\citep[e.g.,][]{Verhamme2015,Yang2017,Gronke2017,Gurung-Lopez2022}.  Statistically successful fits, however, 
do not guarantee accurate recovery of outflow properties. The structure of the  shell model is much simpler than 
actual ISM and multi-phase outflows \citep{Schneider2018,Gronke2020,Fielding2022}.

Low-ionization state (LIS) absorption lines in galaxy spectra unambiguously detect outflowing gas and have provided insight
into how outflow properties vary with galaxy properties  \citep{Heckman2000,Heckman2011,Martin2012,Chisholm2015,Martin2015}.
The outflow speeds derived from the blueshifts of these absorption lines offer an opportunity to test the shell model 
velocities, and the results reveal significant discrepancies both at high-redshift \citep{Kulas2012} and among nearby Green 
Pea galaxies \citep{Yang2017,Orlitova2018}. 
Three major discrepancies are reported in those studies:  (1) the best-fit redshifts are larger by 10--250 km s$^{-1}$ than the spectroscopic redshifts; (2) the best-fit outflow velocities of expanding shell are lower than the outflow velocities derived by LIS lines; (3) the intrinsic \lya\ line widths of shell model are broader than those of Balmer lines.
\citet{Li2022} proposed that those discrepancies might be caused by the degeneracies between model parameters, but no explanation for these puzzles based on observations has been found.

We also draw attention to another limitation of the shell model. A large fraction of Green Peas and higher-redshift 
star-forming galaxies show \lya\ emission line in the bottom of a DLA system \citep[][]{McKinney2019,Reddy2016}.
These profiles cannot be produced by a homogeneous shell model.  The low column density shells that produce double-peaked
profiles contradict the presence of damped absorption which requires  very high column density. 
Even larger peak separations are predicted by a clumpy shell because the fitted shell expansion speed lies between the outflow velocities of the neutral clouds and the hot interclump medium \citep{Li2022}. 

Comparing physical properties derived from \lya\ shell modeling to those measured from other spectral lines can 
therefore provide new insight about the structure of the multi-phase gas. Because these properties determine the LyC 
escape fraction from galaxies, there is an urgent need to understand the puzzling properties of \lya\ profiles
in a sample of EoR analogs. To place the unexpected profile shapes, i.e. the double peaked emission lines in DLA systems,
in the broader context of the full diversity of observed \lya\ profile shapes, requires high-resolution and high S/N ratio
UV spectroscopy of EoR analogs, including, but not limited to, Green Pea galaxies.  The James Webb Space Telescope (JWST) 
observations reveal a diversity of galaxies in the EoR  \citep[e.g.,][]{Tang2023,Labbe2022,Looser2023,Saxena2023,Bunker2023}, 
spanning much wider ranges of galaxy properties than the local Green Peas.

In this paper, we analyze 45 \lya\ line profiles obtained by the COS Legacy Archive Spectrocopy SurveY (CLASSY) 
\citep{Berg2022,James2022}. This UV-surface brightness selected sample includes the lowest redshift Green Pea galaxies,
local Lyman Break Galaxy Analogs (LBAs) \citep{Heckman2011},
and the two local galaxies that are the nearest spectral match to the emission-line spectra of GN-z11 
\citep{Senchyna2023}. Thus the range in metallicity and ionizing continuum properties include 
the extreme conditions that were common during galaxy assembly. We present a uniform analysis of the \lya\ profiles. 
Outflow properties have been determined from the blueshifted components of the LIS resonance lines \citep{Xu2022}
and the excited fine-structure lines \citep{Xu2023}. The results provide new insight into the clumpiness of the ISM, 
as described by the relative covering fractions of high-$N_\mathrm{HI}$ and low-$N_\mathrm{HI}$ gas, yet also strongly suggest that the 
discrepancies between shell model parameters and LIS absorption lines arise from aperture vignetting. Among CLASSY targets, 
the physical size of COS aperture ranges from the scale of star clusters ($\sim 100$ pc) to galaxies  ($\sim 10$ kpc).
The large variations in aperture losses make it possible to view individual \lya\ profile shapes in a broader context.

This paper is organized as follows.  In Sec. 2, we introduce the CLASSY sample of \lya\ profiles,
describe how we remove the damped \lya\ absorption and measure the properties of the high column density
neutral gas, and discuss the large variation in the amount of aperture vignetting across the sample.  In Sec. 3, we use the radiative transfer code \tlac\ to fit shell models to the net \lya\ emission-line profiles, investigating different choices for the continuum level (and hence the line equivalent width). 
In Sec. 4, we discuss the H \textsc{i} column 
density distribution in EoR analogs, the size scale of the holes leaking LyC radiation, and argue that
aperture vignetting biases shell model properties in the directions required to solve the discrepancies
with independently measured outflow properties.

Throughout this paper, we adopt a Flat $\Lambda$CDM cosmology with $\Omega_m=0.3$, $\Omega_\Lambda=0.7$, and $\mathrm{H}_0=70$ 
km s$^{-1}$ Mpc$^{-1}$. We also adopt the Spearman rank method to quantify the correlation strengths $r$.
The data used in this paper is available via the CLASSY high-level science products (HLSP) 
homepage\footnote{
Data will appear at \url{https://archive.stsci.edu/hlsp/classy} after acceptance by the ApJ. The data product
can be found here (\url{https://drive.google.com/drive/folders/1NCUyr1vQ10z4BZuGBqsBuIjL0dWJnmZ1?usp=sharing}) during the review
period.}, including the best-fit DLA systems, the \lya\ emission lines 
after subtracting the DLA and continuum, and the best-fit shell model spectra.

\section{Sample of \lya\ Profiles} \label{sec:2-1}

\begin{figure*}[htbp]
    \centering
    \includegraphics[width=6.5in]{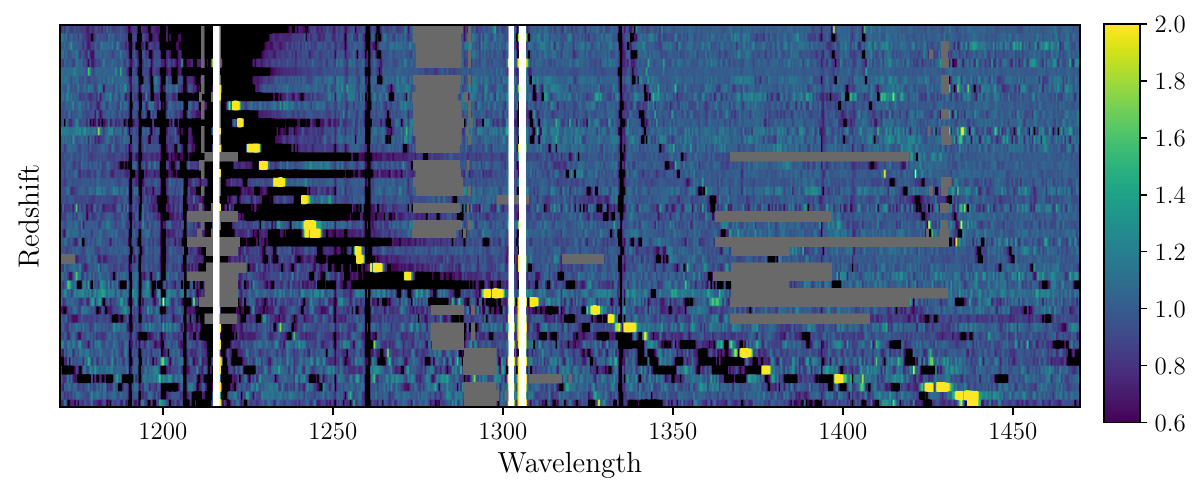}
    \caption{{\label{fig:1}} Overview of the COS G130M spectra obtained in the CLASSY survey with the wavelength coverage of 1170 to 1470 \AA\ on the X-axis.
    The image is built by one continuum-normalized spectrum per image line. 
    The redshift increases from top to bottom along the Y-axis. 
    The dark is low flux and the yellow is high flux.
    The yellow waterfall highlights the redshifted \lya\ emission lines.
    We use vertical white lines to mask the geocoronal \lya\  and O \textsc{ii} emission lines.
    The vertical dark lines indicate the MW absorption lines and the gray strips indicate the CCD gaps.}
\end{figure*}

Here  we present high-S/N \lya\ spectra for the 45 CLASSY targets. Each of these nearby galaxies has a compact, 
far-UV bright star-forming region which was the target of the COS observation.  The sample provides a diverse 
set of local analogs of high-redshift galaxies, including both Green Pea galaxies and LBAs.
Physical conditions in the starburst range cover oxygen abundances  from $12+\log \mathrm{(O/H)\sim 7}$ to 8.8 and 
electron densities from $n_e \sim 10$ to $1120$ cm$^{-3}$. The stellar masses and star formation rates of their host 
galaxies sample the range $\log\ \mathrm{(M_\star/M_\odot)} \sim 6.2$ to 10.1 and
$\log\ \mathrm{(SFR/M_\odot\ yr^{-1}) \sim -2}$ -- $1.6$, respectively \citep{Berg2022}. 
The raw spectroscopic data were reduced using the CALCOS pipeline (v3.3.10), including spectrum extraction, 
wavelength calibration, and vignetting correction, and then coadded using a custom pipeline \citep[][]{James2022}. 
The Galactic foreground extinction was corrected assuming a ratio of total-to-selective extinction
$R_V=3.1$ and a Milky Way (MW) extinction curve \citep{Cardelli1989}.

Fig. \ref{fig:1} shows an overview of the G130M and G160M spectra, ordered by redshift.  The CLASSY spectra 
easily resolve the damping wings of the broad absorption trough imprinted by H \textsc{i} absorption from the Milky Way.  
A large fraction of the spectra show a second damped \lya\ absorber at the redshift of the CLASSY galaxy. In the 
lowest redshift galaxies, the blueshifted damping wing of the target is blended with the redshifted damping wing 
of the Milky Way absorption.  
The yellow waterfall across Fig. \ref{fig:1} highlights the redshifted \lya\ emission.
Surprisingly, the \lya\ emission is frequently detected in the bottom of a damped absorption trough. Profiles of this type cannot be produced by a uniform shell of neutral hydrogen.

In this paper, we adopt an approach that we have not seen used previously.  We fit the damping
wing profile, including a non-unity covering factor. We then extract the net emission-line profile 
relative to the damping trough, as others have done.  The equivalent width of this net \lya\ emission, however, 
has been previously neglected.  We address this in Sec.~\ref{sec:profile} below, where we demonstrate that the best normalization 
for the \lya\ emission is the fraction of the stellar continuum {\it not intercepted} by the high column density neutral hydrogen. 

\begin{figure*}[]
    \centering
    \includegraphics[width=0.49\textwidth]{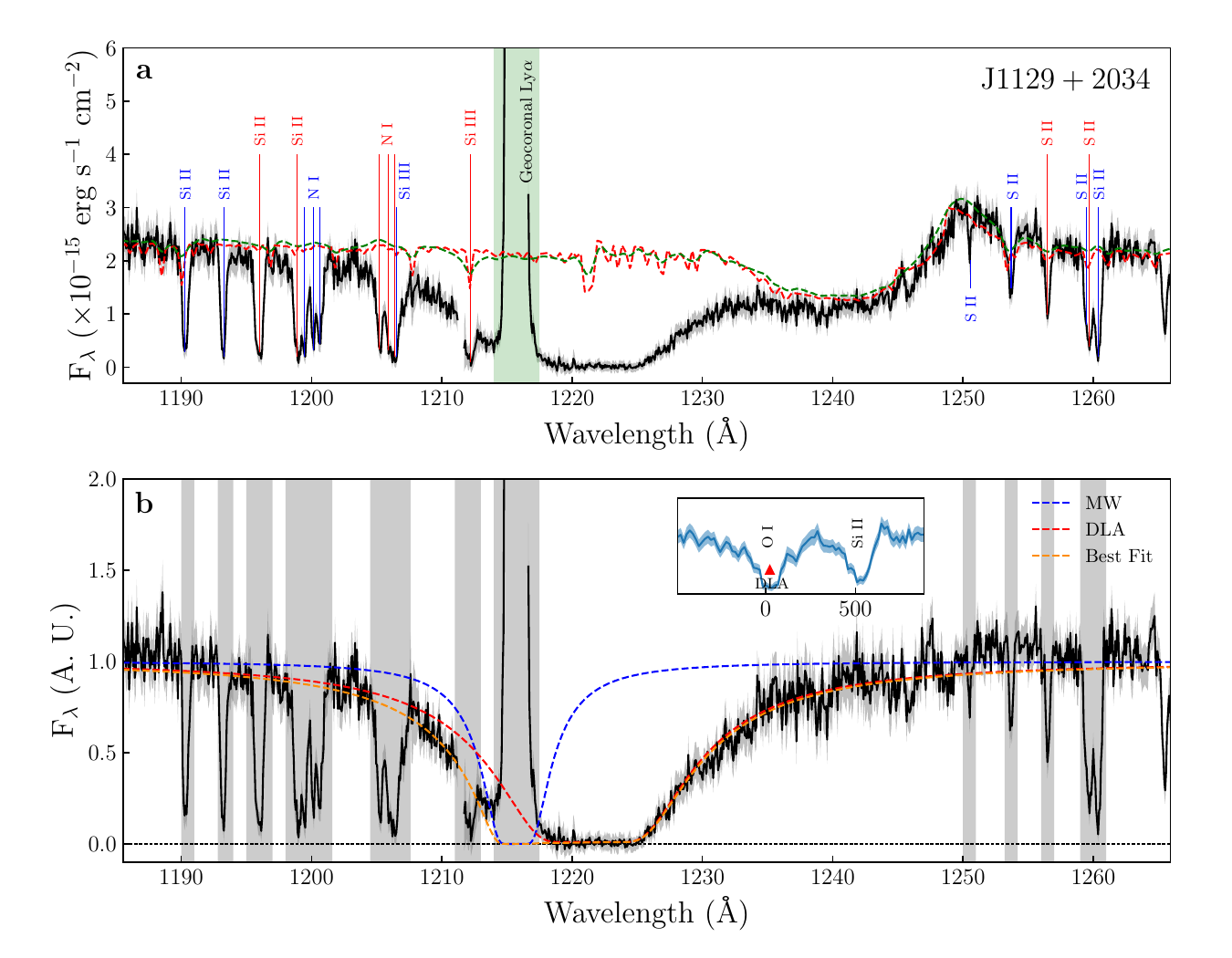}
    \includegraphics[width=0.49\textwidth]{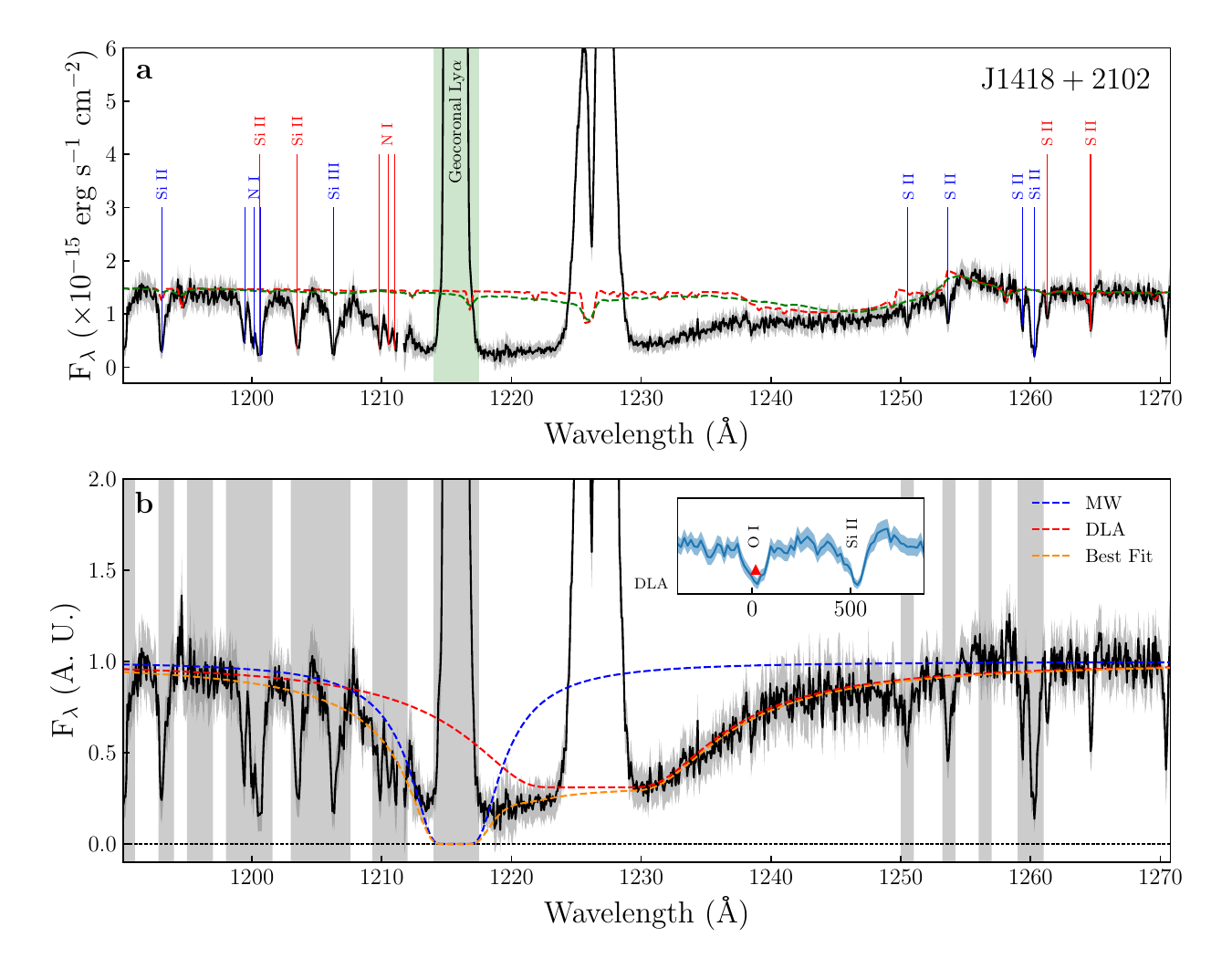}
    \includegraphics[width=0.49\textwidth]{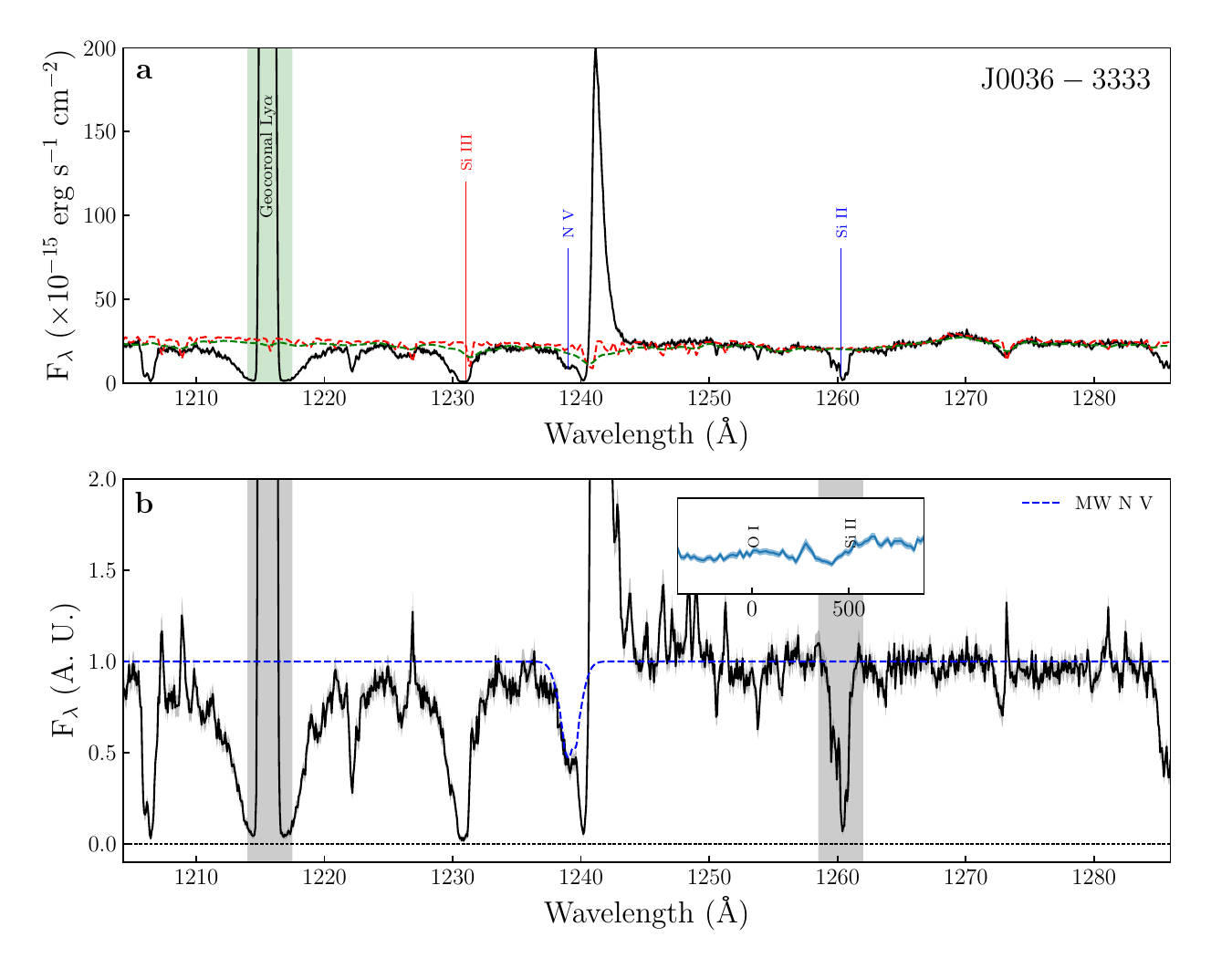}
    \includegraphics[width=0.49\textwidth]{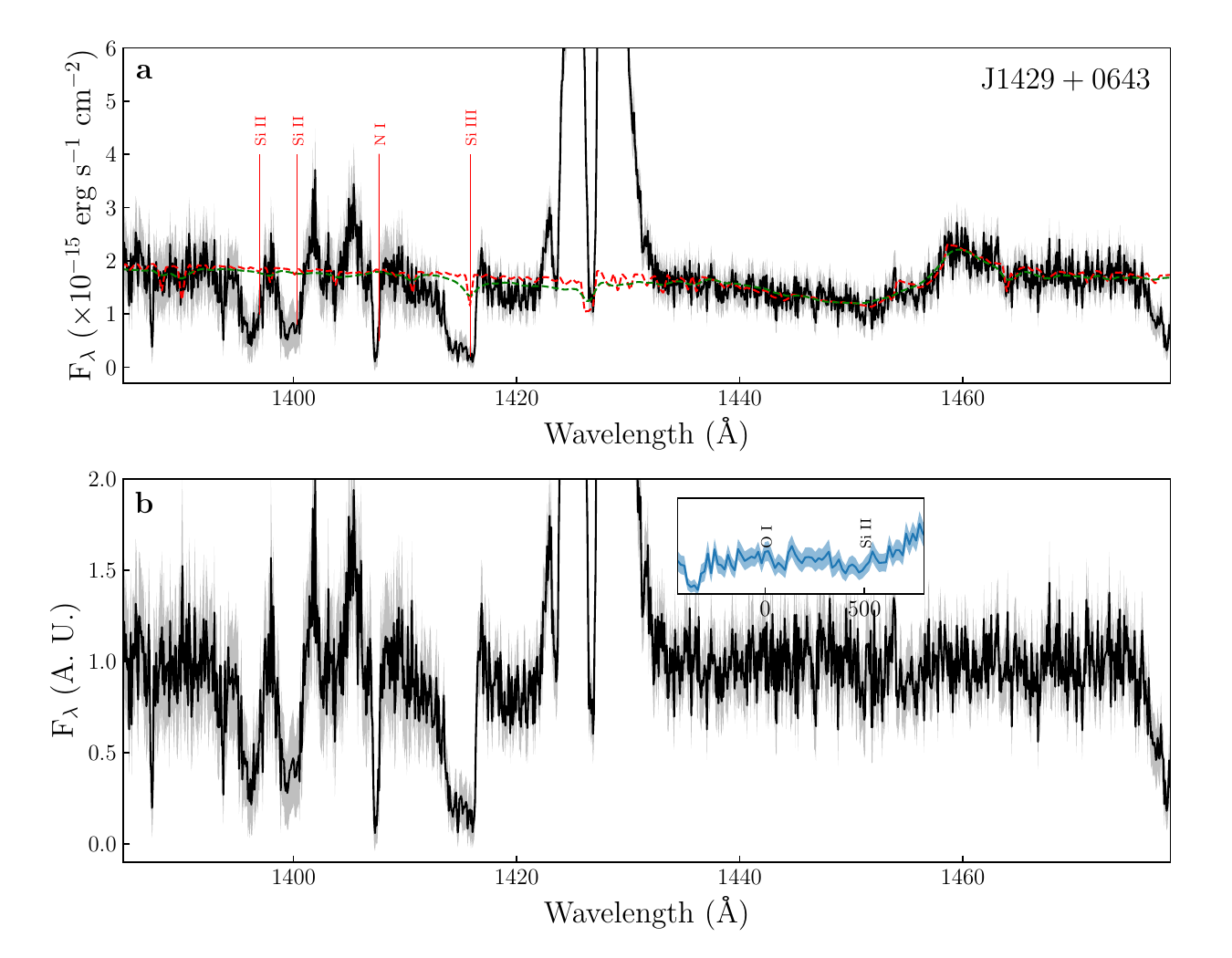}
    \caption{\label{fig:specs} Examples of four typical \lya\ profiles: pure DLA system, \lya\ emission over DLA system, P-Cygni-like \lya\ emission, and pure \lya\ emission. Panel a: the flux spectra (black) and error spectra (gray shade) close to \lya\ line.
    The vertical green shade masks the geocoronal \lya\ emission line.
    The absorption lines are labeled in blue and red for absorption lines of MW and host galaxy, respectively.
    We overplot the best-fit continua in red and green dashed lines for the first and second methods (see Sec. \ref{sec:2-1}), respectively. 
    Panel b: the best-fit absorption profiles (orange). 
    The flux spectrum and error spectrum are normalized by the best-fit continuum (first method, red dashed line in panel a).
    We mask the absorption lines (vertical gray shades) and \lya\ emission line and use two components to fit the absorption profile which corresponds to the MW component (blue) and a host-galaxy DLA component (red).
    For the host-galaxy component, we adopt a Voigt profile.
    For the MW component, we adopt a Voigt profile if it is an MW DLA system with fixed galactic H \textsc{i} column density \citep{Hartmann1997}, or a Gaussian profile if it is an MW metal absorption line.
    The small inset shows the O \textsc{i} $\lambda$1302\AA\ absorption line.
    The \lya\ profiles of the other 41 galaxies are shown in the continued Fig. \ref{fig:allspec}.}
\end{figure*}

We use physical models to define the continuum level near \lya, allowing us to accurately 
model the DLA system in Sec.~\ref{sec:2-2}.
CLASSY provides two models for the continuum (Senchyna et al. in preparation). 
Both models assume the observed continuum can be reconstructed as a
linear combination of a set of single-age, single-metallicity stellar populations \citep{Chisholm2019}, 
and, thus, be fitted using the following relation:
\begin{equation}
    F_{\mathrm{obs}} (\lambda) = 10^{-0.4E(B-V)k(\lambda)}\ \Sigma_i \ X_i M_i({\lambda}),
\end{equation}
where $F_{\mathrm{obs}} (\lambda)$ is the observed spectrum, $k(\lambda)$ is the attenuation law, 
$M_i({\lambda})$ is the spectrum of the $ith$ single stellar population (SSP), and $X_i$ is its coefficient. 
The main difference between the two methods is the stellar population synthesis framework. The top panels of 
Fig. \ref{fig:specs} illustrate each best-fit continuum.  The red dashed line represents the continuum built from 
STARBURST~99  synthesis models \citep{Leitherer1999} and a \citet{Reddy2016} attenuation law, and the green dashed line uses 
the latest version of the \citet{Bruzual2003} model \citep[Charlot \& Bruzual in prep., see also][]{Plat2019} and an 
SMC extinction law \citep{Gordon2003}. These two continua both reproduce the prominent \textsc{N v} $\lambda$1240 
stellar P-Cygni line well. The narrow dip visible at \lya\ in both models is not physical (C. Leitherer, private 
communication), and we interpolate over it. We fit the DLA profiles using both the continuum models and found 
similar parameters. We adopt the first method for the analysis that follows because the STARBURST99 models 
have the higher spectral resolution.

\begin{figure*}[ht]
    \centering
    \includegraphics[width=1\textwidth]{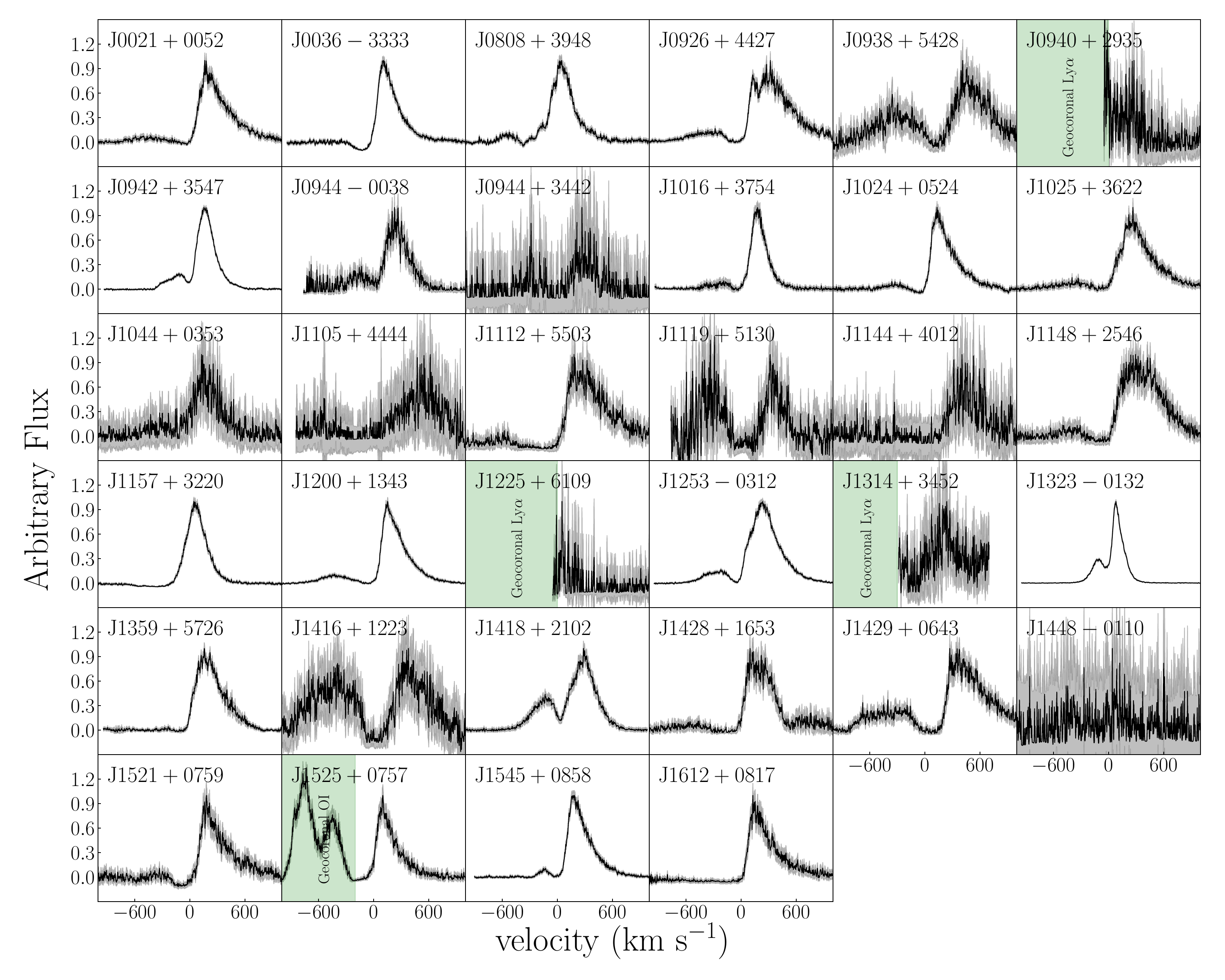}
    \caption{\label{fig:lya} The \lya\ spectra (black) and error spectra (gray) after subtracting the continuum and DLA of 
      34 CLASSY galaxies.     We use green shades to show the masked geocoronal \lya\ and O \textsc{i} emission lines. The 
      continuum-subtracted spectra are normalized by peak flux. The total \lya\ fluxes are measured by integrating fluxes 
      within the wavelength range where \lya\ emission lines meet zero flux.}
\end{figure*}

\subsection{DLA fitting} \label{sec:2-2}

Fig. \ref{fig:specs} illustrates the diversity of CLASSY \lya\ profiles:  pure DLA systems, \lya\ emission in the
bottom of a damping trough (hereafter Abs+EM profile), P-Cygni-like  profiles, and double-peaked \lya\ emission. 
A large fraction of CLASSY spectra ($31/45$) have a DLA system, and 20 out of these 31 galaxies 
show double- or single-peaked \lya\ emission lines. CLASSY spectra offer the high spectral resolution 
and S/N ratio required to remove the contribution of the DLA system  \citep[e.g.,][]{Reddy2016b,McKinney2019} and 
extract the \lya\ emission lines.

Fig. \ref{fig:1} shows that geocoronal emission lines intersect the broad damping wings at low redshift 
and at $z \approx 0.07$.  In addition, the LIS absorption lines from the MW and the target galaxy affect 
the wings of the DLA systems but intersect only a few \lya\ emission lines (see Sec. \ref{sec:notes}). 
We mask these lines as indicated in the second row of  Fig. \ref{fig:specs}. 

To uniquely describe the MW DLA system, we adopt the Galactic H \textsc{i} column density derived from 21 cm emission 
in the direction of the target \citep{Hartmann1997}. The DLA line profile is described by a Voigt profile 
\citep[e.g.,][]{Prochaska2019}, which is defined by a Doppler parameter ($b$) and a column density ($N_\mathrm{HI}$).
We assume a Doppler parameter of 30 km s$^{-1}$ \citep[e.g.,][]{Prochaska2015}, no velocity shift, and complete 
covering of the continuum source.  These steps define the Voigt profile, which we convolve with the instrumental 
resolution, and then subtract from the normalized continuum to uncover the \lya\ profile of the CLASSY target. 

Because DLA systems are optically thick, the bottom of the Voigt profile is completely dark. However, we found 
significant residual intensity in the bottom of the damping troughs. Partial covering of the continuum source 
therefore turns out to be critical for fitting damping profiles.  This partial covering  was sometimes subtle, 
as in the top left panel (J1129+2034) of Fig. \ref{fig:specs}. In contrast, the top right panel (J1418+2102) of 
Fig. \ref{fig:specs} shows strong H \textsc{i} damping wings and prominent residual intensity in the trough. Here we 
adopt a modified Voigt profile which allows a velocity offset, $v$, and a velocity-independent covering fraction, $f_C$.  

We convolve each Voigt profile with a Gaussian line spread function whose width is determined by the spectral
resolution \citep[Table 3 in][]{Berg2022}. Our fitting code then multiplies the normalized continuum by the optical depth 
of each Voigt profile. The error is measured using a Monte Carlo (MC) approach; we add random noise to the observed 
spectra and refit it 1000 times. Leaving all the parameters free provided statistically good fits; however, 
we noticed degeneracies between the fitted velocity $v$ of the DLA and the wings of the damping profile, and 
also the overlaps between the wings of the DLA system and \lya\ emission. 

We broke  these degeneracies by using 
the O \textsc{i} $\lambda1302.2$ \AA\ absorption line to constrain the parameters of the Voigt profile, an approach 
Section~\ref{sec:oi} justifies below.  The profile shape is not sensitive to $b$, and we fixed $b$ to be 30 km s$^{-1}$. 
The second row of Fig. \ref{fig:specs} presents the continuum-normalized spectra, our model for the  MW absorption, 
and the fitted damped \lya\ absorption. Table \ref{tab:1} summarizes the best-fit Voigt parameters for the DLAs.

We  extract the \lya\ emission lines by subtracting the stellar continuum and DLA profile.
Previous works have visually selected a local continuum close to the \lya\ emission. A comparison of common targets 
shows that the resulting \lya\ can be sensitive to the wavelength range used to define the local continuum. 
For example, the beginning of the wavelength range of J0938+5428 used in \citet{Alexandroff2015} is the \lya\ blue 
peak of J0938+5428 in Fig. \ref{fig:specs}. For this same target, \citet{Yang2017} determine the wavelength range from the 
intersection of the \lya\ emission line with the DLA profile. This method recovers the \lya\ blue peak; however, 
it  underestimates the \lya\ flux because the bottom of the DLA system is poorly estimated.

Among 45 CLASSY galaxies, 24 galaxies show significant double-peak \lya\ emission lines, 10 show single-peak \lya\ emission.
Fig. \ref{fig:lya} presents the \lya\ emission line spectra of 34 CLASSY galaxies. {The remaining 11 galaxies show pure 
DLA systems, and are therefore not included in Fig. \ref{fig:lya}.}

\subsubsection{Constraining the DLA Properties with O \textsc{i} Absorption} \label{sec:oi}

\begin{figure*}[htb]
    \centering
    \includegraphics[width=6.5in]{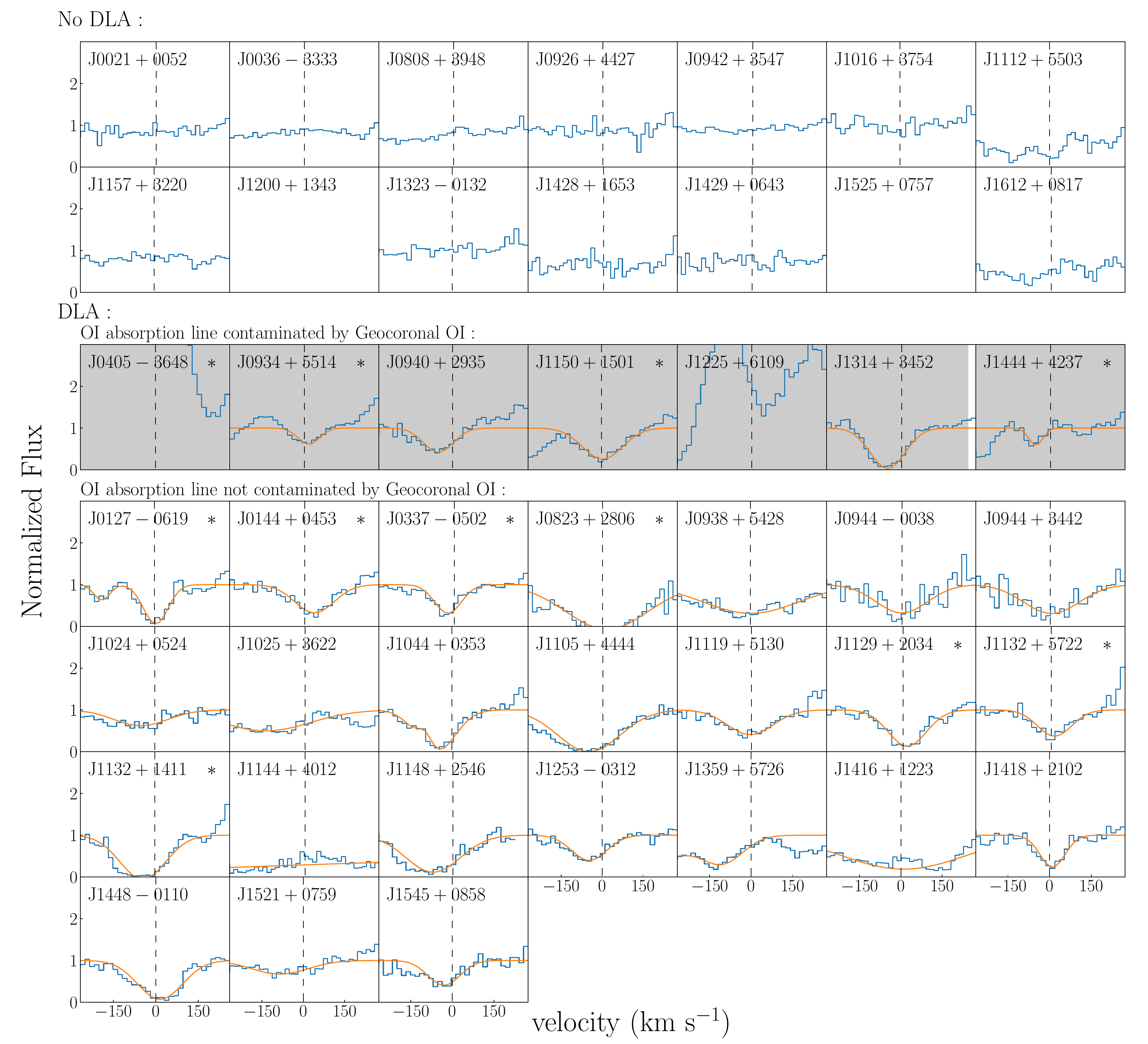}
   \caption{\label{fig:oispec} O \textsc{i} absorption profiles (blue) of CLASSY sample. The orange lines indicate best-fit 
     Gaussian profiles. The first two rows show O \textsc{i} absorption profiles of galaxies without DLA system. The third row 
     shows those O \textsc{i} profiles that might overlap with geocoronal O \textsc{i} emission and the rest shows others.
     The shaded regions show the wavelength range (1302.2--1307.5 \AA) that might be contaminated by geocoronal O\textsc{i} 
     emission lines. An example of geocoronal O \textsc{i} line can be seen in the spectrum of J1525+0757 in Fig. \ref{fig:allspec}.
     The O \textsc{i} absorption lines of 2 galaxies (J0405-3648 and J1225+6109) overlap with geocoronal O \textsc{i} line, thus, we 
     adopt their C \textsc{ii} velocities as DLA velocities. 
     The verticle dashed lines indicate the zero velocity.
     Asterisks mark galaxies which show a pure DLA system with no \lya\ emission.}
\end{figure*}

We use the narrow O \textsc{i} absorption lines to constrain the velocity of the DLA. Since the ionization potentials of O and H 
are very similar, we expect the O \textsc{i} to trace H \textsc{i} gas in the DLA absorber. Fig. \ref{fig:oispec} validates this expectation;
the DLA systems in CLASSY always associate with strong O \textsc{i} absorption. The only O \textsc{i} absorber without a DLA is J1112+5503, which shows a P-Cygni \lya\ profile still suggesting substantial H $\textsc{i}$ gas.

For optically thick O \textsc{i} absorption, the residual flux at the bottom of the fitted Gaussian profile 
determines the covering fraction of O \textsc{i} gas.
The O \textsc{i} optical depth can be measured following 
\begin{equation}
\tau = 0.318\bigg( \frac{\mathrm{N_\mathrm{OI}}}{10^{14}\ \mathrm{cm^{-2}}} \bigg) \bigg(\frac{\mathrm{30\ km\ s^{-1}}}{b}\bigg),
\end{equation}
where $N_\mathrm{OI}$ is the O \textsc{i} column density and $b$ the Doppler parameter \citep{Draine2011}. 
Since the H \textsc{i} column densities of DLAs in the CLASSY sample are $>$ 10$^{20}$ cm$^{-2}$, and the metallicities 
12+$\log\ (\mathrm{O/H})$ are $>7.5$, we find that the O \textsc{i} optical depths are $>10$, and the line is saturated.
We acknowledge that this argument relies on the assumption that O \textsc{i} is uniformly distributed in the neutral gas. 
If the intervening O \textsc{i} clouds have different velocities, the covering fraction derived from O \textsc{i} 
absorption would place a lower limit on the covering fraction of neutral gas \citep{Rivera-Thorsen2015}.

\begin{figure}[htb]
    \centering
    \includegraphics[width=3.3in]{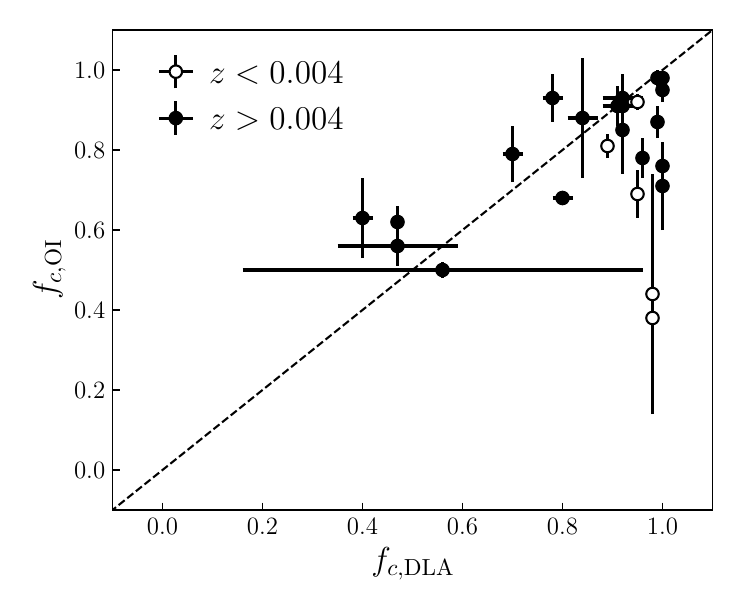}
    \caption{\label{fig:oi} DLA covering fraction versus O \textsc{i} covering fraction. We mark those O \textsc{i} 
      covering fractions which are possibly contaminated by geocoronal O \textsc{i} emission line (at 1302.2 -- 1307.5 \AA) 
      as open circles. Excluding those points, the observed correlation suggests O \textsc{i} covering fraction probe 
      H \textsc{i} covering fraction.}
\end{figure}

Fig. \ref{fig:oi} shows that the O \textsc{i} covering fraction is approximately equal to the covering fraction of the
DLA system. Table \ref{tab:1} collects the best-fit velocities and covering fractions for the  DLA and O \textsc{i} absorption.

\subsubsection{Notes on individual galaxies} \label{sec:notes}

\begin{itemize}

\item
The bottom of the DLA system is hidden under the \lya\ emissions in
J0938+5428, J1024+0524, J1416+1223, and J1521+0759 in Fig. \ref{fig:allspec},
so the residual flux in the damping trough is not directly constrained.
Since the blue wing of the damping profile is contaminated by 
metal absorption lines, the shape of the damping profile is poorly constrained.
Therefore, for these four galaxies, we adopt the O \textsc{i} covering 
fractions to be their DLA covering fractions. 

\item
The covering fractions of four galaxies (J0337-0502, 
J0405-3648, J1132+1411, J1448-0110) are fixed to be a constant measured visually but also in agreement with
their O \textsc{i} covering fractions. The Voigt profile fit for these four galaxies underestimates the covering fraction because
the CLASSY error spectra do not account for the small counts at the trough bottom which produce an 
asymmetric error  \citep{Cash1979}.

\item
The DLA systems of three galaxies (J0127-0619, J1044+0353, J1359+5726) were not fitted well by a single Voigt profile, and
we noticed that their O \textsc{i} absorption lines show a second component. Thus, we adopted two Voigt profiles and matched
their velocities and covering factors to those of the O \textsc{i} components.

\item
In J1105+4444, the \lya\ peak separation is exceptionally broad, $\sim1000$ \kms.
We suggest that the peaks are likely emitted by different regions within the COS aperture.
To test this conjecture, we inspected  HST/COS NUV acquisition image \citep[see Fig. 3 in][]{Berg2022}.
We found that J1105+4444 is not only an elongated object with multiple clumps, but the major axis
of these clumps is along the dispersion direction of the COS observation. Thus, their spatial offset 
in the aperture may cause an apparent velocity shift which is not physical.
This object is excluded in the following analysis.
For completeness, we note that the DLA fit for J1105+4444 failed when constrained by two O \textsc{i} components,
and we used a double-Voigt profile with free velocities. 

\item
The blue part of the J1525+0757 \lya\ line is likely a P-Cygni profile, so the impact of the geocoronal O \textsc{i} 
emission should be negligible. 

\item
We also exclude J1448-0110 from the \lya\ emission analysis
due to the low S/N.

\end{itemize}

\subsection{DLA system and Aperture Loss} \label{sec:aperture}

\begin{figure}
    \centering
    \includegraphics[width=3.1in]{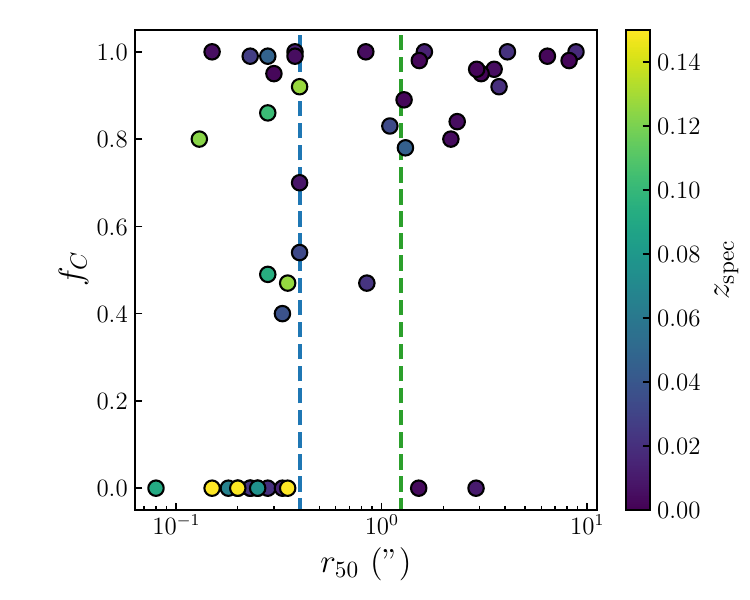}
    \caption{\label{fig:cf_ruv} The DLA covering fraction versus the angular UV half-light radius. 
    A covering fraction of 0 indicates no DLA.
    We color-coded the circles based on their spectroscopic redshift, thereby identifying variations in the physical scale of the aperture. 
    {We also use the blue and green vertical dashed lines to indicate the unvignetted radius (0\farcs4) and the total radius (1\farcs25) for the COS aperture}.
    Roughly a half of higher-redshift galaxies ($z>0.1$) show significant DLA system. 
    This fraction increases to the lower-redshift sample which tends to have a larger UV size (in arcsec) due to the projection effect.}
    
\end{figure}

The fraction of emitted \lya\ photons captured by the 2\farcs5 diameter COS aperture will vary dramatically among the targets because of their large range of distances.  For a typical target, the physical diameter of the aperture is roughly 700 pc, which is larger than the half-light radius of the UV continuum emission core but smaller than the Str\"{o}mgren radius of the nebula.\footnote{To estimate the volume ionized by the stars within the COS aperture we have used the extinction corrected the H$\alpha$ luminosity in the SDSS fiber and assumed a 
volume-average electron density of 1 cm$^{-3}$ and case~B recombination at $10^4$~K.}
The most distant CLASSY targets are LBAs at $z \approx 0.18$. Here, the COS aperture subtends
nearly 8 kpc, vignetting extended halo emission but likely capturing most of the \lya\ luminosity.
CLASSY also includes several very nearby galaxies, where the COS aperture subtends just a few hundred parsecs, 
and damped \lya\ absorption troughs are prominent in their spectra (see Fig. \ref{fig:1} and \ref{fig:specs}).

We suggest that the DLA detections indicate the \lya\ emission is scattered outside the COS aperture.
In support of this claim, Fig. \ref{fig:cf_ruv} shows that 14 of 16 galaxies with UV half-light radii larger 
than the COS aperture \citep{Xu2022} have a DLA in their COS spectrum\footnote{The sizes of compact galaxies with 
$r_\mathrm{50}<0\farcs4$ are measured using COS acquisition images and the sizes of extended galaxies are measured 
using SDSS $u$-band images.}.  The frequency of DLA detections is reduced among the galaxies with half-light radii 
smaller than the COS aperture. The Lyman Break Analog sample has the fewest DLA detections.
Although the physical size of the aperture grows with increasing redshift,
we do not find a one-to-one correlation between DLA detections and redshift.
For $z>0.1$ (yellow circles), the physical scale of COS aperture reaches $\sim5$ -- 10 kpc and is larger than 
the UV sizes of those galaxies; however, a large fraction (4/9) of their spectra still show significant DLA system.
The spectra of high-redshift galaxies observed with similar aperture size sometimes show DLA systems
as well \citep{Steidel2011,Reddy2016b,Reddy2022}. For example, \citet{Steidel2011} reveals a similar fraction 
(40/92) of galaxies at redshift $\sim2.2$ -- 3.2 which shows the DLA system.

We can gain some quantitative insight from the \lya\ imaging studies of \citet[][]{Hayes2014}.
LARS~9 and LARS~14 correspond to the CLASSY galaxies J0823+2806 at $z=0.04722$ and J0926+4427 at $z=0.18067$,
respectively. The closer 
galaxy shows a pure-DLA \lya\ profile, whereas the more distant one has a double-peaked emission profile with no DLA.
Inspection of Figure 1 in \cite{Hayes2014} shows the \lya\ emission comes from a shell around J0823+2806, whereas the
\lya\ emission from J0926+4427 is centrally concentrated.  In the latter example, the COS aperture includes roughly
60\% of the total \lya\ flux \citep[][Fig. 4]{Hayes2014}, showing that the \lya\ luminosity is significantly attenuated 
even in the case of no absorption.  For the DLA, the growth curve shows net emission only when the aperture is enlarged
to a diameter of 9.5~kpc, about four times larger than the COS aperture.  Galaxy-by-galaxy aperture corrections for 
\lya\ are not currently available, but these examples support our conjecture that the galaxies showing damped \lya\ 
absorption would show net emission in spectra obtained through larger apertures.

\subsection{\lya\ Measurements}  \label{sec:measure}

In this section, we present the measurements of the \lya\ emission properties. We measure the continuum and 
DLA-subtracted profiles. We will demonstrate in Sec. \ref{sec:tlac} that the \lya\ emission emerges from
holes between the DLA clouds. Since these parts of the line profile have different origins, they must be
separated to obtain a meaningful analysis.

\subsubsection{\lya\ Kinematics}

We measure the blue peak velocity, $v^\mathrm{Ly\alpha}_\mathrm{blue}$, and red peak velocity, 
$v^\mathrm{Ly\alpha}_\mathrm{red}$, as the position of the local maximum in the \lya\ emission line at 
velocity $v <0$ and $v > 0$, respectively, relative to the systemic velocity. The minimum between the two 
emission peaks defines the \lya\ trough velocity, $v^\mathrm{Ly\alpha}_\mathrm{trough}$. We define the 
peak separation as $\Delta v_\mathrm{Ly\alpha} = v^\mathrm{Ly\alpha}_\mathrm{red} - v^\mathrm{Ly\alpha}_\mathrm{blue}$.

\subsubsection{\lya\ Fluxes and \lya\ Escape Fraction}

For double-peaked \lya\ profiles, we measure the fluxes of the blue and red components
by integrating to the velocity of the \lya\ trough between the components.
We also measure the asymmetry parameter of the red peak of \lya\ emission, defined as $A_f = (\int^\infty_{\lambda^\mathrm{red}_\mathrm{peak}} f_\lambda d \lambda)/(\int^{\lambda^\mathrm{red}_\mathrm{peak}}_{\lambda_\mathrm{trough}} f_\lambda d \lambda)$, where $\lambda^{red}_\mathrm{peak}$ is the wavelength of red peak and $\lambda_\mathrm{trough}$ the wavelength of the trough \citep{Rhoads2003,Kakiichi2021}.
The total \lya\ fluxes are measured by integrating flux between the wavelengths where the profile meets zero flux,
including the central dip in double peaked profiles and the negative flux in P Cygni profiles.
We convert the total \lya\ fluxes to luminosity using luminosity distance from Table 
\ref{tab:1}, which is corrected for the cosmic flow.
The rest-frame equivalent widths, EWs,  are computed using the \lya\ spectra and the {\it total stellar continuum},
 $\mathrm{EW}(\mathrm{Ly\alpha}) = \int F_\mathrm{Ly\alpha}(\lambda)/F_\mathrm{cont}(\lambda)\ d \lambda / (1+z)$.

We estimate the \lya\ escape fractions $f^{\mathrm{Ly\alpha}}_\mathrm{esc}$ based on intrinsic \lya\ fluxes inferred 
through dust-corrected H$\alpha$ (or H$\beta$) fluxes assuming a Case-B recombination \citep{Brocklehurst1971}: 
$f^{\mathrm{Ly\alpha}}_\mathrm{esc} = F_\mathrm{Ly\alpha}/(8.7 \times F_\mathrm{H\alpha})$\footnote{We adopt the factor of 
8.7 to be consistent with previous works. It corresponds to a temperature of 10,000 K and an electron density of 
$\sim300$ cm$^-3$. }. \citet{Mingozzi2022} have measured the H$\alpha$ and H$\beta$ fluxes using optical spectra from 
SDSS, MUSE, KCWI, MMT, and VIMOS. Since the UV spectra and optical spectra are obtained via different instruments with 
different aperture sizes, a scaling factor between UV spectra and optical spectra is needed to correct the different 
aperture losses. \citet{Mingozzi2022} measured the scaling factor by matching the optical spectra to the extrapolation 
of the best-fit UV stellar continuum model (see their Appendix A). The scaling factors for most objects approximate the 
ratio between apertures of different instruments but are not exactly the same because some other effects may also cause 
the flux offsets such as the vignetting.  For example, the median of the scaling factor for SDSS spectra is $\sim0.79$ 
and the aperture size ratio is $(2\farcs5)^2/(3\farcs)^2\sim 0.69$. We refer readers to \citet{Mingozzi2022} for more 
details.  In this work, we adopt the corrected H$\alpha$ fluxes. Since the H$\alpha$ for J0934+5514 and J1253-0312 are 
unavailable, we convert their H$\beta$ fluxes to H$\alpha$ fluxes using a factor of 2.86, by assuming Case-B 
recombination with a temperature of 10,000 K and electron density of 100 cm$^{-3}$.

\subsubsection{\lya\ trough flux density}

The flux density at the \lya\ trough velocity defines the trough flux density, $f_\mathrm{trough}$. The $f_\mathrm{trough}$ of J0926+4427 and J1429+0643 have also been measured in \citet{Gazagnes2020} based on the spectra obtained by 
HST/COS G140L with a resolution of 1,500. \citet{Gazagnes2020} measured the \lya\ trough flux density based on the 
continuum-unsubtracted spectra but our measurements are based on the continuum-subtracted \lya\ spectrum. Thus, our 
$F_\mathrm{trough}/F_\mathrm{cont}$ should be lower by 1 than those in \citet{Gazagnes2020}.  
{Here $F_\mathrm{cont}$ is the flux density of \textit{total stellar continuum} estimated from STARBURST99.}

However, accounting for this difference, the $F_\mathrm{trough}/F_\mathrm{cont}$ of J0926+4427 and J1429+0643 in our measurements are still lower. 
Particularly for J0926+4427, we do not see the net residual \lya\ trough flux density (i.e., $F_\mathrm{trough}<0$). 
This is because the CLASSY spectra have much higher resolution, ranging from $\sim$ 2,200 to 15,000 with a median of 
5,000 \citep[measured from MW absorption line;][]{Berg2022}. The high-resolution spectra resolve the small structures at 
the central trough, which were smoothed due to the lower resolution in \citet{Gazagnes2020}. We note that the resolution 
around \lya\ emission line might be lower than the resolution for the continuum as \lya\ emission often subtends a 
larger solid angle than the continuum.

\subsubsection{Aperture Effects on \lya\ Measurements}

The COS aperture, therefore, attenuates the \lya\ emission relatively more than the UV emission due 
to the scattering of \lya\ photons.  Thus, even though \lya\, H$\alpha$,  and the UV continuum are measured 
locally in the same aperture, we expect \lyaesc\ and the \lya\ EW to be underestimated. In our example of
J0823+2806, see discussion in Sec. 2.2, the attenuation is severe because most of the \lya\ emission
is scattered outside the COS aperture. If scattering outside the COS aperture produces the large fraction 
of DLA systems in CLASSY, then the \lyaesc\ and \lya\ EW of these galaxies are significantly underestimated.

A more subtle bias that we will examine is the possibility that this vignetting modifies the shape of
the \lya\ emission-line profile. \citep{Zheng2010} predicted that the blue-to-red peak ratio (hereafter B/R ratio) 
would increase with increasing impact parameters because the front-scattered \lya\ photons (blue peak) are closer 
to the resonance center of the outflowing gas and thus, tend to be scattered to larger impact parameters, compared 
with the back-scattered \lya\ photons (red peak).  Integral field spectroscopy confirms this trend in a few \lya\
halos \citep{Erb2018,Erb2022}.  Another possible interpretation is that the average projected outflow velocity 
decreases with the increasing radius \citep{Li2022b}.

\section{Radiative Transfer Modeling} \label{sec:tlac}

The high fraction of DLA systems in CLASSY was not anticipated. More surprising,
however, was the discovery of double-peaked \lya\ emission in the bottom 
of the broad absorption profiles. We have drawn attention to an important property of 
these DLA systems; the high-column density gas only partially covers the continuum 
source (see Sec. \ref{sec:2-2}). The residual intensity in the continuum-normalized spectra indicates the 
uncovered fraction of the continuum emission (within the COS aperture). 
{In this section, we explore what continuum is linked to the net \lya\ emission profile, the total continuum or the uncovered fraction.}

{ Specifically, we utilize the shell model to fit the \lya\ profiles which are normalized by the STARBURST99 continuum or the DLA continuum (hereafter normalized \lya\ profile\footnote{ To avoid confusion, we define the emergent \lya\ profile as the observed \lya\ emission line with the underlying continuum and DLA, the net \lya\ profile as the \lya\ profile after removing the underlying continuum and DLA, and the normalized \lya\ profile as the net \lya\ profile after being normalized by the underlying continuum.
}).}
The model \lya\ line profile is computed using the Monte Carlo radiative 
transfer code \tlac\ \citep{Gronke2014,Gronke2015}. 
This technique has been used to successfully reproduce the observed profiles of \lya\ 
emission lines \citep[e.g.,][]{Yang2017,Orlitova2018,Gurung-Lopez2022}.
The shell model can produce \lya\ emission when the dust optical depth 
is low, or a DLA system when  there is a substantial neutral hydrogen column 
with a moderate dust optical depth \citep[e.g., see Fig. 1 of][]{Gronke2015}.
However, the homogeneous shell model cannot produce a \lya\ emission line in the DLA 
trough,  i.e., the Abs+Em profiles seen in our CLASSY sample (see Fig .\ref{fig:specs}). 
The \lya\ emission line requires low-$N_\mathrm{HI}$ channels (with low 
dust optical depth), which contradicts the presence of damped absorption 
which requires very high column density. This requires a non-uniform shell 
model to describe the multi-component ISM. Although radiative transfer 
through clumpy media has been explored \citep{Hansen2006,Verhamme2015,Gronke2016,Gronke2017,Li2022}, 
a non-uniform shell is beyond the scope of this work. 

Here, we adopt an alternative method to fit the \lya\ profiles of the CLASSY sample.
We fit and remove the DLA system to extract the net \lya\ profile, as described in Sec. \ref{sec:2-2}, 
and then we fit the \tlac\ model to the normalized \lya\ profile using two different approaches described in Sec. \ref{sec:profile}.
The variant fitting results could reveal the physical links between the gasses probed by \lya\ emission and DLA absorption, as discussed in Sec. \ref{sec:imp}.

Some properties of the shell model have been mapped to those of more realistic 
outflows \citep{Gronke2016,Li2022}. However, the shell-model parameters are found to
have systematic discrepancies with independently measured outflow 
velocities and the velocity dispersion of the intrinsic line profile \citep[e.g.,][]{Orlitova2018}. 
To understand the origin of the discrepancies, we perform more fittings with constrained redshift priors and compare it with the previous results in Sec. \ref{sec:redshift}.

\subsection{Shell Model}

\tlac\ computes \lya\ resonant scattering through a uniform, expanding shell, 
which is composed of  dust and neutral hydrogen gas.   
The shell model used in \tlac\ has 6 free parameters, including 2 parameters 
for the central radiation source: intrinsic line width $\sigma_i$ and intrinsic 
equivalent width EW$_i$, and 4 parameters for the expanding shell: neutral 
hydrogen column density $N_\mathrm{H\textsc{i}}$, dust optical depth $\tau_\mathrm{d}$, 
shell velocity $v_\mathrm{exp}$, and temperature $T$. 
In addition to these six parameters, a redshift parameter $z_{tlac}$ is 
also applied to shift the rest-frame of the \lya\ profile relative to the systemic redshift of the galaxy.

The \lya\ photons and underlying continuum photons are generated from the 
central source with an intrinsic width of $\sigma_i$ and intrinsic equivalent 
width of EW$_i$. The photon is then emitted into the H \textsc{i} shell with 
a random direction and travels a distance before being absorbed or resonantly scattered.
The distance that a photon can travel is calculated using the total optical 
depth of dust $\tau_\mathrm{d}$ and neutral hydrogen $N_\mathrm{H\textsc{i}}$ 
in the expanding shell with velocity $v_\mathrm{exp}$ and temperature $T$.
The probability that a photon is resonantly scattered or absorbed at a 
specific position is estimated by comparing the optical depth of neutral 
hydrogen with the total optical depth at that position. If the photon is 
resonantly scattered, a new direction and a new frequency are drawn from 
the proper phase function and the frequency redistribution function, respectively. 
The previous steps are repeated until the photon escapes from the simulation 
domain or is absorbed by the dust. If the photons escape from the simulation 
domain, their frequency, and other properties are recorded. This simulation 
has been run thousands of times over a discrete grid of ($v_\mathrm{exp}$, 
$N_\mathrm{H\textsc{i}}$, $T$) and then been post-processed with a continuous 
grid of ($\sigma_i$, $\tau_\mathrm{d}$, EW$_i$) to generate the simulated 
\lya\ spectra for different parameter values.
To fit the observed \lya\ spectrum, a likelihood function is constructed 
based on the noise and flux spectra. The best-fit spectrum is derived by 
maximizing the likelihood function using the Markov Chain Monte Carlo (MCMC) 
and nonlinear optimization methods. 

We highlight the importance of the intrinsic \lya\ equivalent width (EW$_i$) in the shell model, 
a parameter excluded by studies that fit the continuum-subtracted \lya\ line 
profiles, because the continuum photons are also involved in resonant scattering and 
can dominate the normalized \lya\ profile for low-EW$_i$ cases.

\subsection{Profile fitting} \label{sec:profile}

Our profile-fitting approach draws attention to ambiguity about the appropriate continuum level for normalization. 
When a DLA system is present in the spectrum, the underlying continuum 
could be the total stellar continuum (red lines in panel a of Fig. \ref{fig:specs}), 
and thus, the normalized spectrum is:
\begin{equation} \label{eq:1}
    I^\mathrm{EW}_{\lambda} = (f^\mathrm{Ly\alpha}_{\lambda} + f^\mathrm{cont}_{\lambda}) / (f^\mathrm{cont}_{\lambda}),
\end{equation}
or the residual stellar continuum in the bottom of DLA system (red line in panel b) and thus, the normalized spectrum is:
\begin{equation} \label{eq:2}
    I^\mathrm{EW}_{\lambda} = (f^\mathrm{Ly\alpha}_{\lambda} + f^\mathrm{cont}_{\lambda} \times (1-f_C)) / (f^\mathrm{cont}_{\lambda} \times (1-f_C)),
\end{equation}
where $f^\mathrm{Ly\alpha}_{\lambda}$ is the \lya\ emission line, $f^\mathrm{cont}_{\lambda}$ 
the best-fit total stellar continuum (see Sec. \ref{sec:2-1}), $f_C$ the covering fraction of DLA system.
The choice of the underlying continuum  will change the equivalent width 
of the \lya\ line and, thus, the contribution of continuum photons on the emergent \lya\ profile.
Here we perform profile fittings, assuming each continuum level in turn, and then discuss the results.
We present the best-fit spectra in Append. \ref{app:a}. 
We also present the best-fit model parameters of the second profile fitting in Table. \ref{tab:tlac2}.

The fitted parameters somewhat degenerate with each other. 
For example, in the case of outflowing shells, \cite{Li2022} demonstrate 
that various combinations of shell velocity, column density, temperature, 
and redshift can produce very similar line profiles, for example, ($v_\mathrm{exp}$, 
$\log N_\mathrm{HI}$, $\log T$, $0$), and $\sim$ (2$v_\mathrm{exp}$, $\log 
N_\mathrm{HI}-0.5$ dex, $\log T+1$ dex, $\Delta v$). Nonetheless, the spectra 
generated by these parameters show clear differences at the red peak and 
our high-S/N spectra should be able to distinguish between the degeneracies.

\subsubsection{First attempt: total stellar continuum} \label{sec:fs}
Overall, the quality of the first fitting using the total stellar continuum (Eq. \ref{eq:1}) is quite good, given the simplicity of the model.
However, in a subset of spectra, the results are unsatisfactory,
especially J0938+5428, J0944+3442, J1044+0353, J1119+5130, J1144+4012, J1416+1223, J1521+0759, as presented in Fig. \ref{fig:tlacfail}, of which the best-fit spectra show a very sharp dip around zero velocity compared to the observed \lya\ profile.

Looking at their original spectra (see. Fig. \ref{fig:specs}), we find that all these poorly fitted \lya\ profiles correspond to spectra that show significant DLA systems compared with the successful sample. 
This result motivated us to investigate whether the sharp dips might be caused by an inappropriate underlying continuum, which underestimated the \lya\ EW spectra $I^\mathrm{EW}_\lambda$. 
Thus, we performed a second profile fitting using the residual stellar continuum as described by Eq. \ref{eq:2}.

\begin{figure}
    \centering
    \includegraphics[width=3.1in]{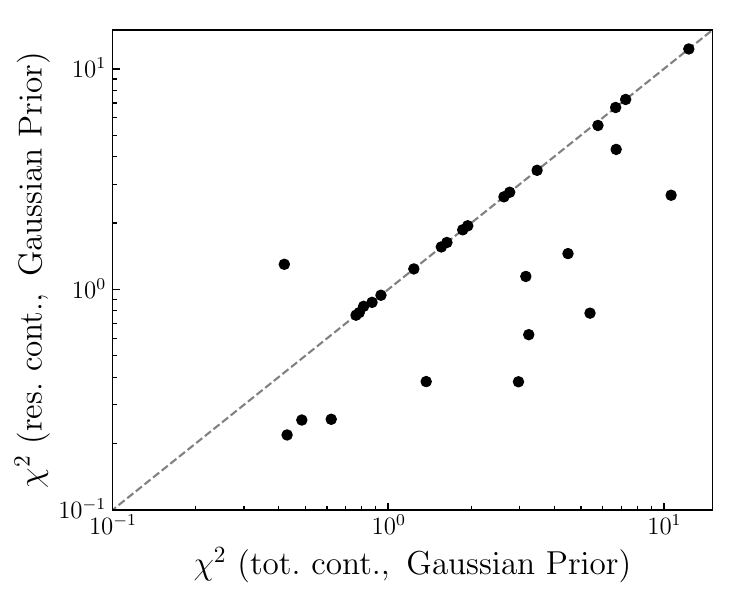}
    \caption{\label{fig:chi3}  Reduced $\chi^2$ values of best-fit \lya\ profiles derived by the first attempt of profile fitting (using total stellar continuum and Gaussian redshift prior) and the second attempt of profile fitting (using residual stellar continuum and Gaussian redshift prior). 
    We notice that except for one object (J0944-0038), the fitting results for objects with DLA are significantly improved if adopting the residual continua which are corrected for the DLA covering fraction (1-$f_c$). For objects without DLA, the corrected continuum equals to the total stellar continuum. Thus, the reduced $\chi^2$ does not change.
    We note the reduced $\chi^2$ of some galaxies (J0944+3442, J1044+0353, J1144+4012, J1148+2546, J1416+1223) are much smaller than unity. This might be due to their error spectra are overestimated (comparing to their flux fluctuation, see Fig. \ref{fig:lya}).}
\end{figure}

\subsubsection{Second attempt: residual stellar continuum}  \label{sec:secondtlac}
In Fig. \ref{fig:tlac}, we present the best-fit \tlac\ models for \lya\ profiles normalized by the residual stellar continua ($1-f_C$).
For the {unsatisfactory sample} in the first attempt, normalizing the \lya\ spectra by the residual stellar continuum significantly improved the best-fit results. The sharp dips seen in the models of Sec. \ref{sec:fs} no longer exist in the new model spectra. 
In Fig. \ref{fig:chi3}, we compare the reduced $\chi^2$ for the first and second attempts. Clearly, most results are significantly improved if adopting the normalization of the residual stellar continuum.
Thus, we can conclude that the dip was caused by an inappropriate continuum level. 
In further analysis, the first attempt of fitting will not be considered.

For the galaxies without DLA systems, $f_C = 0$, the residual continuum rises to the level of the total continuum. It is therefore not surprising that every  \lya\ profile is successfully fitted when the residual continuum is used. 
We conclude that the residual continuum, $1 - f_C$, is the more physical normalization for the emergent \lya\ emission line.
In other words, the DLA covering fraction $f_C$ gives a good indication of the fraction of the intrinsic \lya\ emission that is blocked by the high-column density clouds.

\subsubsection{Implication: Scattering Outside COS Aperture Reveals Low-$N_\mathrm{HI}$ Channels} \label{sec:imp}

We have shown that successful radiative transfer modeling of CLASSY \lya\ 
spectra, in the context of the shell model, requires: (1)  separating the 
\lya\ emission profile from the DLA system, and (2) normalizing the \lya\ 
emission by the leaked continuum, i.e. the residual flux in the DLA system. 
This approach divides the COS aperture into two groups of sightlines, hereafter
channels, distinguished by their column density.  In the schematic picture of a 
thin shell, these two channels represent clouds and the intercloud medium.  More 
generally, for the targets with $C_f > 0$, the \lya\ photons entering the 
high-$N_\mathrm{HI}$ channel do not emerge from the galaxy at radii within the 
COS aperture. If they did, then the  best continuum normalization would be the 
total stellar continuum, which is inconsistent with our fitting results. 

The \lya\ photons entering the high-$N_\mathrm{HI}$ channel must be scattered 
to radii larger than the COS aperture before they escape. The alternative
is that they are absorbed by dust grains which seems less likely for two reasons.  
Most CLASSY galaxies whose COS spectra detect DLA systems have low metallicities
and are relatively dust poor. In addition, substantial amounts of dust in the 
scattering clouds would boost the transmitted \lya\ equivalent width \citep{Hansen2006},
but we do not measure unusually large EW.  

When the \lya\ emission is separated from the DLA system, what do the shell model parameters 
fitted to the emission component represent?  Perhaps the line photons entering low-$N_\mathrm{HI}$ channels 
scatter off both the low $N_\mathrm{HI}$ clouds and the walls of the DLA channels. In the limit
of no intercloud medium, the kinematics of the dense clouds would determine the shape of the \lya\ 
line profile \citep{Neufeld1990,Hansen2006}, so we might expect the kinematics of both the
low- and high-$N_\mathrm{HI}$ channels to impact the \lya\ profile. If \lya\ photons entering
DLA sightlines are scattered outside the spectroscopic aperture, then vignetted apertures may
have one advantage, namely providing a direct view of the properties in low-$N_\mathrm{HI}$ 
channels.

\subsection{Discrepancies between the Shell Model and Observations} \label{sec:redshift}

\begin{figure}
    \centering
    \includegraphics[width=3.1in]{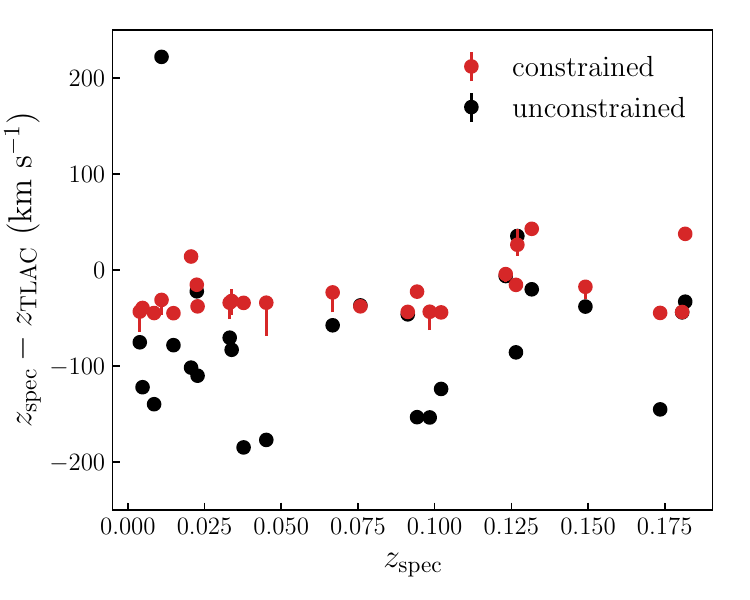}
    \caption{\label{fig:redshift}  Redshift difference between \tlac\ redshift $z_{tlac}$ and spectroscopic redshift $z_\mathrm{spec}$. 
    The black circles indicate the results of the second attempt of profile fitting with a Gaussian prior with $\sigma =120$ km s$^{-1}$ for redshift (unconstrained fitting).
    The red circles indicate the results of the third attempt of profile fitting with a Tophat prior with a width of $\pm44$ km s$^{-1}$ for redshift (constrained fitting). }
\end{figure}

We have presented that whether the shell model can well-fit the observed \lya\ profile is critical to infer the ISM properties.
However, the three discrepancies reported in \citet[][]{Orlitova2018} (see also Sec. \ref{sec:intro}) might suggest a limited physical meaning of the model parameters.
These discrepancies are also observed in the CLASSY sample with high significance, as shown in Fig. \ref{fig:redshift} (black circles).
The best-fit redshifts are always larger by 0 -- 200 \kms\ than the spectroscopic redshift, consistent with \citet{Orlitova2018}.
One possible origin of the discrepancies is the degeneracies between the model parameters, suggested in \citet{Li2022}. 
To test this scenario and gain more insight into the discrepancies, we perform a third profile fitting following \citet{Li2022} which constrains the range of redshift parameters to break the degeneracies. 

\subsubsection{Third attempt: constraining the redshift}

\begin{figure*}
    \centering
    \includegraphics[width=1\textwidth]{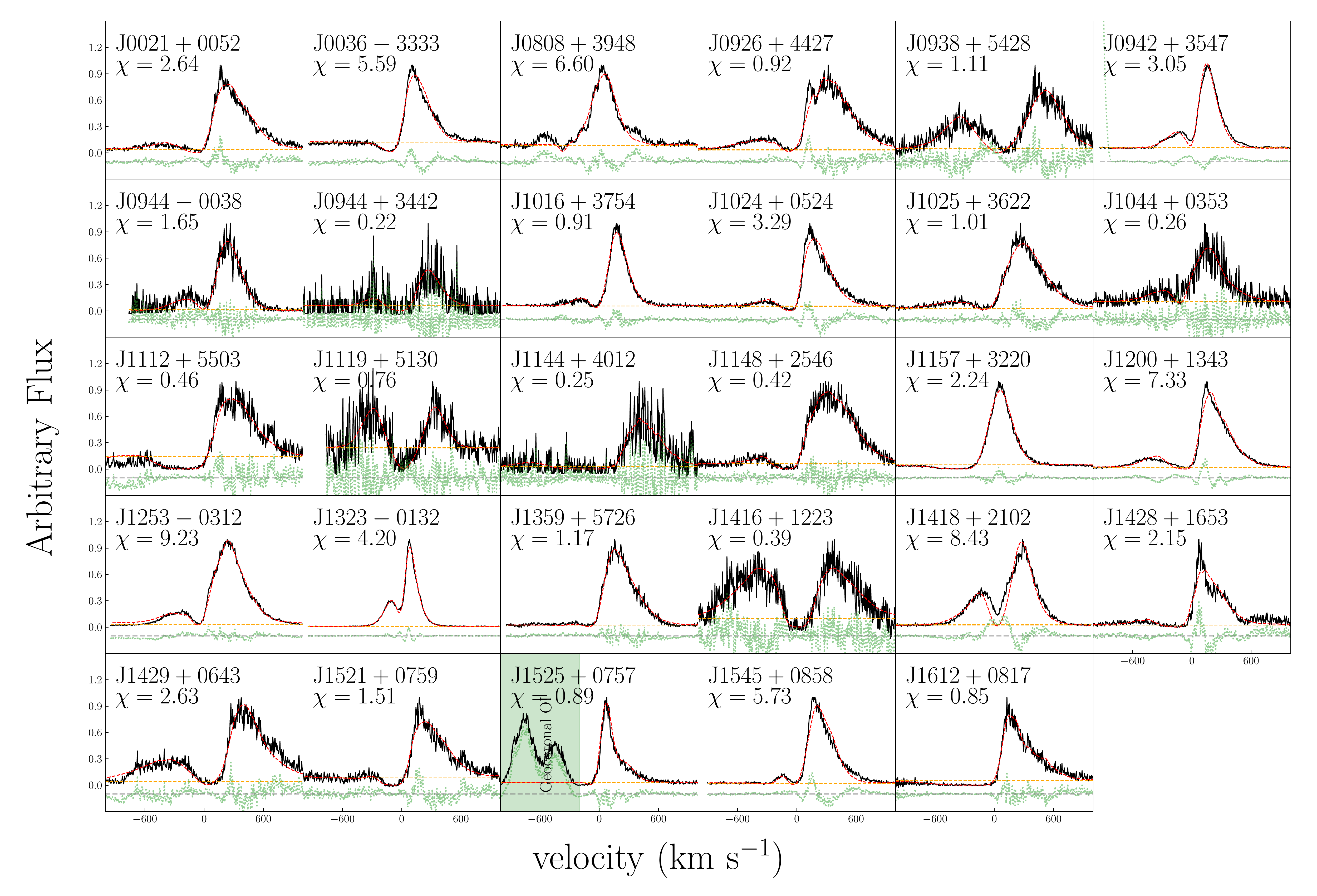}
    \caption{\label{fig:tlac_fixedz} Best-fit \lya\ spectra (red) for 29 CLASSY galaxies using the residual stellar continuum and a Tophat redshift prior with a width of $\pm$48 \kms\ (third attempt). The spectra are normalized by the peak flux and the orange dashed lines indicate the continuum level for each object.
    We use the green lines to show the residual and manually shift it by $-0.1$ for better illustration. The gray dashed lines indicate the zero level of the residual.}
\end{figure*}

The CLASSY redshifts derived from  UV nebular lines agree well with those derived from optical lines; 
the standard deviation of velocity difference is $\sim22$ km s$^{-1}$ \citep{Mingozzi2022}.  A spatial offset between the scattered \lya\ 
emission and the optically-thin emission lines would introduce an additional redshift error if, and only if, the offset were along
the dispersion axis of the spectrograph.  Based on the radius of the unvignetted aperture (0\farcs4), non-perfect alignment
could shift the \lya\ wavelength scale by as much as $\pm 44$ km s$^{-1}$.   For a redshift-constrained fit, we adopted a narrow Tophat 
probability distribution of width  $\pm$  44 km s$^{-1}$ as the prior on redshift. 
The best-fit spectra are presented in Fig. \ref{fig:tlac_fixedz}.
In contrast, in the second attempt at profile fitting (Sec. \ref{sec:secondtlac}),
we adopted a Gaussian prior on redshift, and this broad distribution with $\sigma(z_{tlac}) = 120$ \kms\ serves as the unconstrained fit.

\begin{figure}
    \centering
    \includegraphics[width=3.1in]{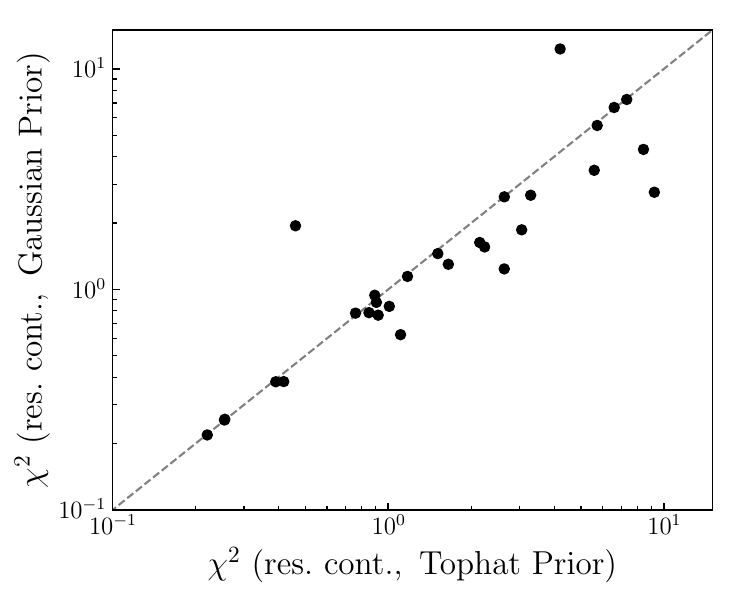}
    \caption{\label{fig:chi2} Reduced $\chi^2$ values of best-fit \lya\ profiles derived by the second attempt (using the residual continuum and a Gaussian redshift prior) and the third attempt (using the residual continuum and a Tophat redshift prior). The second attempt adopts a Gaussian probability distribution with $\sigma$ of 120 km s$^{-1}$ as prior and the third attempt adopts a Tophat probability distribution with a width of $\pm$48 km s$^{-1}$ as prior.}
\end{figure}

\subsubsection{Can constrained fitting alleviate the discrepancies?}

The redshift differences are apparently improved when adopting the constrained fitting (red circles in Fig. \ref{fig:redshift}).
This is because the constrained redshift prior sets a hard limit of the difference to be 44 \kms.
On the other hand, comparing the best-fit \lya\ profiles of constrained fitting (see Fig. \ref{fig:tlac_fixedz}) with those of unconstrained fitting (see Fig. \ref{fig:tlac}), it is hard to distinguish the difference between them by visual inspection.
We compare the reduced $\chi^2$ of two profile fittings in Fig. \ref{fig:chi2}, which shows that the results of constrained profile fitting are slightly worse than those of unconstrained profile fitting, but still acceptable\footnote{
We notice two best-fit spectra (J1112+5503, J1323-0132) of constrained profile fitting are improved compared with the unconstrained profile fitting. This might be because the unconstrained profile fitting for these two objects is trapped in a local maximum of the likelihood.}.
Thus, our test confirms that adopting a constrained redshift prior for profile fitting can somewhat alleviate the redshift discrepancy 
observed in previous works \citep{Yang2016,Orlitova2018}. 
We present the best-fit parameters of the third profile fitting in Table \ref{tab:tlac}.

However, the best-fit redshift remains systematically larger than the spectroscopic redshift as most of the red circles are still below zero velocity. This indicates that the constrained fitting does not fully resolve the observed discrepancies.
We return to this topic in Sec. \ref{sec:vtrough}, where we combine the comparison between the \tlac\ shell velocity and spectral measurements of outflow velocity.

We do not discuss the line width discrepancy in this work because a clumpy model is needed to resolve this discrepancy.
By comparing the \lya\ profiles generated by the uniform shell model and a clumpy model, \citet{Gronke2017} and \citet{Li2022} find that a larger line width is always required for the shell model to produce a similar \lya\ profile as the clumpy model. The intrinsic difference between the two models is that the clumpy model includes the turbulent velocity dispersion of the clumps while the shell model does not. 
Thus, the line width of the shell model needs to be artificially broadened to compensate for the omission of turbulent motion in the shell model.

\section{Properties of the Neutral ISM} \label{sec:lyaprop}

In this section, we discuss the relationship between the H \textsc{i} column 
densities inferred from the \lya\ absorption and emission components of 
the line profile. We then discuss indirect evidence LyC leakage. Finally, we
return to the problem of why the shell model systematically mispresents outflow 
properties, finding that the problem lies in the spectroscopic aperture.

\subsection{Structure of the ISM in CLASSY galaxies} \label{sec:ismstruct}

In Sec. \ref{sec:tlac}, we found evidence that the neutral ISM consists 
of several components with different column densities. The DLA system requires 
high-$N_\mathrm{HI}$ clouds with $N_\mathrm{HI}>10^{20}$ cm$^{-2}$. In 
Sec. \ref{sec:profile}, the \tlac\ fitting revealed that the observed \lya\ 
emission line requires low-$N_\mathrm{HI}$ holes with $10^{18}<N_\mathrm{HI}<10^{20}$ cm$^{-2}$.
Combining these two results demonstrates the existence of sightlines with different H 
\textsc{i} column densities in individual galaxies.  We have argued that
scattering of a significant fraction of the \lya\ photons out of the COS aperture 
makes the high-$N_\mathrm{HI}$ channels visible via \lya\ absorption, whereas their 
damping profiles would be filled in by scattered emission in spectra obtained through
larger apertures.  Apparently, the \lya\ halos of many CLASSY galaxies are much larger than
the COS aperture, and the scattering of \lya\ photons out of the COS aperture
provides a unique opportunity to describe the structure of the neutral ISM, as we
show here.

New insight into how LyC radiation escapes from local analogs of EoR galaxies
may be obtained by comparing the structure of the ISM in hydrodynamical simulations 
to the column density distribution we derive. Feedback from massive stars is widely 
believed to shape the pathways for LyC escape, but the mechanism is debated.
For example, \citet{Ma2020} argue that positive feedback, essentially propagating 
star formation triggered by the mechanical feedback from massive stars, is essential
to shift the production of LyC radiation away from the densest region of a galaxy.
In contrast, in H \textsc{ii} regions too young to have produced supernova explosions,
turbulence driven by ionization fronts may open channels for LyC escape \citep{Kakiichi2021}.

One difference between these two mechanisms is the size of the channels. Whereas the channels 
opened by turbulence are individually small, the low-$N_\mathrm{HI}$ bubbles driven by mechanical 
feedback have scales reaching hundreds of pc \citep{Ma2020}. Thus, the size of the channels provides
insight of particular interest for understanding the escape pathways.

In this section, we adopt the column density estimation from the third profile fitting (see Sec. \ref{sec:redshift}), because it incorporates more constraints from the observation. 
However, adopting the estimation from the second profile fitting does not change the conclusion of this section.

\subsubsection{Column Density Distribution of Neutral ISM} \label{sec:nhdist}

\begin{figure}[tbp]
    \centering
    \includegraphics[width=3.in]{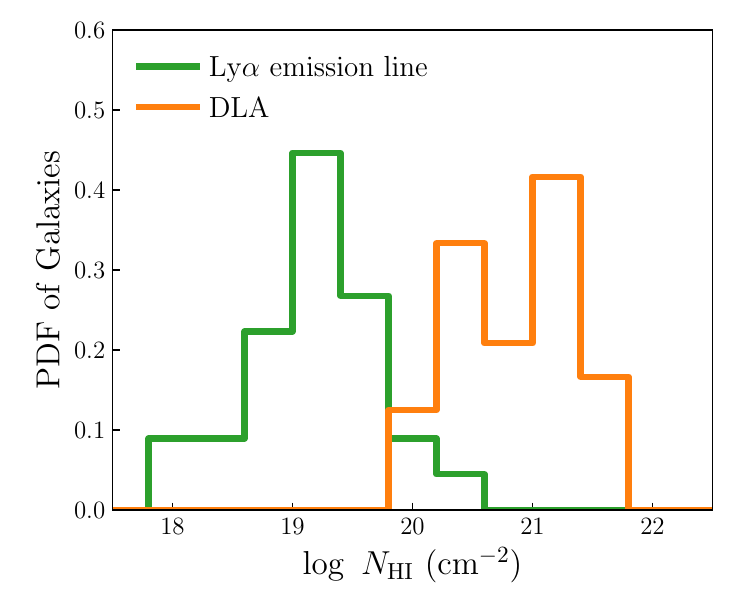}\\
    \includegraphics[width=3.in]{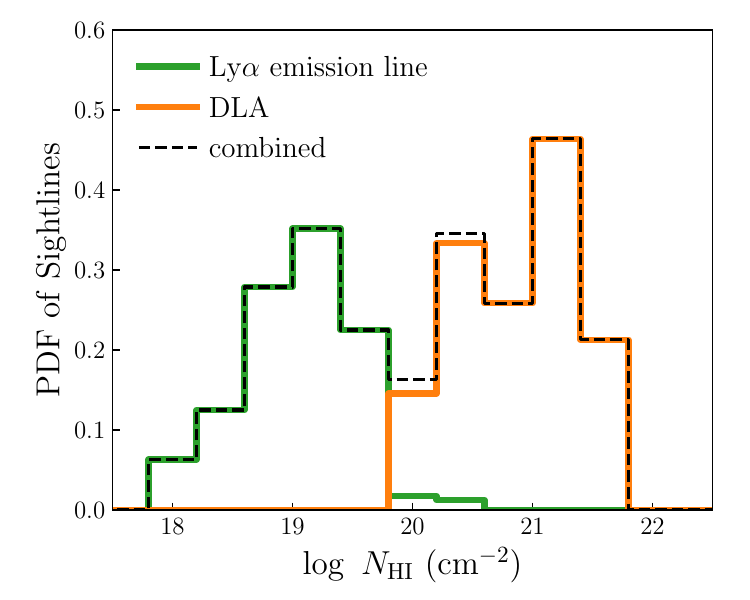}\\
    \includegraphics[width=3.in]{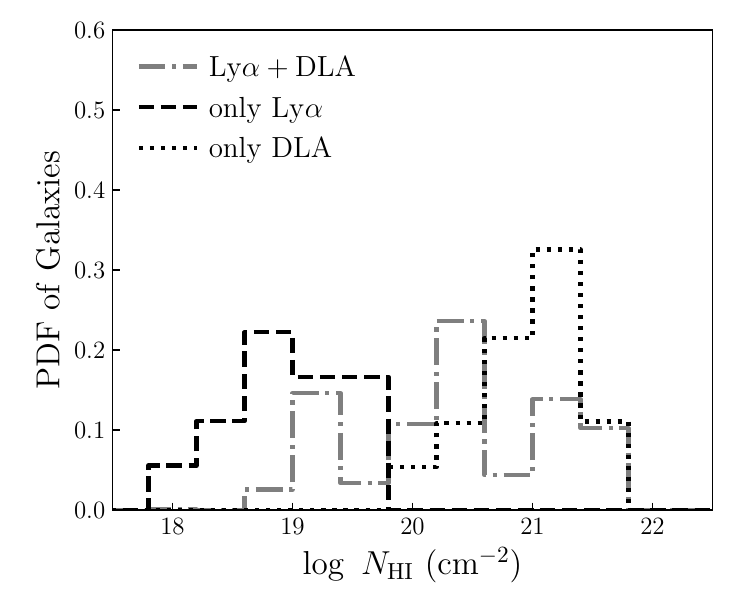}\\
    \caption{\label{fig:nhdist} The probability distribution functions 
of column densities derived by the DLA fitting (orange line) and the \lya\ 
profile fitting (green line).    In the top panel, the histograms are normalized 
by the total numbers of galaxies showing DLA system (31) or \lya\ emission line (28
after excluding the J0808+3948 AGN), respectively.    In the middle panel, 
the histogram is weighted by the covering fraction of low-$N_\mathrm{HI}$ 
and high-$N_\mathrm{HI}$ paths: $(1-f_C)$ and $f_C$, respectively. The black
dashed line represents the combined distribution over the $\sim 10^{18}$ --$10^{22}$ cm$^{-2}$ 
range for CLASSY galaxies. In the bottom panel, we present the distribution 
of column densities for three types of \lya\ profiles. The black dashed 
line indicates the galaxies with only \lya\ emission, the gray dash-dot 
line indicates the galaxies with \lya\ emission in the bottom of DLA system, 
and the dotted line indicates the galaxies with only DLA systems. The resulting 
distribution remains bimodal regardless of the details of the weighting and subsample.}
\end{figure}

Fig. \ref{fig:nhdist} compares the distribution of low-$N_\mathrm{HI}$ 
channels returned by the shell model fits and the high-$N_\mathrm{HI}$
column densities measured from the damping wing absorption.
In the top panel, the histograms are normalized by the total number of 
galaxies showing a DLA system or \lya\ emission line, respectively.
Their combined distribution has two peaks: one at $N_\mathrm{HI} 
\approx 10^{19}$ cm$^{-2}$ which represents the path of the escaping 
\lya\ photons\footnote{{If adopting the $N_\mathrm{HI}$ from the second profile fitting, the peak shifts to lower by 0.4 dex.}}
, and a second peak representing the typical DLA system
at $N_\mathrm{HI} \approx 10^{21}$ cm$^{-2}$. We recognize that the 
DLA sightlines and the pathways of the scattered \lya\ emission 
select specific channels through a turbulent, multiphase ISM. 
Nonetheless, their combined distribution may represent a large fraction
of all sightlines because we found that these components cover complementary 
fractions of the UV continuum (see Sections \ref{sec:2-1} and \ref{sec:tlac}).

We weight the column densities by the covering fraction of each system
in the middle panel of Fig. \ref{fig:nhdist}. This normalization indicates 
how many sightlines are covered by the low-$N_\mathrm{HI}$ or high-$N_\mathrm{HI}$ 
paths. After accounting for the covering fraction, the peak of the distribution 
of low-$N_\mathrm{HI}$ paths shifts to lower column density;
the lower column densities have higher weights, i.e., larger covering fraction 
of low-$N_\mathrm{HI}$ paths.  In other words, the galaxies with only \lya\ 
emission line observed have lower column densities compared to those showing 
\lya\ emission in the bottom of DLA system.

In the bottom panel, we present the combined distribution of column densities 
in galaxies with only \lya\ emission, only DLA system, or \lya\ emission 
in the bottom of DLA system. Similar to the middle panel, the distributions 
are weighted by the covering fraction.  Clearly, the column densities increase 
with the presence of DLA system, consistent with the middle panel. Overall,
however, the distribution remains bimodal, consistent with the argument that
the distribution includes a large fraction of all sightlines. At a qualitative level, the bimodal 
distributions in Figure~\ref{fig:nhdist} confirm a structural similarity between 
the ISM in CLASSY galaxies and the ISM in hydrodynamical simulations focusing on
the star -- gas interplay \citep{Kakiichi2021,Ma2020}. In detail, however, we
recognize several quantitative differences.

\subsubsection{Column Density Distribution in Simulations} \label{sec:nhdist}

In the H \textsc{ii} region simulations of  \citet{Kakiichi2021}, turbulence driven by 
ionization fronts creates a bimodal distribution of column densities. In their Figure 6, 
the higher column-density peak covers $N_\mathrm{HI}$ values similar to our Figure~\ref{fig:nhdist}. 
The  simulated column density distribution actually reaches a minimum around $10^{19}$~cm$^{-2}$, 
however, right where where Figure~\ref{fig:nhdist} shows a maximum. The lower column-density peak 
is offset to $10^{17}$~cm$^{-2}$ in the simulated distribution.  These simulations 
zoom in on individual H \textsc{ii} region, and it is possible that placing 
the H \textsc{ii} region in a more realistic galactic environment would shift the distribution.

Comparing the histogram in Fig. \ref{fig:nhdist} 
to those Fig. 11 of \citet[][]{Ma2020}, we find the high-$N_\mathrm{HI}$ 
gas spread over a similar range in column density. In those simulations, 
 the fraction of high-$N_\mathrm{HI}$ is sensitive to galaxy mass; 
 for their $10^7$ -- $10^8$ $M_\odot$ sample, the fraction of sightlines 
with high-$N_\mathrm{HI}$ {to the total H \textsc{i} sightlines} is about 
one-third as large seen in Fig. \ref{fig:nhdist}. Since their histograms exclude 
the gas  within $0.2 R_{vir}$ of the starburst, it is possible that the addition 
of the starburst region would eliminate, or at least mitigate, the discrepancy. 
Another difference is the column density of the lower-density peak. 
This peak is seen at $N \approx 18-20$ cm$^2$ in CLASSY, whereas \citet{Ma2020} find
the low-$N_\mathrm{HI}$ channels spread, primarily, over the $N \approx 16-18$ cm$^2$ range.
This result may indicate that the feedback in \citet{Ma2020} is too efficient and removes 
too much neutral hydrogen. 

Integral-field spectroscopy is clearly needed to address two observational biases.
The histograms in Figure~\ref{fig:nhdist} combine measurements made on different physical 
scales because the physical size of the aperture changes with galaxy distance. It is not
fully understood how the aperture affects the column density derived by shell-model fitting.
In addition, we emphasize that the lowest colum density sightlines may be missing from 
Figure~\ref{fig:nhdist}. The shell model returns a column density that represents the 
total column of clouds plus an intercloud medium \citep{Li2022}; it follows that the 
lowest (and highest) column density sightlines may not be represented in Figure~\ref{fig:nhdist}.
The low-$N_\mathrm{HI}$ channels may therefore include lower column-density 
pathways, and we aim to understand whether CLASSY galaxies have sightlines optically thin
to LyC radiation.

\subsection{Pathways for LyC Leakage}

\begin{figure}[tbp]
    \centering
    \includegraphics[width=3.3in]{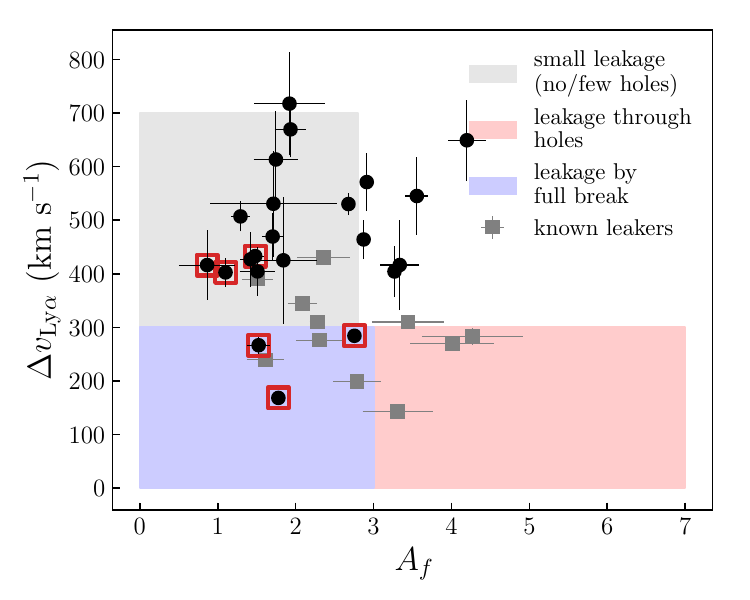}
    \caption{\label{fig:asym} \lya\ red peak asymmetry versus \lya\ peak 
separation. The gray, blue, and pink shaded regions correspond to three 
different regimes of LyC leakage: small leakage (gray), leakage through low-$N_\mathrm{HI}$ 
holes (small holes, red), and leakage by the full break (large density-bounded 
holes, blue), respectively. We overplot the known leakers from \citet{Izotov2016,Izotov2018,Izotov2018b} 
as gray squares.  The red squares highlight six galaxies (J0942+3547, J0944-0038, 
J1253-0312, J1323-0132, J1418+2102, J1545+0858) which have net \lya\ trough 
flux, suggesting they might be LyC leakers.}
\end{figure}

In this section, we will investigate the LyC-thin sightlines\footnote{Column densities lower than $10^{18}$ cm$^2$ corresponds to LyC escape fractions
$\gtrsim1\%$ \citep{Kakiichi2021}.} with $N_\mathrm{HI}<10^{18}$ cm$^{-2}$ in CLASSY sample by analyzing the positive 
residual \lya\ trough fluxes and small \vsep.

\subsubsection{Peak Separation, Trough Flux, and Red Asymmetry} \label{sec:dv_trough}

Peak separation is a good, empirical tracer of LyC escape \citep{Izotov2020}, and 
the shell model provides a theoretical basis for this relation \citep{Dijkstra2016,Eide2018}.
In galaxies where there are few holes through which LyC can escape (low LyC leakage), the scattered \lya\ photons traverse 
optically thick channels, leading to a broad peak separation. Whereas in galaxies with
high LyC leakage, the density-bounded channels result in a small peak separation.
However, the peak separation does not distinguish how the \lya\ photons escape \citep{Kakiichi2021}, as many small holes in 
a turbulent medium can produce a narrow peak separation just like a large, wind-blown cavity. 

The \lya\ asymmetry parameter $A_f$ help to quantify the multiphase nature of the turbulent H \textsc{ii} regions. 
It is originally introduced by \citet{Rhoads2003} to measure the attenuation imprinted by intergalactic medium at high redshift. Here, we apply it in a different context recently introduced by \citet{Kakiichi2021}. 
The two dominant types of \lya\ escape (single flight or excursion) tend to produce a symmetric \lya\ line. 
Thus, when the medium is dominated either by ionization- or density-bounded channels as in the blue or gray region in Fig. \ref{fig:asym}, the asymmetry of the emergent line is low.
However, when the two channels coexist as in the red region in Fig. \ref{fig:asym}, the asymmetry is high.

In Fig. \ref{fig:asym}, we plot \lya\ peak separation against red peak asymmetry.  
We divide the diagram into three distinct regions: (gray) low LyC leakage, 
 (red) significant leakage through low-$N_\mathrm{HI}$ channels (ionization-bounded, 
$f^\mathrm{LyC}_\mathrm{esc}>10\%$), and (blue) significant through large 
holes (density-bounded, $f^\mathrm{LyC}_\mathrm{esc}>10\%$). The boundaries
come from Figure 13 of \citet{Kakiichi2021}, which shows these regions in the 
\vsep\ -- $f_\mathrm{esc}^\mathrm{LyC}$ and $A_f$ -- $f_\mathrm{esc}^\mathrm{LyC}$ planes.
We find that the strongest LyC leakers in CLASSY are the three galaxies -- J0942+3547, 
J1323-0312, J1545+0858 -- in the blue region. 

The \lya\ profiles of these three galaxies also show residual fluxes  at \lya\ 
trough: $F_\mathrm{trough}/F_\mathrm{cont}= 1.36\pm0.07, 19.62\pm0.42, 0.17\pm0.12$, respectively.
Their net \lya\ trough flux supports the conclusion that these galaxies have LyC-thin sightlines 
\citep{Verhamme2015,Gazagnes2020}. Based solely on their \lya\ profile properties then, these galaxies 
are likely strong LyC leakers. When we compare their location in Fig. \ref{fig:asym} to directly
confirmed LyC leakers, we find that their peak separation is as small as the smallest values
measured among directly confirmed LyC leakers \citep{Izotov2016,Izotov2018,Izotov2018b}.

Many CLASSY galaxies are located in the gray-shaded region of Fig. \ref{fig:asym}, 
suggesting they have lower LyC escape fractions than the three galaxies in the 
blue zone. Three known leakers from \citet{Izotov2016,Izotov2018,Izotov2018b} also
lie in the gray zone of Fig. \ref{fig:asym}, just 100 \kms\ above the blue -- gray boundary. 
Based on this comparison to the \lya\ properties of the known leakers, we suggest that
the CLASSY sample contains more LyC leakers than the (blue) shaded region indicates.
The \citet{Kakiichi2021} simulations zoom in on individual H \textsc{ii} 
regions, so perhaps the boundary might shift 100 \kms\ in more realistic 
environments, i.e. those composed of multiple H \textsc{ii} regions.

To gain insight into the empirical boundary, we inspect the positions 
of the other three CLASSY galaxies with net \lya\ trough flux. Non-zero 
trough flux in the emission-line profile requires a low-$N_\mathrm{HI}$ 
column at the systemic velocity. We find three more galaxies with net trough 
flux, and each has \vsep\ $<$ 400 km s$^{-1}$.  The galaxies are
J0944-0038, J1253-0312, and J1418+2102; their trough fluxes are 
$F_\mathrm{trough}/F_\mathrm{cont}=0.47\pm0.29, 0.22\pm0.07, 1.46\pm0.17$, respectively.

We acknowledge that \lya\ trough fluxes are sensitive to the spectral resolution,
which is not precisely known for the \lya\ emission.  We therefore
compared the \lya\ trough width to the width of the red peak which represents an 
upper limit on the unresolved linewidth. Four galaxies (0942+3547, 
J1323-0312, J1545+0858, J1418+2102) show broader \lya\ trough widths than 
the \lya\ peak widths, so these troughs are clearly resolved. For the other two 
objects, J0944-0038 and J1253-0312, their \lya\ trough widths are similar to 
\lya\ peak widths, so higher-resolution spectroscopy might find that we possibly
over-estimate their residual trough flux.
Consequently, we identified at least four CLASSY galaxies containing density-bounded channels.

We conclude that the empirical boundary between the blue and gray zones 
lies closer to a peak separation of 400 \kms, roughly 100 \kms\ larger 
than the blue-gray boundary suggested by the simulations.  
Based solely on the properties of  \lya\ line profiles, we conclude that 
four to six of the CLASSY galaxies (highlighted by red squares in Fig. \ref{fig:asym}) 
are strong LyC leakers.  Their red peaks have a low asymmetry, $A_f < 3$, which
indicates they are best described as density-bounded galaxies. In contrast,
even though they span the same range of peak separations, half of the directly 
confirmed leakers have $A_f > 3$, suggesting their leakage is through ionization-bounded channels
in a multiphase medium.

\subsubsection{Combining Perspectives from  \lya\ and O32}

\begin{figure}[tbp]
    \centering
    \includegraphics[width=3.3in]{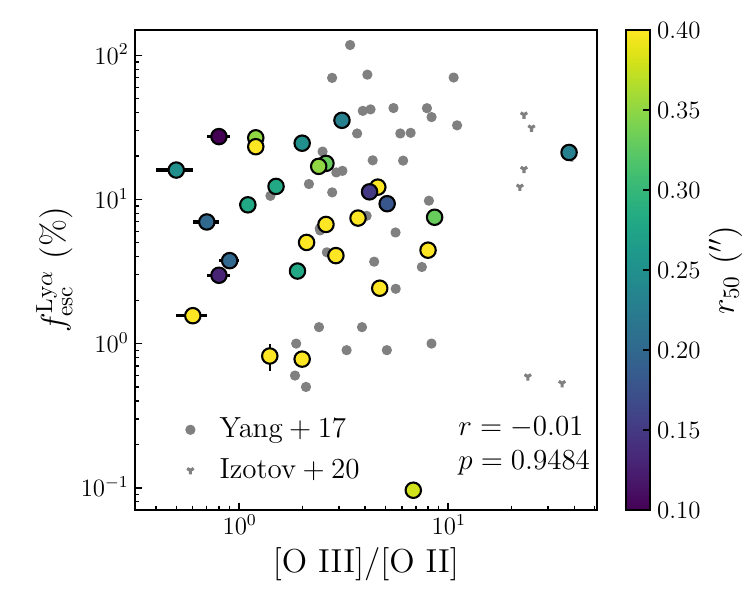}
    \\ \includegraphics[width=3.3in]{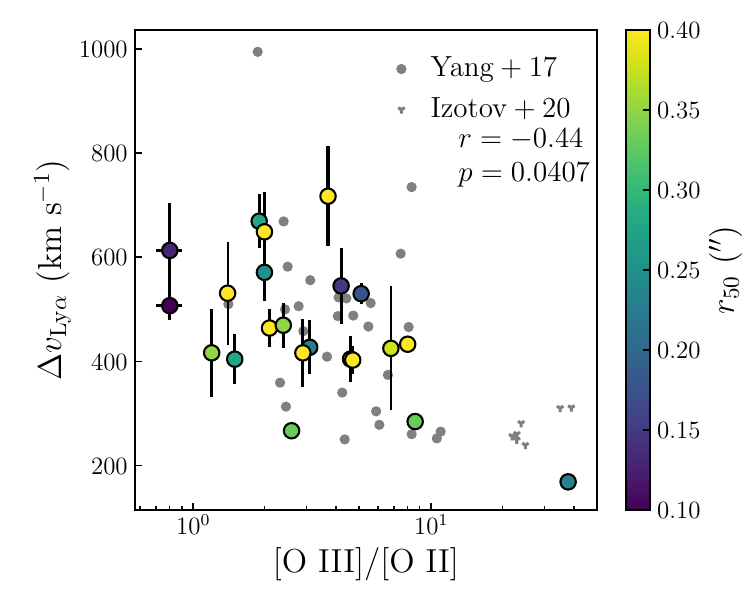}
    \caption{\label{fig:o32} \lyaesc\ (top) and \vsep\ (bottom) versus O32 ratio. The O32 ratio is defined as [O \textsc{iii}] $\lambda5007/$[O \textsc{ii}] $\lambda3727$.
    We color-code the circles based on their UV half-light radius. The yellow circles have a UV half-light radius larger than 0\farcs4, which corresponds to the COS unvignetting aperture.
    We overplot the data in the literature to expand the dynamic range. 
    The gray dots indicate the dwarf galaxies at $z\sim0.1-0.4$ from \citet{Yang2017}.
    The gray Y-shape markers indicate the local dwarf galaxies at $z\sim0.02-0.07$ with extreme O32 ratios from \citet{Izotov2020}.}
\end{figure}

In the previous section, we have shown that the \lya\ trough flux, \lya\ 
peak separation, and \lya\ red peak asymmetry converge at the same selection 
of galaxies with density-bounded holes in their neutral ISM.  
Here we examine the ionization structure of these galaxies, as measured by optical nebular emission lines, to reveal the underlying relation between LyC leaking channels and ionization.
We adopt [O III] $\lambda$5007/[O II] $\lambda$3727 (O32) ratio, one of the most important ionization diagnostics \citep{Kewley2019}, where a high O32 ratio can indicate a density-bounded 
galaxy\footnote{Two of our three best candidates for density-bounded galaxies, J1323-0312 and 
J1545+0858, have the largest O32 ratios among the CLASSY sample (37.8 and 8.6, respectively). 
On the other hand, J0942+3547 has a lower O32 ratio of 2.6.} \citep{Jaskot2013,Nakajima2014,Izotov2016,Flury2022}.

\paragraph{\lyaesc\ and O32}
Intuitively, we expect a high escape fraction of \lya\ photons from density-bounded galaxies.
Yet, in the top panel of Fig. \ref{fig:o32}, the O32 ratio shows no correlation 
with \lyaesc\ (Spearman coefficient $\sim$ 0.04), contradicting the correlation observed among high-redshift galaxies \citep{Trainor2019} and among local dwarf galaxies \citep[][]{Hayes2023}. 
We argue here that the lack of correlation in our sample might result from the scattering of \lya\ photons outside the COS aperture,
an effect that we argued produces DLA systems in many CLASSY spectra (see Sec. \ref{sec:aperture}). 

The slits used to observe high-redshift galaxies in \citet{Trainor2019} typically subtend 5 to 10 kpc, much larger than the physical scale subtended by the COS aperture for the lowest redshift targets.  
Although the \emph{Lyman alpha Spectral Database} \citep[LASD,][]{Runnholm2021,Hayes2023} includes some low-redshift galaxies, the CLASSY sample has a lower median redshift than LASD, so scattering outside the COS aperture plausibly introduces a more serious bias. 
To test this explanation, we restrict the analysis to the subsample with UV radius $<$ 0\farcs4, the radius of the unvignetted COS aperture and find a positive correlation; among the yellow points in Fig. \ref{fig:o32}, the Spearman coefficient of 0.22.

However, the galaxy distance might not be the only factor influencing scattering outside the spectroscopic aperture.
The \lya\ escape fraction of higher redshift galaxies may also be significantly affected. 
In the top panel of Fig. \ref{fig:o32}, we overplot measurements for Green Pea galaxies at redshift 0.1 to 0.4  \citep{Yang2017}. We
add LyC leakers from \citet{Izotov2020} with extreme O32 ratios (ranging from 22 -- 39).
Although the joint sample has a similar redshift range as \citet{Hayes2013}, it also shows no correlation between \lyaesc\ and O32 ratio.
A subset of the joint targets with a large O32 ratio has modest \lyaesc\ of $\sim1\%$. 
Thus, using \lyaesc\ to probe the density-bounded channels should always be aware of those exceptions, not only the aperture loss.

\paragraph{\vsep\ and O32}
Consistent with previous studies  \citep{Yang2017,Jaskot2019,Izotov2020,
Hayes2023}, the \lya\ peak separation  \vsep\ among CLASSY galaxies  
is anti-correlated with the O32 ratio, as shown in the {bottom panel} of Fig. \ref{fig:o32}.
Excluding the galaxies with large UV radius ($>0\farcs4$) does not change the correlation strength, and thus,
we conclude that the \lya\ peak velocity measurements are only weakly affected by the aperture loss.
The \lya\ profiles of LyC leakers with extreme O32 ratios of 22 -- 39 from 
\citet{Izotov2020} show \vsep\ $\approx 250$ \kms, similar to the J1545+0858 and J0942+3547 in CLASSY sample, consistent
with a minimum \vsep\ around 250 \kms. The only data in Fig. \ref{fig:o32} 
with lower \vsep\ is our new data point for J1323-0312. 

The joint sample shows that high O32 galaxies always have narrow peak separations, {while the low O32 galaxies spread a large range of \vsep}. We argue 
that a high O32 ratio traces a large global covering fraction of LyC-thin sightlines,
whereas the narrow \lya\ peak separations appear when the covering fraction of LyC-thin sightlines 
in our direction is high. Variations in the direction of LyC-thin sightlines relative
to our viewing angle, therefore, produce the scatter observed in the \vsep\ vs. O32 ratio 
diagram. 

\vspace{1em}

The correlation between \vsep\ and O32, and the non-correlation between \lyaesc\ and O32 might hint that the different \lya\ features probe \lya\ photons from different channels.
This speculation is in line with the simulations of \citet{Kakiichi2021}.
When the pathways for LyC escape  have  a low covering fraction, the majority of \lya\ photons 
still need to escape through low-$N_\mathrm{HI} \approx 10^{18} - 10^{20}$cm$^{-2}$ 
channels,  the \lya\ emission line emerges with a broad width and large peak separation.
Meanwhile, a smaller fraction  of  \lya\ photons  will pass through {the remaining} columns {which are} optically thin to 
the LyC ($<10^{18}$ cm$^{-2}$), and these sightlines contribute \lya\ emission 
with narrow lines and small peak-separation \citep[see Fig. 12 in][]{Kakiichi2021}.
It follows that the transition from ionization-bounded leakers to
density-bounded leakers is accompanied by a change in the shape of the 
\lya\ profiles (namely, the relative strength of the narrow and broad lines). 
As the covering fraction of LyC thin holes increases, more 
of the emergent \lya\ flux is contributed by the component with narrow peaks. 
When the intensities of two narrow peaks are larger than those of broad peaks, these LyC thin channels can determine \vsep\ while the covering fraction of channels with $> 10^{18}$ remains significant and continues to produce broad peaks with a wider separation.
Thus, in the case of a significant covering fraction of LyC-thin holes, 
the peak separation is probing the H \textsc{i} column in LyC-thin holes and we expect O32 to increase as \vsep\ decreases.
However, in the case of no LyC leakage or a small LyC leakage, the \lya\ 
photons that pass through the columns $> 10^{18}$~cm$^{-2}$ dominate the \lya\ profile (peaks and wings).

\subsection{Outflow Velocity of Neutral ISM} \label{sec:ismvel}

\citet{Chevalier1985} described an adiabatic galactic wind that could reach speeds of roughly 1000 \kms. Theoretical models that explain the relation of this hot phase to the widely observed cool outflows have been a subject of studies for an extended period of time \citep{Klein1994,Schneider2018,Fielding2022}.   Photoionization modeling of the LIS absorption lines in CLASSY spectra indicates the outflowing component traces gas in which hydrogen is mostly ionized \citep{Xu2022}.   
Yet, the combined neutral and molecular phases transport as much (or more) mass than does the warm-ionized phase in the outflow from M82 \citep{Martini2018,Yuan2022}.

Since \lya\ probes the portion of the outflow where hydrogen is neutral, outflow detection using \lya\ complement studies of the highly-ionized outflow.
When \lya\ photons scatter in outflowing gas, the resonance center will move blueward with respect to the rest-frame \lya\ line center.  
Consequently, the outflow velocity is imprinted on the \lya\ profile.
Here, we suggest that the \vtrough\ indicates the average outflow velocity $v$ of neutral clouds, where  $-v$ corresponds to the largest optical depth \citep{Orlitova2018,Michel-Dansac2020,Li2022}. 
In this section, we first compare \lya\ trough velocity \vtrough\ against the Doppler shift of LIS lines.
Then we compare the outflow speeds of neutral ISM in low-$N_\mathrm{HI}$ channels, \lya\ trough velocity \vtrough, 
to tracers of high-$N_\mathrm{HI}$ clouds. 
Finally, adopting \vtrough\ as a direct measurement of the mean Doppler shift of the neutral gas, we revisit why radiative-transfer modeling is typically driven towards a shell velocity faster than \vtrough.

\subsubsection{\lya\ Trough Velocity and LIS Velocity} \label{sec:siii}

\begin{figure}
    \centering
    \includegraphics[width=3.3in]{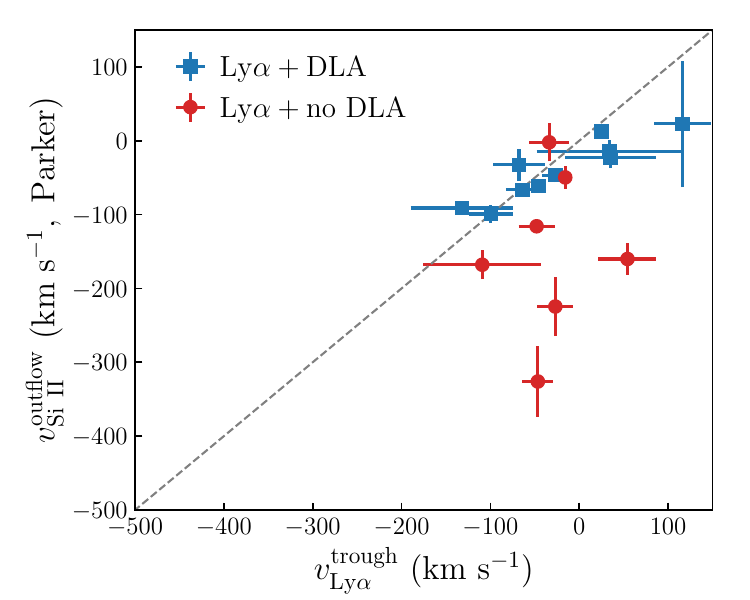}\\ 
    \includegraphics[width=3.3in]{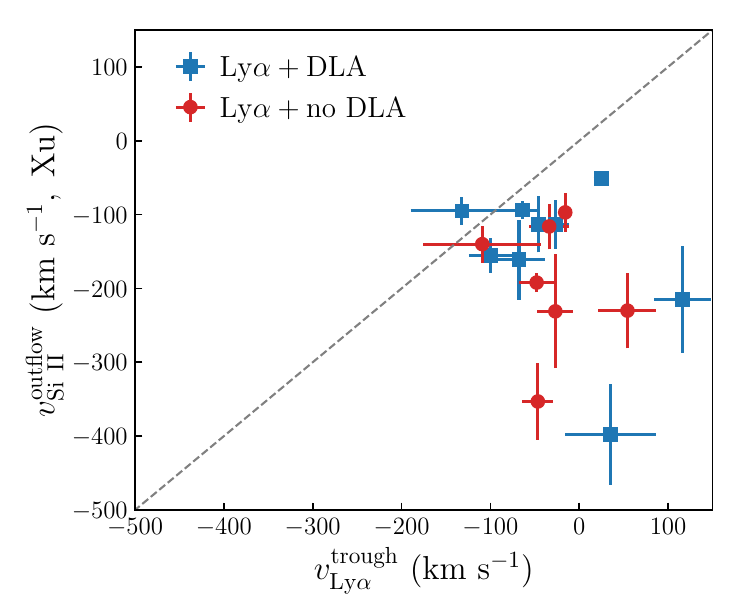}
    \caption{\label{fig:outflow} \vsiii\ vs. \vtrough.
    The top panel adopts the \vsiii\ measured by the single-Voigt fitting (Parker et al. in prep) and the bottom panel adopts the \vsiii\ measured by a double-Gaussian fitting \citep{Xu2022}.
     The dashed lines indicate the 1:1 relationship. the Si \textsc{ii} absorbers are multi-phase, including neutral hydrogen in addition to the mostly-ionized phase. By comparing \vtrough\ and \vsiii, we are able to distinguish ionization status of the ISM.}
\end{figure}

Resonance UV absorption lines, e.g., Si \textsc{ii} and C \textsc{ii}, have been extensively used to measure outflow speeds \citep[e.g.,][; Hayes et al. 2023]{Reddy2016b,Henry2015,Orlitova2018}.
Here we focus on Si \textsc{ii} $\lambda1260$, which is well measured by the CLASSY collaboration.

The Doppler shifts of Si \textsc{ii} in the CLASSY sample have been measured using two different methods.
\citet{Xu2022} use a double-Gaussian profile to deblend the outflow component from the static ISM component of Si \textsc{ii} and find that the outflow component is mostly ionized.
On the other hand, Parker et al. (in prep) fit a single-Voigt profile to determine the average velocity of all LIS absorbers.
As the LIS lines can also arise from the neutral ISM, the Parker measurements should include the contribution of neutral ISM.
Conceptually, if the LIS absorber is dominated by the static ISM, the Parker measurement, which is close to 0 \kms, should be distinct from the Xu measurement. 
But if the LIS absorber is dominated by the outflow component, the Parker measurement should be similar to the Xu measurement.

Fig. \ref{fig:outflow} presents the comparisons of \vtrough to both LIS outflow measurements derived by the two methods\footnote{J0808+3948 is excluded because its polycyclic aromatic hydrocarbon feature suggests it might be an AGN \citep{Xie2014}.}.  
Directly comparing the two LIS outflow measurements in the top and bottom panels, we notice the positions of two objects (J1416+1223, J0938+5428) shift significantly.
Parker et al. (in prep) derive a velocity close to 0 km s$^{-1}$, but the outflow velocities derived in \citet{Xu2022} can reach several hundred km s$^{-1}$, suggesting the LIS absorber of these two galaxies are mainly static.
Galaxies that shift between the two panels have substantial absorption at v=0,  which we attribute to the static ism.

Secondly, we see that the Si \textsc{ii} velocity measured by Parker et al. (in prep) shows a better agreement with \vtrough, particularly the galaxies with both \lya\ and DLAs (blue squares) in the top panel of Fig. \ref{fig:outflow}.
This suggests that in those galaxies, the Si \textsc{ii} absorbers (in both outflow and static) contain a significant fraction of neutral hydrogen, though \citet{Xu2022} suggest that the Si \textsc{ii} in the outflow traces mostly ionized gas.

However, looking at the galaxies with no DLAs (red circles in Fig. \ref{fig:outflow}), their Si \textsc{ii} velocities disagree with \vtrough\ in both two panels.
The three most deviant circles (J0021+0052, J0926+4427, J1429+0643) show that their \vtrough\ are close to 0 \kms\ but the Si \textsc{ii} velocities are $\le -200$ \kms. 
We find that their Si \textsc{ii} line profiles are dominated by the outflow component: they 
have very little absorption at the systemic velocity, so the velocity is not sensitive to the measurement method.
This suggests that the Si \textsc{ii} absorption comes mostly from the ionized gas in these galaxies, similar to \citet{Xu2022}.

\subsubsection{\lya\ Trough Velocity and DLA Velocity} \label{sec:vdla}

\begin{figure}
    \centering
    \includegraphics[width=3.3in]{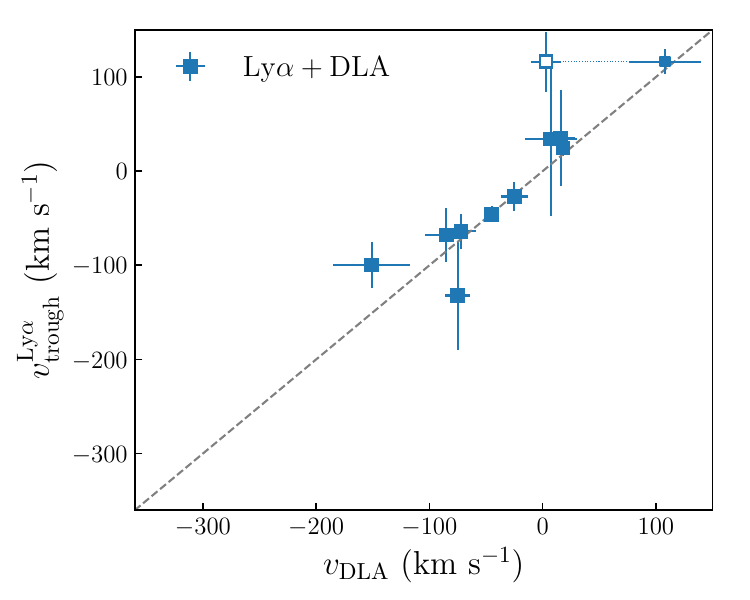}\\
    \includegraphics[width=3.3in]{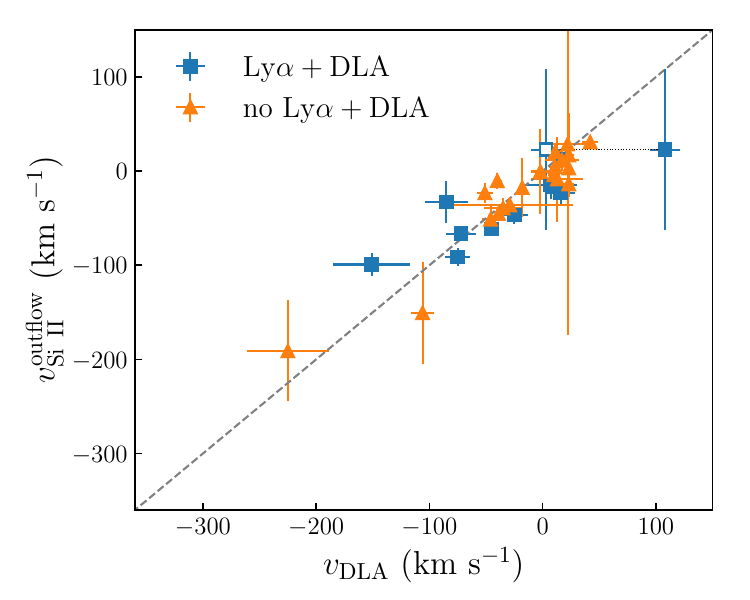}
    \caption{$v_\mathrm{DLA}$ vs. \vtrough\ (top) and \vsiii\ (bottom). 
    Here we adopt the \vsiii\ measured by Parker et al. (in prep).
    The gray dashed lines indicate the 1:1 relationship. 
    In the bottom panel, we separate the sample based on whether their spectra have \lya\ emission lines. 
    We notice the O \textsc{i} absorption line of one galaxy (J0938+5428, open square) is contaminated by the refilling O \textsc{i} emission line.
    Thus, we adopt the velocity of C \textsc{ii} absorption line and we use a black dashed line to connect them.
    {The good agreement suggests that the gas in low-$N_\mathrm{HI}$ has same velocity as the gas in high-$N_\mathrm{HI}$ channels.}}
    \label{fig:vdla}
\end{figure}

In the top panel of Fig. \ref{fig:vdla}, we compare the \lya\ trough velocity \vtrough\ and the velocity of high-$N_\mathrm{HI}$ clouds (i.e., DLA system velocity probed by O \textsc{i} absorption line).
It is intriguing to see such a good agreement between these two independent measurements, suggesting that the low-$N_\mathrm{HI}$ channels have the same velocity as the high-$N_\mathrm{HI}$ channels.

In the bottom panel of Fig. \ref{fig:vdla}, we further find that DLA velocity agrees with Si \textsc{ii} velocity.
Here, we include the galaxies, which do not have \lya\ emission lines, as the circles. 
They are consistent with those galaxies which have both \lya\ emission lines and DLA systems.
This hints that, for the galaxies with DLA systems, the intrinsic reason for the correlation between Si \textsc{ii} velocity and \vtrough\ is that the Si \textsc{ii} mainly traces the high-$N_\mathrm{HI}$ clouds and the high-$N_\mathrm{HI}$ clouds have similar velocity as the low-$N_\mathrm{HI}$ clouds.

\subsubsection{Revisiting Outflow Velocity Discrepancy} \label{sec:vtrough}

\begin{figure}
    \centering
    \includegraphics[width=3.1in]{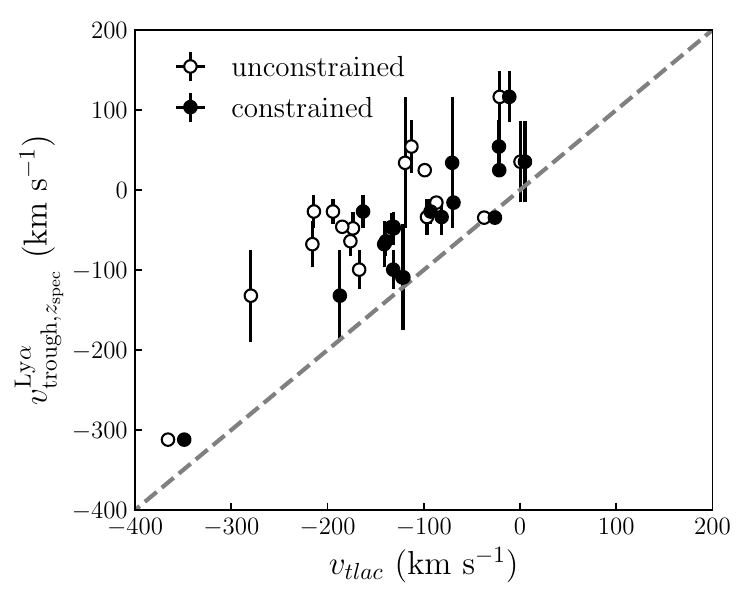}\\
    \includegraphics[width=3.1in]{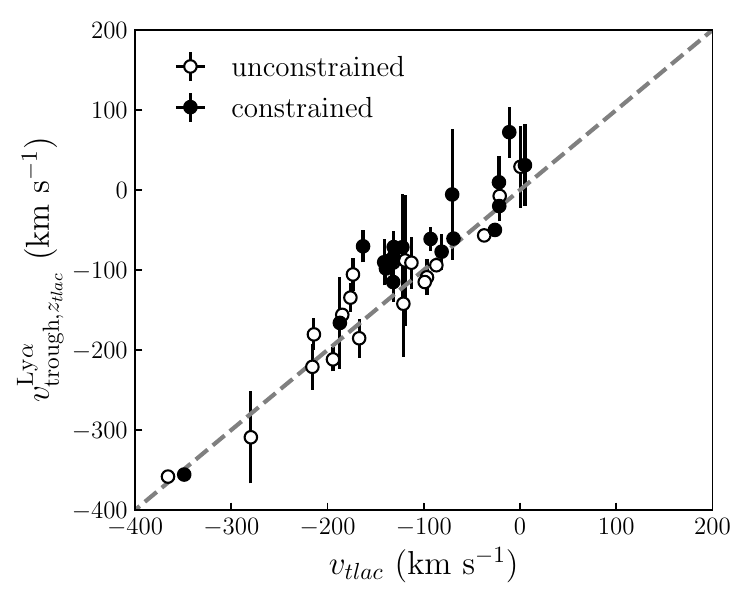}
    \caption{\label{fig:vtlac} The \lya\ trough velocity measured from spectrum vs. the outflow velocity estimated by the shell model. Top panel: the \vtrough\ is measured based on the spectroscopic redshift. Bottom panel: the \vtrough\ is measured based on the redshift from \tlac\ profile fitting. We plot the measurements from both the $z$-unconstrained fitting (open circle, Sec. \ref{sec:secondtlac}) and the $z$-constrained fitting (solid circle, Sec. \ref{sec:redshift}). The dashed line indicates the 1:1 relationship.}
\end{figure}

In this section we discuss the outflow velocity discrepancy using the same profile fittings as Sec. \ref{sec:redshift} and propose a new explanation of the discrepancies.
Here we adopt the \vtrough\ as the intrinsic outflow velocity since it traces the neutral ISM which scatters the \lya\ photons.

First, we directly compare the \vtrough\ measured based on the spectroscopic redshift 
and the outflow velocities $v_{tlac}$ estimated by the shell model. 
In the top panel of Fig. \ref{fig:vtlac}, we present the comparison for both 
the second profile fitting (redshift-unconstrained) and the third profile 
fitting (redshift-constrained). Though a clear correlation between 
\vtrough\ and $v_{tlac}$ can be seen, $v_{tlac}$ is larger by 0 -- 200 
\kms\ and 0 -- 140 \kms\ than the \vtrough\ for the second and third fittings, 
respectively. 
We speculate that the reason for this discrepancy is a `redshift error' required by the model fitting. To test this idea, we shift our \vtrough\ measurements to the fictitious reference frame chosen by the fitted \tlac\ redshift.

The bottom panel of Fig \ref{fig:vtlac} shows the \vtrough\ measurements in the \tlac\ reference frames defined by the second and third fittings.
We have shifted the measurements by 
\begin{equation}
    v^\mathrm{Ly\alpha}_{\mathrm{trough},z_{tlac}} = v^\mathrm{Ly\alpha}_{\mathrm{trough}} - (z_{tlac} - z_{spec})\times c,
\end{equation}
where $c$ is the speed of light.
The new correlations are significantly improved and close to the 1:1 relationship.
Especially for the redshift-unconstrained fitting (second attempt), \vtrough\ and $v_{tlac}$ agree well with each other. 
These results confirm that we should compare \vtrough\ and $v_{tlac}$ in a common redshift frame. This also confirms that the outflow velocity and the redshift are coupled in the shell model:
\begin{equation}
    \label{eq6}
    \frac{\lambda^\mathrm{Ly\alpha}_\mathrm{trough}}{(1+z)\times \lambda_\mathrm{Ly\alpha}}-1 = -\frac{v^\mathrm{outflow}}{c},
\end{equation}
where $\lambda^\mathrm{Ly\alpha}_\mathrm{trough}$ is the wavelength of \lya\ trough and $\lambda_\mathrm{Ly\alpha}$ the rest-frame wavelength of \lya.
Once the redshift of the shell model is fixed, the model outflow velocity is also determined by the Doppler offset of the observed \lya\ trough with respect to the model redshift.
Thus, the redshift and outflow velocity discrepancies are the "two sides of the same coin".

The preferred larger outflow velocity by \tlac\ may hint that the observed B/R ratio is lower than the intrinsic B/R ratio. 
Moreover, as we discussed in Sec. \ref{sec:aperture}, the observed B/R ratio can be biased by the aperture loss. 
Thus, this inspires us to connect the discrepancies to the aperture loss.

We, therefore, propose an explanation for the discrepancies.
Since the aperture loss modifies the B/R ratio to a lower value and the 
B/R ratio is tightly anti-correlated with outflow velocity, to achieve 
the smaller observed B/R ratio, the shell model will suggest a larger outflow velocity.
Meanwhile, a higher systematic redshift is required to match the \lya\ trough velocity to the outflow velocity (Eq. \ref{eq6}).
Thus, the best-fit redshift and outflow velocity from the shell model are larger than that observed from the spectra.
The aperture loss has a non-negligible impact on the \lya\ profile and should always be considered when interpreting \lya\ profile.

\subsection{A Schema of the Neutral ISM}

\begin{figure*}
    \centering
    \includegraphics[width=6.5in]{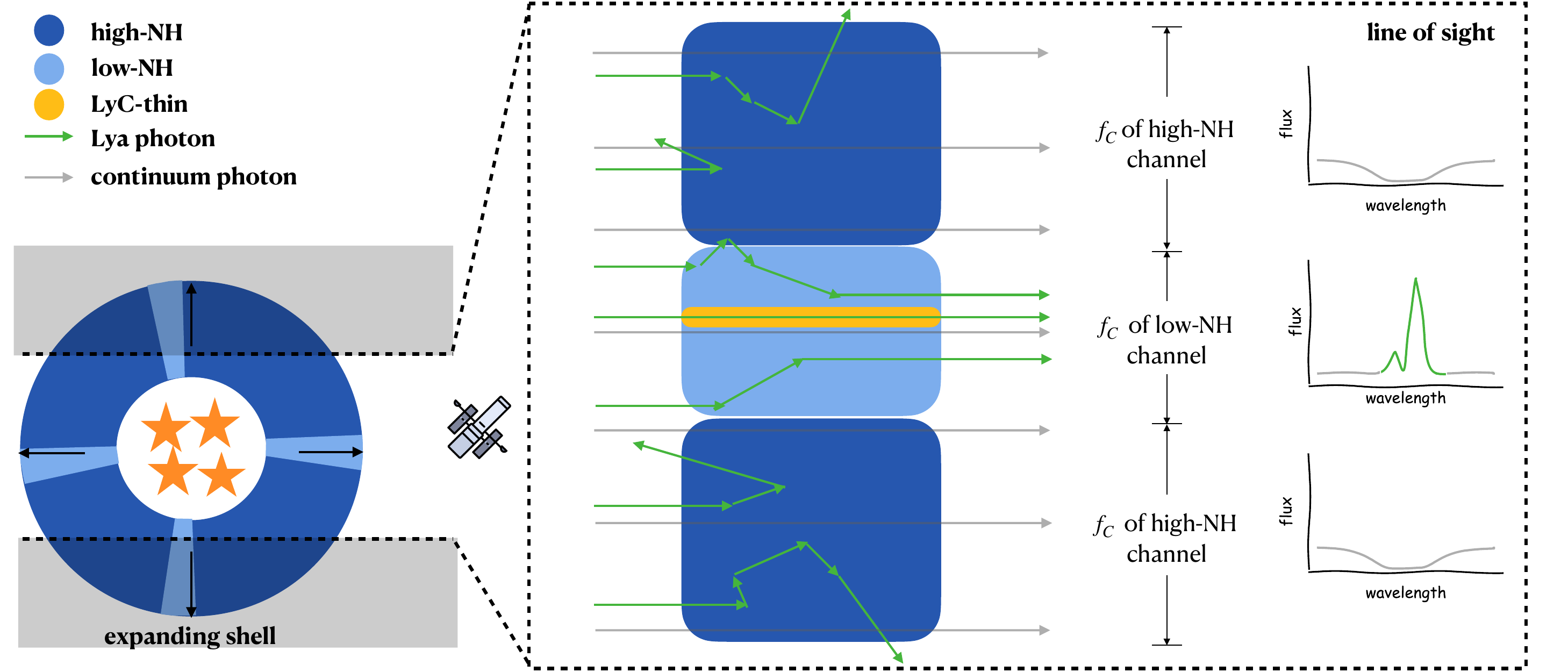}
    \caption{\label{fig:illustration} 
    {The left panel shows the side view of HST/COS observation. 
    We adopt a continuous shell geometry, which contains different column-density regions as dark blue and light blue. This is somewhat an oversimplification as the ISM can be clumpy, but should be the simplest way to illustrate the ISM in galaxies.
    We note the scale length of different regions in this figure does not indicate their physical scale length.
    We use the gray shaded regions to indicate the parts that are missed by the HST/COS aperture.}
    The right panel shows the zoom-in structure of a small slab of the outflowing shell. The yellow, light blue, and dark blue indicate the LyC-thin, low-$N_\mathrm{HI}$, and high-$N_\mathrm{HI}$ paths, respectively. The gray and green lines show the radiative transfer processes of the continuum and \lya\ photons. 
    We mark the covering fraction of different column density channels. 
    The right-most spectra represent the emergent \lya\ profiles along different regions.
    The low-$N_\mathrm{HI}$ channels lead to the \lya\ emission line but the high-$N_\mathrm{HI}$ channels lead to the DLA system.
    The observed \lya\ spectrum is a combination of those \lya\ profiles and thus, the residual flux in the bottom of the DLA profile
    equals the covering fraction of low-$N_\mathrm{HI}$ channels.}
\end{figure*}

In this section, we summarize our interpretation of ISM structure from the previous sections.
We have demonstrated that the ISM in CLASSY galaxies is inhomogeneous, consisting of high-$N_\mathrm{HI}$, low-$N_\mathrm{HI}$, and even LyC-thin regions, based on the clear separation between the DLA and \lya\ emission (see Sec. \ref{sec:profile}), the non-zero residual flux at \lya\ trough, and small peak separation (see Sec. \ref{sec:ismstruct}). 
In the left panel of \ref{fig:illustration}, we plot a schema of the neutral ISM for illustration. For simplicity, we adopt a continuous shell model.
The low-$N_\mathrm{HI}$ and high-$N_\mathrm{HI}$ paths are shown as light blue and dark blue, respectively.
We also use two gray shades to indicate the \lya\ halos missed due to the aperture effect.
In the right panel, we zoom in to show the \lya\ radiative transfer in a small slab.
The green lines indicate the \lya\ photons and the gray lines indicate the continuum photons.

Although the \lya\ radiative process is highly non-linear and non-additive, the radiative transfer fitting results suggest that we can take the \lya\ emission and DLA system apart.
The DLA system can be well fitted by a partial-covering Voigt profile with a high-$N_\mathrm{HI}$ and the \lya\ emission normalized by the uncovered continuum can be well fitted by the shell model with a low-$N_\mathrm{HI}$.
This clear separation between \lya\ emission and DLA system indicates that the \lya\ exchange between low-$N_\mathrm{HI}$ path and high-$N_\mathrm{HI}$ path should be negligible, as we discussed in Sec. \ref{sec:profile}.
Only very few \lya\ photons which are injected into one region can travel 
to another region and thus,  the \lya\ radiative processes in two different regions are independent.
This is feasible because of two reasons: (1) the possibility of a \lya\ photon traveling from low-$N_\mathrm{HI}$ path to high-$N_\mathrm{HI}$ path is very small, as most of which are just ``reflected'' by the surface between two channels \citep{Hansen2006}; (2) the \lya\ photons including the underlying continuum photons which are injected into high-$N_\mathrm{HI}$ paths are mostly scattered to much larger impact parameters \citep[i.e., the extended \lya\ halo,][]{Zheng2011,Steidel2011}, thus, most of which are missed due to the aperture effect and leave a DLA system.
Thus, only \lya\ photons escape through low-$N_\mathrm{HI}$ regions can be observed and the emergent \lya\ profile is a combination of \lya\ spectra from two regions, as illustrated by the green and gray lines in Fig. \ref{fig:illustration}, and has a profile of \lya\ emission in the bottom of DLA system.

In the left panel of Fig. \ref{fig:illustration}, we plot several low-$N_\mathrm{HI}$ channels in different directions. 
Although all of those low-$N_\mathrm{HI}$ channels can allow the escape of \lya\ photons, only the channels exposed to the COS aperture (i.e., horizontal one in Fig. \ref{fig:illustration}) can contribute to the observed \lya\ emission line. Because for the \lya\ photons which are initially injected into low-$N_\mathrm{HI}$ channels in other directions, they still need to penetrate the high-$N_\mathrm{HI}$ paths before reaching us.

We have proposed a scenario that the aperture loss is responsible for those unexpected profiles of \lya\ emission in the bottom of DLA system in the CLASSY sample.
In this work, we also find that the DLA absorber (neutral gas in high-$N_\mathrm{HI}$ paths) has a similar systematic velocity as the neutral gas in the low-$N_\mathrm{HI}$ paths.
However, the ionized gas, traced by the outflowing component of Si \textsc{ii} absorption line, has a generally larger velocity compared with the neutral gas in the low-$N_\mathrm{HI}$ paths.

Using three LyC leakage diagnostics, we find that at least three galaxies in the CLASSY sample are LyC leaker candidates. 
Thus, in the right panel of Fig. \ref{fig:illustration}, we use yellow to indicate the possible LyC-thin channels in the ISM, through which the \lya\ photons can easily escape without much resonant scattering.
By comparing the \vsep\ with O32 ratio, we conclude that the O32 ratio is tracing the covering fraction of LyC-thin channels, consistent with those known LyC leakers \citep{Flury2022}. The covering fraction increases as the O32 ratio increases, and thus, the probability of observing small \vsep\ increases.

\section{Summary \& Conclusions}

In this paper, we extracted high-resolution \lya\ line profiles from CLASSY spectra of 45 EoR analogs.
These HST COS/G130M spectra show a wide variety of \lya\ profiles, including damped absorption,
\lya\ emission in damped \lya\ absorption (DLA) profiles,  P-Cygni profiles, and pure \lya\ emission. 
We attribute the damped absorption to \lya\ photons being scattered out of the spectroscopic aperture, and
we argue that the especially large diversity among CLASSY \lya\ profiles can be largely attributed to
large range of physical scales subtended by the COS aperture,  a little over 100 pc up to nearly 8 kpc.

We separated the DLA and \lya\ emission components of the profiles.   Specifically, we adopted the precisely measured
Doppler shifts of the O \textsc{i} absorption components as priors for the Doppler shift of each broad DLA
profile, and we fitted the damped \lya\ absorption with modified Voigt profiles. After subtracting the
stellar continuum and the DLA profile, we modeled the \lya\ emission profile and the appropriate
underlying continuum using the shell model. For the first time, we measure the properties in the neutral
shell traversed by the
emergent \lya\ emission, and the conditions in the high column density clouds, in the same sample of galaxies.
For double-peaked \lya\ emission line profiles, we defined the Doppler shift of the minimum between the two
emission lines as the {\it trough velocity}, which we compared to the Doppler shifts of LIS absorption lines
and the DLA. Our results are summarized below:

\begin{itemize}
    \item The \lya\ emission in the bottom of the DLA profile reveals the inhomogeneity of the ISM and the
      outflows. The DLA profile and \lya\ emission line can be surprisingly well fitted by simply splitting
      a geometric covering factor between the high-column density sightlines and the lower-$N_\mathrm{HI}$ channels through which \lya\ photons escape.   This suggests little \lya\ exchange between high- and
      low-$N_\mathrm{HI}$ paths.  Combining the sightlines probed by \lya\ emission lines with those
      producing damped absorption, the net distribution of column densities is bimodal and therefore
      qualitatively similar to the distributions predicted by numerical simulations of H \textsc{i} regions
      \citep{Ma2020,Kakiichi2021}. It is important to note, however, that this observed distribution
      is offset to higher $N_\mathrm{HI}$ compared with the simulations. This discrepancy could arise
      from gas on larger spatial scales than the simulations include, or from structural differences
      in the star-forming complexes; but, whatever its origin, an understanding of the offset will
      better inform our understanding of the channels through which not only \lya\, but also LyC,
      photons escape from galaxies.

    \item
      We find that the Doppler shift of the \lya\ trough velocity matches that of the Si \textsc{ii} velocity in most galaxies with DLAs, suggesting that the Si \textsc{ii} absorber in those galaxies are mainly in neutral phase. 
      However, for galaxies without DLA systems, the \lya\ trough velocity is always smaller than the
      Si \textsc{ii} velocity, suggesting Si \textsc{ii} tracing a more ionized phase of the outflow, consistent with \citet{Xu2022}.
      Thus, the Si \textsc{ii} absorbers are multi-phase, including neutral hydrogen in addition to the mostly-ionized phase.
      Combining the \lya\ and Si \textsc{ii}, we are able to identify the ionization of Si \textsc{ii} absorbers.
      Our comparison also suggests that the \lya\ trough velocity directly measures the average velocity of neutral gas in the static ISM and outflows.

    \item
      In spectra with a DLA, the \lya\ trough velocity agrees well with the DLA velocity (O \textsc{i} velocity), suggesting that 
      the high-$N_\mathrm{HI}$ clouds have similar kinematics as low-$N_\mathrm{HI}$ clouds. 
      Further, the Si \textsc{ii} also agrees well with the DLA velocity, even for galaxies without \lya\ emission.
      Thus, we conclude that Si \textsc{ii} mainly traces the neutral gas in high-N$_\mathrm{HI}$ columns if the galaxies show DLAs.

    \item
      Motivated by the numerical simulations of \citet{Kakiichi2021},
      we combine the measurements of \lya\ peak separation and \lya\ red peak asymmetry in a diagnostic
      diagram that differentiates the type of channels for LyC leakage. Comparing the diagram
      with the known LyC leakers, we suggest that the boundary for distinguishing substantial leakage
      from small leakage is a peak separation less than  $\sim400$ \kms.  In the case of leakage, or
      equivalently small peak separation, then the red peak asymmetry parameter distinguishes holes,
      where $A_f > 3$, from the more symmetric profiles generated by full breaks.
     Six CLASSY galaxies are identified as the density-bounded LyC leakers by this technique, agreeing with
     the selection of net \lya\ trough flux.
    The inferred properties of the LyC-thin sightlines depend on galaxy orientation, whereas
    the [O \textsc{iii}]/[O \textsc{ii}] ratio offers a sightline-independent perspsective.
    We confirm the presence of an inverse relation between \lya\ peak separation and the
    [O \textsc{iii}]/[O \textsc{ii}] ratio, as has been noted previously \citep{Jaskot2019,Flury2022}.

    \item
      Similar to \citet{Orlitova2018}, we find that the fitted redshift is always larger than the
      spectroscopic redshift and the fitted outflow velocity is larger by 10 -- 200 \kms\ than the \lya\
      trough velocity.  The connection between the \lya\ trough velocity and the outflow velocity offers
      new insight into the origin of those discrepancies, which we suggest are not adequately explained
      by parameter degeneracies  \citet{Li2022}. We argue instead that aperture vignetting is the primary
      source of the discrepancies.  The COS aperture vignets the blue-shifted peak more than the
    red-shifted peak, resulting in a lower blue-to-red peak ratio.    To match the lower blue-to-red peak ratio,
    the radiative transfer model requires a higher outflow velocity and thus, a larger redshift to match the
    outflow velocity to \lya\ trough velocity.

\end{itemize}

Our results underline the sensitivity of \lya\ profiles to aperture vignetting.
The COS aperture not only excludes a large fraction of \lya\ photons, it modifies the \lya\ profile.
Like many CLASSY targets, the composite \lya\ spectra of star-forming galaxies at $z \sim 1.8$ -- 3.5 show DLA systems as well
\citep{Reddy2016,Reddy2022}.  An important difference, however, is that the typical slit width used in
ground-based spectroscopy, 1\farcs2, corresponds to $\sim10$ kpc. The COS aperture subtends a
comparable physical scale only for the most distant Lyman Break Analogs in CLASSY, and their
COS spectra  do not show DLAs.  Nonetheless, our analysis suggests the DLAs appear in the $z\sim 2$
spectra because the \lya\ escape on spatial scales is larger than the slit width.
An important implication of this paper is that aperture vignetting could strongly affect
{recent JWST observation of EoR galaxies using Near Infrared Spectrograph (NIRSPec) slit mode}, of which the
slit width is just 0\farcs2, corresponding to only $\sim 1$ kpc.

In this paper, we leveraged these aperture effects, recognizing an opportunity to characterize
the properties of the low-$N_\mathrm{HI}$ channels and high-$N_\mathrm{HI}$ clouds in the same
set of galaxy sightlines. To fully understand the connection between the observed \lya\ profile and
LyC leakage, the radiative transfer simulations will need to predict the spatial variations in
profile shape. The extracted \lya\ profiles used in this work, including
the DLA profiles and the best-fit shell model spectra, can be downloaded from the CLASSY High Level Science
Products database, which is developed and maintained at STScI, Baltimore, USA.\footnote{
Data will appear at \url{https://archive.stsci.edu/hlsp/classy} after acceptance by the ApJ. The data product
can be found here (\url{https://drive.google.com/drive/folders/1NCUyr1vQ10z4BZuGBqsBuIjL0dWJnmZ1?usp=sharing}) during the review
period.}

\section*{acknowledgements}

The CLASSY team is grateful for the support that was provided by NASA through  grant  HST-GO-15840,
from the Space Telescope Science Institute,  which is
operated by the Associations of Universities for Research in Astronomy, Incorporated, under NASA
contract NAS5-26555. CLM thanks the NSF for support through AST-1817125.
BLJ thanks support from the European Space Agency (ESA).
The CLASSY collaboration extends special gratitude to the Lorentz Center for
useful discussions during the "Characterizing Galaxies with Spectroscopy with a view for JWST" 2017 workshop that
led to the formation of the CLASSY collaboration and survey.\\

\facilities{HST (COS)}
\software{ astropy (The Astropy Collaboration 2013, 2018), CalCOS (STScI), python}

\appendix

\section{Best-fit \lya\ spectra} \label{app:a}

\begin{figure*}
    \centering
    \includegraphics[width=1\textwidth]{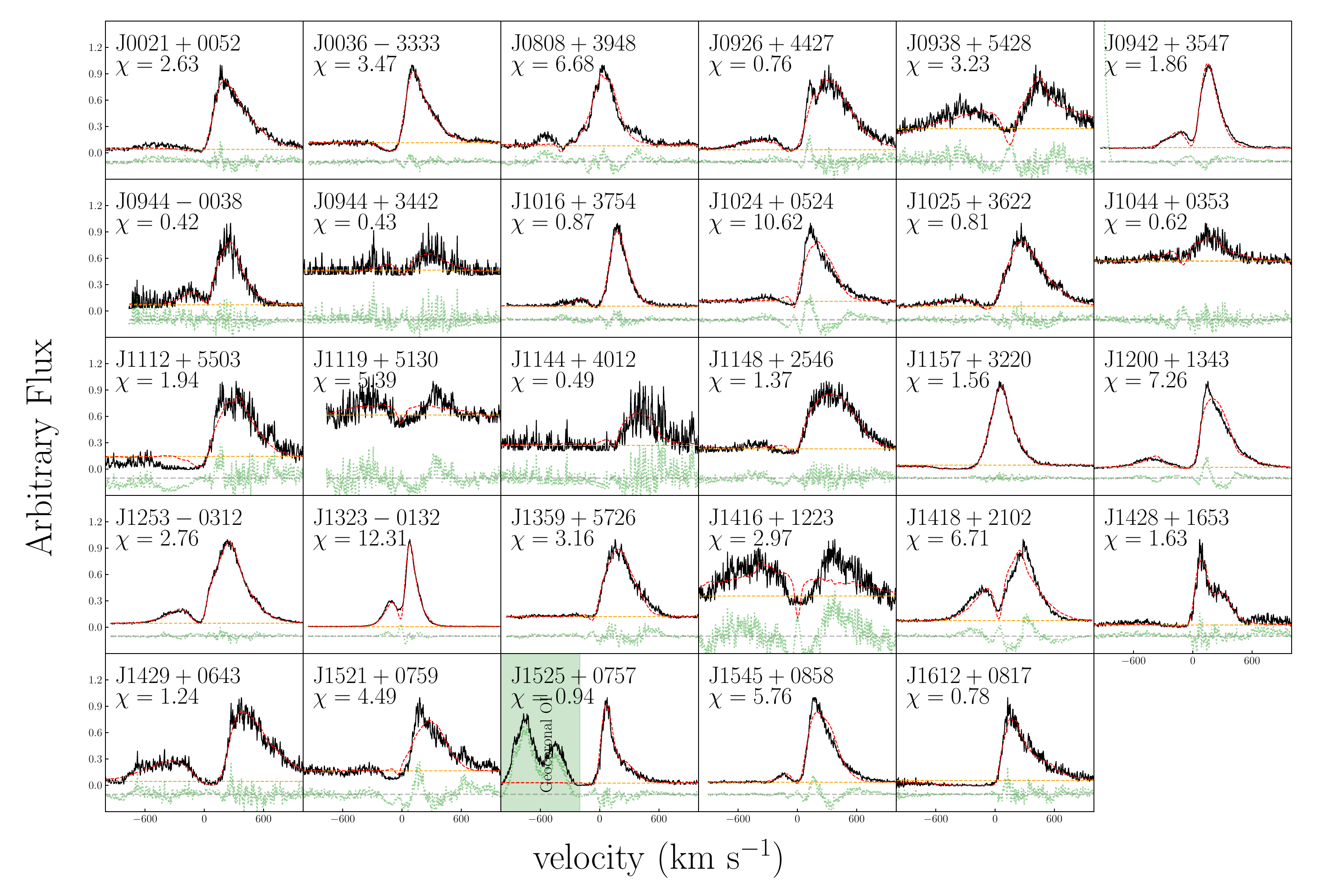}
    \caption{\label{fig:tlacfail} Best-fit \lya\ spectra (red) for 29 CLASSY galaxies using the total stellar continuum and a Gaussian redshift prior with $\sigma=120$ km s$^{-1}$ (first attempt). The spectra are normalized by the peak flux and the orange dashed lines indicate the continuum level for each object. Clearly, the \lya\ spectra of J0938+5428, J0944+3442, J1044+0353, J1119+5130, J1144+4012, J1416+1223, and J1521+0759 are failed to be reproduced.
    We use the green lines to show the residual and manually shift it by $-0.1$ for better illustration. The gray dashed lines indicate the zero level of the residual.}
\end{figure*}

Fig. \ref{fig:tlacfail} and \ref{fig:tlac} present the best-fit \lya\ spectra obtained using approaches described in Sec. \ref{sec:fs} and \ref{sec:secondtlac}, respectively.

\begin{figure*}
    \centering
    \includegraphics[width=1\textwidth]{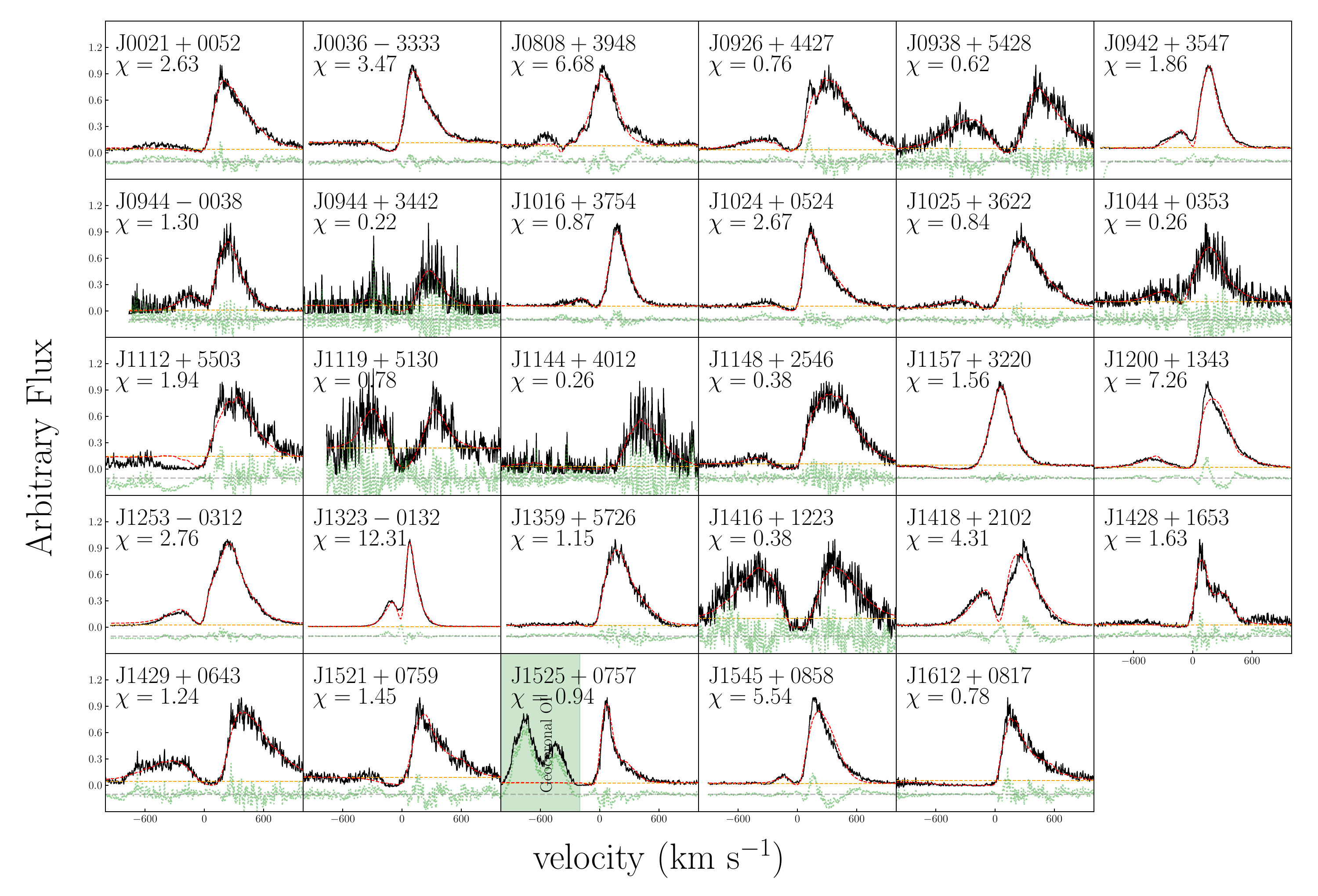}
    \caption{\label{fig:tlac} Best-fit \lya\ spectra (red) for 29 CLASSY galaxies using the residual stellar continuum and a Gaussian redshift prior with $\sigma=120$ km s$^{-1}$ (second attempt). The spectra are normalized by the peak flux and the orange dashed lines indicate the continuum level for each object. Every input \lya\ profile is successfully reproduced by \tlac\ model.
    We use the green lines to show the residual and manually shift it by $-0.1$ for better illustration. The gray dashed lines indicate the zero level of the residual.}
\end{figure*}

\bibliography{main}
\bibliographystyle{aasjournal}

\clearpage

\begin{table*}
    \centering
    \caption{DLA Measurements \label{tab:1}}
    \begin{threeparttable}
    \begin{tabular}{c c c c c c c c}
        \hline
        \hline
        object & $z_\mathrm{spec}$ & D$_\mathrm{L}$$^a$ & $f_{c,\mathrm{DLA}}$ & $f_{c,\mathrm{OI}}$ & $\log N_\mathrm{H}$ & $v$ & COS Aperture Size \\
         & & Mpc & & & cm$^{-2}$ & km s$^{-1}$ & kpc \\
         (1) & (2) & (3) & (4) & (5) & (6) & (7) & 8\\
        \hline
        J0021+0052 & 0.09839 & 452 & ... & ... & ... & ... & 4.5 \\
        J0036-3333 & 0.02060 & 80 & ... & ... & ... & ... & 0.93 \\
        J0127-0619 & 0.00540 & 18 & $0.08\pm{0.04}$ & $0.38\pm0.15$ & $21.21\pm{0.27}$ & $-189\pm{14}$ & 0.22 \\
         & & & $0.92\pm{0.04}$ & $0.93\pm0.06$ & $21.04\pm{0.03}$ & $11\pm{5}$ \\
        J0144+0453 & 0.00520 & 17 & $0.96\pm{0.01}$ & $0.78\pm0.05$ & $20.18\pm{0.02}$ & $42\pm{7}$ & 0.2 \\
        J0337-0502 & 0.01352 & 51 & $1.00$ & $0.76\pm0.03$ & $21.81\pm{0.00}$ & $-18\pm{3}$ & 0.61 \\
        J0405-3648 & 0.00280 & 11 & $>0.99$ & ...$^d$ & $20.80\pm{0.02}$ & $-35\pm{17}$$^b$ & 0.13 \\
        J0808+3948 & 0.09123 & 417 & ... & ... & ... & ... & 4.2 \\
        J0823+2806 & 0.04722 & 210 & $0.99\pm{0.01}$ & $0.98\pm0.02$ & $21.61\pm{0.03}$ & $22\pm{20}$ & 2.3 \\
        J0926+4427 & 0.18067 & 875 & ... & ... & ... & ... & 7.6 \\
        J0934+5514 & 0.00250 & 12 & $0.98\pm{0.00}$ & $0.38\pm0.02$$^d$ & $21.24\pm{0.00}$ & $23\pm{3}$ & 0.14 \\
        J0938+5428 & 0.10210 & 471 & $0.80\pm{0.05}$$^c$ & $0.80\pm0.05$ & $20.34\pm{0.08}$ & $3\pm{13}$ & 4.7 \\
        J0940+2935 & 0.00168 & 10 & $0.95\pm{0.01}$ & $0.69\pm0.06$$^d$ & $21.26\pm{0.01}$ & $-51\pm{7}$ & 0.12 \\
        J0942+3547 & 0.01486 & 65 & ... & ... & ... & ... & 0.76 \\
        J0944-0038 & 0.00478 & 24 & $0.84\pm{0.03}$ & $0.88\pm0.15$ & $21.67\pm{0.05}$ & $7\pm{24}$ & 0.29 \\
        J0944+3442 & 0.02005 & 86 & $0.92\pm{0.01}$ & $0.85\pm0.11$ & $21.51\pm{0.03}$ & $13\pm{23}$ & 1 \\
        J1016+3754 & 0.00388 & 20 & ... & ... & ... & ... & 0.24 \\
        J1024+0524 & 0.03319 & 143 & $0.54\pm{0.05}$$^c$ & $0.54\pm0.05$ & $20.58\pm{0.07}$ & $-72\pm{13}$ & 1.6 \\
        J1025+3622 & 0.12650 & 593 & $0.47\pm{0.12}$ & $0.56\pm0.05$ & $20.69\pm{0.19}$ & $-151\pm{34}$ & 5.7 \\
        J1044+0353 & 0.01287 & 60 & $0.09\pm{0.03}$ & $0.50\pm0.09$ & $20.08\pm{0.44}$ & $-101\pm{75}$ & 0.71 \\
         & & & $0.91\pm{0.03}$ & $0.91\pm0.05$ & $21.84\pm{0.03}$ & $-29\pm{9}$ \\
        J1105+4444$^e$ & 0.02154 & 92 & $0.42\pm{0.07}$ & ... & $20.88\pm{0.09}$ & $-121\pm{31}$ & 1.1 \\
         & & & $0.58\pm{0.07}$ & ... & $21.87\pm{0.07}$ & $-54\pm{85}$ &  \\
        J1112+5503 & 0.13164 & 619 & ... & ... & ... & ... & 5.9 \\
        J1119+5130 & 0.00446 & 22 & $0.80\pm{0.02}$ & $0.68\pm0.04$ & $20.77\pm{0.03}$ & $-2\pm{8}$ & 0.26 \\
        J1129+2034 & 0.00470 & 27 & $0.99\pm{0.01}$ & $0.87\pm0.04$ & $21.11\pm{0.01}$ & $23\pm{5}$ & 0.32 \\
        J1132+5722 & 0.00504 & 24 & $1.00\pm{0.01}$ & $0.71\pm0.11$ & $21.24\pm{0.02}$ & $17\pm{15}$ & 0.29 \\
        J1132+1411 & 0.01764 & 74 & $1.00$ & $0.98\pm0.01$ & $20.53\pm{0.01}$ & $-40\pm{4}$ & 0.87 \\
        J1144+4012 & 0.12695 & 595 & $0.92\pm{0.03}$ & $0.91\pm0.03$ & $20.52\pm{0.06}$ & $-225\pm{36}$ & 5.7 \\
        J1148+2546 & 0.04512 & 195 & $0.78\pm{0.02}$ & $0.93\pm0.06$ & $21.19\pm{0.03}$ & $-75\pm{11}$ & 2.2 \\
        J1150+1501 & 0.00245 & 11 & $0.89\pm{0.00}$ & $0.81\pm0.03$$^d$ & $21.04\pm{0.01}$ & $10\pm{5}$ & 0.13 \\
        J1157+3220 & 0.01097 & 52 & ... & ... & ... & ... & 0.62 \\
        J1200+1343 & 0.06675 & 300 & ... & ... & ... & ... & 3.2 \\
        J1225+6109 & 0.00234 & 10 & $0.96\pm{0.01}$ & ...$^d$ & $21.26\pm{0.01}$ & $11\pm{5}$$^b$ & 0.12 \\
        J1253-0312 & 0.02272 & 100 & $0.47\pm{0.01}$ & $0.62\pm0.04$ & $21.41\pm{0.04}$ & $-45\pm{6}$ & 1.2 \\
        J1314+3452 & 0.00288 & 12 & $0.95\pm{0.01}$ & $0.92\pm0.02$$^d$ & $20.71\pm{0.01}$ & $-39\pm{2}$ & 0.15 \\
        J1323-0132 & 0.02246 & 93 & ... & ... & ... & ... & 1.1 \\
        J1359+5726 & 0.03383 & 140 & $0.56\pm{0.4}$ & $0.50\pm0.02$ & $20.07\pm{0.11}$ & $-306\pm{17}$ & 1.6 \\
         & & & $0.34\pm0.1$ & $0.82\pm0.04$ & $21.44\pm0.15$ & $-106\pm10$ & \\
        J1416+1223 & 0.12316 & 576 & $0.80\pm{0.05}$$^c$ & $0.80\pm0.05$ & $20.19\pm{0.05}$ & $16\pm{13}$ & 5.5 \\
        J1418+2102 & 0.00855 & 40 & $0.70\pm{0.02}$ & $0.79\pm0.07$ & $21.30\pm{0.03}$ & $18\pm{6}$ & 0.48 \\
        J1428+1653 & 0.18167 & 881 & ... & ... & ... & ... & 7.6 \\
        J1429+0643 & 0.17350 & 837 & ... & ... & ... & ... & 7.4 \\
        J1444+4237 & 0.00230 & 9 & $0.98\pm{0.00}$ & $0.44\pm0.3$$^d$ & $21.57\pm{0.01}$ & $-46\pm{8}$ & 0.11 \\
        J1448-0110 & 0.02741 & 111 & $>0.99$ & $0.95\pm0.03$ & $21.56\pm{0.01}$ & $23\pm{5}$ & 1.3 \\
        J1521+0759 & 0.09426 & 432 & $0.49\pm{0.06}$$^c$ & $0.49\pm0.06$ & $20.42\pm{0.11}$ & $-85\pm{19}$ & 4.4 \\
        J1525+0757 & 0.07579 & 343 & ... & ... & ... & ... & 3.6 \\
        J1545+0858 & 0.03772 & 159 & $0.40\pm{0.02}$ & $0.63\pm0.1$ & $21.54\pm{0.04}$ & $-25\pm{12}$ & 1.8 \\
        J1612+0817 & 0.14914 & 709 & ... & ... & ... & ... & 6.5 \\
        \hline
    \end{tabular}
    \tablecomments{(1) object name; (2) spectroscopic redshift from \citep{Berg2022}; (3) luminosity distance; (4) covering fraction of DLA absorber; (5) covering fraction of O \textsc{i} absorption; (6) column density of DLA absorber; (7) velocity of O \textsc{i} absorption line (except J1105+4444); (8) physical size of COS aperture.
    \begin{itemize}
        \item[$^a$] The luminosity distances have been corrected for cosmic flow using \texttt{Cosmicflows-3} model\footnote{\url{http://edd.ifa.hawaii.edu/}} \citep{Kourkchi2020}.
        We adopt the CF3 model, as it considers several mass concentrations, including the Virgo Cluster, the Great Attractor, etc., and provides distance-velocity relation for every random galaxies at distance within 200 Mpc.
        \item[$^b$] The velocity of O \textsc{i} absorber is measured using C \textsc{ii} absorption line.
        \item[$^c$] The covering fraction of DLA absorber is measured using O \textsc{i} absorption line.
        \item[$^d$] The covering fraction of O \textsc{i} might be underestimated due to contamination of geocoronal O \textsc{i} emission line.
        \item[$^e$] The Voigt parameters of J1105+4444 are fitted using two free velocities. For more details, see Sec. \ref{sec:notes}.
    \end{itemize}
    }
\end{threeparttable}
\end{table*}

\begin{longrotatetable}
\begin{deluxetable*}{c c c c c c c c c c c c}
    \tablecaption{\lya\ Measurements \label{tab:2}}
    \tablewidth{900pt}
    \tabletypesize{\scriptsize}
    \tablehead{
        \colhead{object} & \colhead{$f_\mathrm{Ly\alpha}$} & \colhead{$\log L_\mathrm{Ly\alpha}$} & \colhead{$\mathrm{EW_{Ly\alpha}}$} & \colhead{$f^{\mathrm{Ly\alpha}}_{\mathrm{esc}}$} & \colhead{A$_f$} & \colhead{$\Delta v_\mathrm{Ly\alpha}$} & \colhead{$v^\mathrm{blue}_\mathrm{Ly\alpha}$} & \colhead{$v^\mathrm{red}_\mathrm{Ly\alpha}$} & \colhead{$v^\mathrm{trough}_\mathrm{Ly\alpha}$} & \colhead{$f^\mathrm{blue}_\mathrm{Ly\alpha}$} & \colhead{$f^\mathrm{red}_\mathrm{Ly\alpha}$} \\
        \colhead{} & \colhead{$10^{-15}$ erg s$^{-1}$ cm$^{-2}$} & \colhead{erg s$^{-1}$} & \colhead{\AA} & \colhead{$\%$} & \colhead{} & \colhead{km s$^{-1}$} & \colhead{km s$^{-1}$} & \colhead{km s$^{-1}$} & \colhead{km s$^{-1}$} & \colhead{$10^{-15}$ erg s$^{-1}$ cm$^{-2}$} & \colhead{$10^{-15}$ erg s$^{-1}$ cm$^{-2}$} \\
         (1) & (2) & (3) & (4) & (5) & (6) & (7) & (8) & (9) & (10) & (11) & (12)\\
}
        \startdata
        \multicolumn{12}{c}{Double Peaks} \\
        \hline
        J0021+0052 & $144.56\pm{1.43}$ & $42.5$ & $29.04\pm{0.29}$ & $25\pm{0.45}$ & $2.91\pm0.04$ & $571\pm{54}$ & $-419\pm{53}$ & $152\pm{12}$ & $-27\pm{20}$ & $8.4\pm{0.5}$ & $136.4\pm{1.4}$\\
        J0808+3948 & $64.31\pm{0.52}$ & $42.1$ & $15.02\pm{0.12}$ & $27\pm{0.22}$ & $1.29\pm0.12$ & $507\pm{28}$ & $-470\pm{26}$ & $37\pm{9}$ & $-312\pm{6}$ & $3.2\pm{0.2}$ & $61.2\pm{0.5}$\\
        J0926+4427 & $64.64\pm{0.56}$ & $42.8$ & $40.65\pm{0.35}$ & $35\pm{0.67}$ & $1.42\pm0.13$ & $427\pm{52}$ & $-203\pm{45}$ & $224\pm{25}$ & $-47\pm{17}$ & $7.5\pm{0.2}$ & $57.1\pm{0.5}$\\
        J0938+5428 & $21.14\pm{0.52}$ & $41.7$ & $4.06\pm{0.10}$ & $3.2\pm{0.083}$ & $1.93\pm0.19$ & $669\pm{52}$ & $-296\pm{41}$ & $373\pm{31}$ & $116\pm{32}$ & $7.4\pm{0.3}$ & $13.8\pm{0.4}$\\
        J0942+3547 & $97.61\pm{0.31}$ & $40.7$ & $17.95\pm{0.06}$ & $18\pm{0.093}$ & $1.53\pm0.15$ & $267\pm16$ & $-113\pm14$ & $154\pm7$ & $-16\pm7$ & $14.6\pm0.3$ & $82.6\pm0.4$ \\
        J0944-0038 & $20.47\pm{0.43}$ & $39.2$ & $9.95\pm{0.21}$ & $4.1\pm{0.085}$ & $0.86\pm0.35$ & $416\pm{65}$ & $-150\pm{61}$ & $267\pm{23}$ & $34\pm{82}$ & $3.4\pm{0.7}$ & $17.1\pm{0.8}$\\
        J0944+3442 & $0.43\pm{0.09}$ & $38.6$ & $0.56\pm{0.11}$ & $0.82\pm{0.17}$ & $1.71\pm0.59$ & $531\pm{99}$ & $-273\pm{67}$ & $257\pm{76}$ & $-20\pm{150}$ & $0.1\pm{0.1}$ & $0.4\pm{0.1}$\\
        J1016+3754 & $146.06\pm{1.74}$ & $39.8$ & $15.42\pm{0.18}$ & $12\pm{0.16}$ & $1.51\pm0.21$ & $404\pm{45}$ & $-230\pm{43}$ & $175\pm{13}$ & $-34\pm{22}$ & $12.2\pm{0.9}$ & $133.9\pm{1.4}$\\
        J1024+0524 & $54.13\pm{0.50}$ & $41.1$ & $8.72\pm{0.08}$ & $5\pm{0.097}$  & $2.87\pm0.08$ & $464\pm{36}$ & $-338\pm{35}$ & $126\pm{8}$ & $-64\pm{18}$ & $2.1\pm{0.3}$ & $52.1\pm{0.5}$\\
        J1025+3622 & $53.28\pm{0.62}$ & $42.3$ & $21.95\pm{0.25}$ & $17\pm{0.25}$ & $1.70\pm0.15$ & $469\pm{43}$ & $-263\pm{39}$ & $206\pm{17}$ & $-100\pm{25}$ & $4.5\pm{0.2}$ & $48.8\pm{0.6}$\\
        J1044+0353 & $1.55\pm{0.08}$ & $38.8$ & $0.90\pm{0.05}$ & $0.096\pm{0.005}$ & $1.84\pm0.36$ & $425\pm{119}$ & $-293\pm{111}$ & $132\pm{45}$ & $-123\pm{98}$ & $0.2\pm{0.1}$ & $1.3\pm{0.1}$\\
        J1105+4444 & $2.52\pm{0.18}$ & $39.4$ & $0.53\pm{0.04}$ & $0.072\pm{0.0051}$ & ... & $999\pm{109}$ & $-517\pm{76}$ & $482\pm{76}$ & $-204\pm{244}$ & $0.4\pm{0.2}$ & $2.2\pm{0.2}$\\
        J1119+5130 & $2.30\pm{0.16}$ & $38.1$ & $0.71\pm{0.05}$ & $0.78\pm{0.053}$ & $4.20\pm0.27$ & $649\pm{76}$ & $-337\pm{67}$ & $312\pm{34}$ & $-67\pm{108}$ & $1.2\pm{0.2}$ & $1.1\pm{0.1}$\\
        J1148+2546 & $12.34\pm{0.23}$ & $40.7$ & $5.62\pm{0.10}$ & $7.4\pm{0.14}$  & $1.92\pm0.46$ & $717\pm{97}$ & $-450\pm{73}$ & $268\pm{63}$ & $-132\pm{58}$ & $0.4\pm{0.1}$ & $12.0\pm{0.2}$\\
        J1200+1343 & $80.54\pm{0.44}$ & $41.9$ & $56.87\pm{0.31}$ & $9.3\pm{0.11}$ & $2.68\pm0.05$ & $530\pm{21}$ & $-394\pm{20}$ & $136\pm{5}$ & $-48\pm{20}$ & $9.0\pm{0.2}$ & $71.6\pm{0.4}$\\
        J1253-0312 & $338.94\pm{1.21}$ & $41.6$ & $32.14\pm{0.12}$ & $4.4\pm{0.032}$ & $1.48\pm0.11$ & $433\pm{14}$ & $-214\pm{11}$ & $219\pm{8}$ & $-46\pm{9}$ & $37.7\pm{0.5}$ & $301.1\pm{1.1}$\\
        J1323-0132 & $190.14\pm{0.56}$ & $41.3$ & $81.19\pm{0.24}$ & $21\pm{0.1}$ & $1.78\pm0.06$ & $168\pm{12}$ & $-95\pm{12}$ & $74\pm{1}$ & $-35\pm{7}$ & $45.0\pm{1.5}$ & $142.7\pm{1.5}$\\
        J1416+1223 & $11.70\pm{0.47}$ & $41.7$ & $3.26\pm{0.13}$ & $3\pm{0.13}$ & $1.75\pm0.28$ & $613\pm{91}$ & $-315\pm{80}$ & $298\pm{43}$ & $35\pm{51}$ & $6.0\pm{0.4}$ & $5.7\pm{0.3}$\\
        J1418+2102 & $26.76\pm{0.22}$ & $39.7$ & $19.14\pm{0.16}$ & $2.4\pm{0.021}$ & $1.10\pm0.18$ & $403\pm{26}$ & $-122\pm{25}$ & $281\pm{9}$ & $25\pm{9}$ & $7.6\pm{0.2}$ & $19.0\pm{0.2}$\\
        J1428+1653 & $38.94\pm{0.82}$ & $42.6$ & $12.32\pm{0.26}$ & $27\pm{1}$ & $3.34\pm0.34$ & $416\pm{84}$ & $-328\pm{76}$ & $89\pm{38}$ & $-109\pm{66}$ & $2.6\pm{0.3}$ & $36.4\pm{0.8}$\\
        J1429+0643 & $67.53\pm{0.90}$ & $42.8$ & $33.22\pm{0.44}$ & $11\pm{0.18}$ & $3.56\pm0.15$ & $545\pm{72}$ & $-292\pm{67}$ & $253\pm{25}$ & $54\pm{33}$ & $15.2\pm{0.5}$ & $52.4\pm{0.7}$\\
        J1448-0110 & $0.43\pm{0.13}$ & $38.8$ & $0.09\pm{0.03}$ & $0.027\pm{0.0083}$ & ... & $395\pm{153}$ & $-354\pm{124}$ & $40\pm{91}$ & $-283\pm{134}$ & $0.1\pm{0.1}$ & $0.3\pm{0.1}$\\
        J1521+0759 & $27.21\pm{0.64}$ & $41.8$ & $4.85\pm{0.11}$ & $12\pm{0.52}$ & $3.27\pm0.09$ & $404\pm{48}$ & $-247\pm{45}$ & $157\pm{16}$ & $-68\pm{29}$ & $-1.0\pm{0.4}$ & $28.4\pm{0.7}$\\
        J1545+0858 & $160.58\pm{1.02}$ & $41.7$ & $29.26\pm{0.19}$ & $7.5\pm{0.048}$ & $2.75\pm0.08$ & $284\pm{14}$ & $-122\pm{10}$ & $163\pm{10}$ & $-27\pm{15}$ & $6.4\pm{0.3}$ & $154.2\pm{1.0}$\\
        \hline
        \multicolumn{12}{c}{Single Peak / P-Cygni}\\
        \hline
        J0036-3333 & $175.93\pm{1.20}$ & $41.1$ & $7.13\pm{0.05}$ & $9.2\pm{0.063}$&...& ... & ... & $100\pm{5}$ & ... & ... & $175.9\pm{1.2}$ \\ 
        J0940+2935 & $0.58\pm{0.06}$ & $36.8$ & $0.34\pm{0.03}$ & $0.46\pm{0.044}$&...& ... & ... & $273\pm{57}$ & ... & ... & $0.6\pm{0.1}$ \\ 
        J1112+5503 & $12.43\pm{0.45}$ & $41.8$ & $5.45\pm{0.20}$ & $3.8\pm{0.15}$&...& ... & ... & $143\pm{47}$ & ... & ... & $12.4\pm{0.4}$ \\ 
        J1144+4012 & $2.50\pm{0.21}$ & $41.0$ & $1.75\pm{0.15}$ & $1.6\pm{0.14}$&...& ... & ... & $318\pm{70}$ & ... & ... & $2.5\pm{0.2}$ \\ 
        J1157+3220 & $389.02\pm{2.53}$ & $41.1$ & $21.63\pm{0.14}$ & $23\pm{0.2}$&...& ... & ... & $50\pm{13}$ & ... & ... & $389.0\pm{2.5}$ \\ 
        J1225+6109 & $0.50\pm{0.21}$ & $36.8$ & $0.04\pm{0.02}$ & $0.016\pm{0.0065}$&...& ... & ... & $52\pm{63}$ & ... & ... & $0.5\pm{0.2}$ \\ 
        J1314+3452 & $0.85\pm{0.06}$ & $37.2$ & $0.21\pm{0.01}$ & $0.027\pm{0.0018}$&...& ... & ... & $210\pm{47}$ & ... & ... & $0.8\pm{0.1}$ \\ 
        J1359+5726 & $72.71\pm{0.68}$ & $41.2$ & $8.71\pm{0.08}$ & $6.7\pm{0.081}$&...& ... & ... & $148\pm{16}$ & ... & ... & $72.7\pm{0.7}$ \\ 
        J1525+0757 & $67.41\pm{1.21}$ & $42.0$ & $14.92\pm{0.27}$ & $16\pm{0.43}$&...& ... & ... & $82\pm{7}$ & ... & ... & $67.4\pm{1.2}$ \\ 
        J1612+0817 & $36.97\pm{0.83}$ & $42.3$ & $12.71\pm{0.29}$ & $7\pm{0.18}$&...& ... & ... & $114\pm{14}$ & ... & ... & $37.0\pm{0.8}$ \\ 
        \hline
    \enddata
    \tablecomments{(1) object name; (2) \lya\ flux; (3) \lya\ luminosity; (4) \lya\ equivalent width; (5) \lya\ escape fraction; (6) \lya\ red peak asymmetry; (7) \lya\ peak separation; (8) \lya\ blue peak velocity offset; (9) \lya\ red peak velocity offset; (10) \lya\ trough velocity offset; (11) \lya\ blue peak flux; (12) \lya\ red peak flux. We note that the luminosity distances of some galaxies used in this work are different with those in \citep{Berg2022} because of the correction of cosmic flow. The properties (e.g., stellar mass, star formation rate)  of those galaxies which rely on the luminosities are scaled accordingly.}
\end{deluxetable*}
\end{longrotatetable}

\begin{deluxetable*}{c c c c c c c c c}
\tablecaption{Ancillary data \label{tab:ancillary}}
    \tablehead{
        \colhead{object} & \colhead{$f_\mathrm{1500}$} & \colhead{M$_\mathrm{1500}$} & \colhead{Z$_\mathrm{neb}$} & \colhead{$\log M_\star$} & \colhead{E(B-V)} & \colhead{O32} & \colhead{$v^\mathrm{outflow}_\mathrm{Si\ II}$} & \colhead{$r_\mathrm{50}$} \\
        \colhead{} & \colhead{$10^{-15}$ erg s$^{-1}$ cm$^{-2}$} & \colhead{} & \colhead{} & \colhead{$M_\odot$} & \colhead{} & \colhead{} & \colhead{km s$^{-1}$} & \colhead{$\arcsec$}  \\
         (1) & (2) & (3) & (4) & (5) & (6) & (7) & (8) & (9) \\
}
\startdata
J0021+0052 & 3.94 & -20.55 & $8.17\pm0.07$ & $9.09^{+0.18}_{-0.38}$ & $0.13\pm0.006$ & $2.0\pm0.1$ & $231^{+77}_{-77}$ & 0.25 \\
J0036-3333 & 16.60 & -18.34 & $8.21\pm0.17$ & $9.09^{+0.26}_{-0.23}$ & $0.30\pm0.012$ & $1.1\pm0.1$ & $157^{+22}_{-22}$ & 0.28 \\
J0127-0619 & 4.04 & -13.58 & $7.68\pm0.02$ & $8.63^{+0.18}_{-0.15}$ & $0.48\pm0.006$ & $1.1\pm0.1$ & ... & 0.15 \\
J0144+0453 & 1.87 & -12.63 & $7.76\pm0.02$ & $7.52^{+0.24}_{-0.29}$ & $0.04\pm0.030$ & $2.1\pm0.1$ & $48^{+16}_{-16}$ & 3.54 \\
J0337-0502 & 7.99 & -16.60 & $7.46\pm0.04$ & $7.01^{+0.24}_{-0.21}$ & $0.05\pm0.006$ & $6.2\pm0.2$ & ... & 1.62 \\
J0405-3648 & 0.96 & -10.90 & $7.04\pm0.05$ & $6.60^{+0.28}_{-0.28}$ & $0.11\pm0.005$ & $0.6\pm0.1$ & ... & 6.43 \\
J0808+3948 & 3.42 & -20.23 & $8.77\pm0.12$ & $9.12^{+0.30}_{-0.17}$ & $0.24\pm0.070$ & $0.8\pm0.1$ & $646^{+65}_{-65}$ & 0.08 \\
J0823+2806 & 3.85 & -18.86 & $8.28\pm0.01$ & $9.38^{+0.33}_{-0.19}$ & $0.21\pm0.004$ & $2.0\pm0.1$ & $136^{+45}_{-45}$ & 0.28 \\
J0926+4427 & 1.14 & -20.64 & $8.08\pm0.02$ & $8.76^{+0.30}_{-0.26}$ & $0.10\pm0.008$ & $3.1\pm0.1$ & $353^{+52}_{-52}$ & 0.23 \\
J0934+5514 & 15.10 & -14.05 & $6.98\pm0.01$ & $6.25^{+0.15}_{-0.20}$ & $0.07\pm0.007$ & $8.7\pm0.1$ & $112^{+37}_{-37}$ & 1.53 \\
J0938+5428 & 3.56 & -20.53 & $8.25\pm0.02$ & $9.15^{+0.18}_{-0.29}$ & $0.13\pm0.006$ & $1.9\pm0.1$ & $215^{+72}_{-72}$ & 0.28 \\
J0940+2935 & 1.45 & -11.14 & $7.66\pm0.07$ & $6.80^{+0.23}_{-0.40}$ & $0.06\pm0.010$ & $0.7\pm0.1$ & $102^{+34}_{-34}$ & 3.06 \\
J0942+3547 & 3.80 & -16.30 & $8.13\pm0.03$ & $7.56^{+0.21}_{-0.29}$ & $0.06\pm0.011$ & $2.6\pm0.1$ & $97^{+26}_{-26}$ & 0.33 \\
J0944-0038 & 1.40 & -13.07 & $7.83\pm0.01$ & $6.89^{+0.44}_{-0.25}$ & $0.16\pm0.010$ & $2.9\pm0.1$ & $64^{+21}_{-21}$ & 2.34 \\
J0944+3442 & 0.69 & -15.06 & $7.62\pm0.11$ & $8.19^{+0.40}_{-0.23}$ & $0.16\pm0.013$ & $1.4\pm0.1$ & ... & 3.74 \\
J1016+3754 & 7.07 & -14.43 & $7.56\pm0.01$ & $6.77^{+0.27}_{-0.22}$ & $0.07\pm0.012$ & $4.6\pm0.2$ & $116^{+31}_{-31}$ & 1.52 \\
J1024+0524 & 4.50 & -18.20 & $7.84\pm0.03$ & $7.88^{+0.37}_{-0.24}$ & $0.10\pm0.016$ & $2.1\pm0.1$ & $94^{+12}_{-12}$ & 0.40 \\
J1025+3622 & 1.81 & -20.30 & $8.13\pm0.01$ & $8.87^{+0.25}_{-0.27}$ & $0.09\pm0.006$ & $2.4\pm0.1$ & $155^{+24}_{-24}$ & 0.35 \\
J1044+0353 & 1.70 & -15.25 & $7.45\pm0.03$ & $6.84^{+0.41}_{-0.26}$ & $0.08\pm0.007$ & $6.8\pm0.1$ & $52^{+12}_{-12}$ & 0.38 \\
J1105+4444 & 4.68 & -17.28 & $8.23\pm0.01$ & $8.98^{+0.29}_{-0.24}$ & $0.17\pm0.005$ & $2.0\pm0.1$ & $115^{+23}_{-23}$ & 4.11 \\
J1112+5503 & 1.91 & -20.45 & $8.45\pm0.06$ & $9.59^{+0.33}_{-0.19}$ & $0.23\pm0.016$ & $0.9\pm0.1$ & $349^{+107}_{-107}$ & 0.20 \\
J1119+5130 & 2.63 & -13.54 & $7.57\pm0.04$ & $6.81^{+0.15}_{-0.28}$ & $0.10\pm0.008$ & $2.0\pm0.1$ & $65^{+22}_{-22}$ & 2.18 \\
J1129+2034 & 1.87 & -13.62 & $8.28\pm0.04$ & $8.20^{+0.37}_{-0.27}$ & $0.23\pm0.011$ & $1.8\pm0.1$ & $51^{+17}_{-17}$ & 0.38 \\
J1132+5722 & 2.57 & -13.69 & $7.58\pm0.08$ & $7.32^{+0.23}_{-0.26}$ & $0.10\pm0.008$ & $0.8\pm0.1$ & ... & 0.84 \\
J1132+1411 & 1.75 & -15.75 & $8.25\pm0.01$ & $8.67^{+0.28}_{-0.19}$ & $0.13\pm0.008$ & $2.7\pm0.1$ & $60^{+10}_{-10}$ & 8.86 \\
J1144+4012 & 1.20 & -19.86 & $8.43\pm0.20$ & $9.89^{+0.18}_{-0.29}$ & $0.22\pm0.010$ & $0.6\pm0.1$ & $246^{+33}_{-33}$ & 0.40 \\
J1148+2546 & 2.07 & -18.03 & $7.94\pm0.01$ & $8.13^{+0.34}_{-0.24}$ & $0.10\pm0.021$ & $3.7\pm0.1$ & $95^{+19}_{-19}$ & 1.31 \\
J1150+1501 & 12.60 & -13.71 & $8.14\pm0.01$ & $6.83^{+0.28}_{-0.30}$ & $0.04\pm0.004$ & $2.3\pm0.1$ & $67^{+22}_{-22}$ & 1.29 \\
J1157+3220 & 14.40 & -17.27 & $8.43\pm0.02$ & $9.08^{+0.32}_{-0.18}$ & $0.08\pm0.006$ & $1.2\pm0.1$ & $238^{+49}_{-49}$ & 2.89 \\
J1200+1343 & 1.38 & -18.53 & $8.26\pm0.02$ & $8.12^{+0.47}_{-0.42}$ & $0.15\pm0.006$ & $5.1\pm0.1$ & $192^{+13}_{-13}$ & 0.18 \\
J1225+6109 & 9.50 & -13.28 & $7.97\pm0.01$ & $7.09^{+0.34}_{-0.24}$ & $0.11\pm0.005$ & $4.7\pm0.1$ & $51^{+17}_{-17}$ & 2.91 \\
J1253-0312 & 9.11 & -18.19 & $8.06\pm0.01$ & $7.66^{+0.51}_{-0.23}$ & $0.16\pm0.008$ & $8.0\pm0.2$ & $113^{+38}_{-38}$ & 0.85 \\
J1314+3452 & 3.72 & -12.65 & $8.26\pm0.01$ & $7.53^{+0.30}_{-0.21}$ & $0.14\pm0.006$ & $2.3\pm0.1$ & $62^{+21}_{-21}$ & 0.30 \\
J1323-0132 & 1.33 & -15.94 & $7.71\pm0.04$ & $6.29^{+0.26}_{-0.10}$ & $0.13\pm0.042$ & $37.8\pm3.0$ & ... & 0.23 \\
J1359+5726 & 6.34 & -18.53 & $7.98\pm0.01$ & $8.39^{+0.31}_{-0.26}$ & $0.09\pm0.006$ & $2.6\pm0.1$ & $161^{+23}_{-23}$ & 1.10 \\
J1416+1223 & 2.62 & -20.63 & $8.53\pm0.11$ & $9.59^{+0.32}_{-0.26}$ & $0.25\pm0.008$ & $0.8\pm0.1$ & $398^{+68}_{-68}$ & 0.13 \\
J1418+2102 & 1.17 & -13.99 & $7.75\pm0.02$ & $6.26^{+0.49}_{-0.35}$ & $0.08\pm0.006$ & $4.7\pm0.1$ & $51^{+7}_{-7}$ & 0.40 \\
J1428+1653 & 1.25 & -20.75 & $8.33\pm0.05$ & $9.56^{+0.15}_{-0.23}$ & $0.14\pm0.008$ & $1.2\pm0.1$ & $140^{+25}_{-25}$ & 0.35 \\
J1429+0643 & 1.62 & -20.92 & $8.10\pm0.03$ & $8.80^{+0.35}_{-0.21}$ & $0.12\pm0.012$ & $4.2\pm0.2$ & $230^{+51}_{-51}$ & 0.15 \\
J1444+4237 & 2.08 & -11.33 & $7.64\pm0.02$ & $6.39^{+0.17}_{-0.17}$ & $0.08\pm0.053$ & $4.1\pm0.1$ & $54^{+18}_{-18}$ & 8.20 \\
J1448-0110 & 4.08 & -17.55 & $8.13\pm0.01$ & $7.58^{+0.41}_{-0.24}$ & $0.15\pm0.005$ & $8.0\pm0.1$ & $145^{+43}_{-43}$ & 0.23 \\
J1521+0759 & 3.52 & -20.33 & $8.31\pm0.14$ & $9.00^{+0.29}_{-0.30}$ & $0.15\pm0.008$ & $1.5\pm0.1$ & $161^{+54}_{-54}$ & 0.28 \\
J1525+0757 & 3.52 & -19.83 & $8.33\pm0.04$ & $10.06^{+0.28}_{-0.42}$ & $0.25\pm0.008$ & $0.5\pm0.1$ & $408^{+28}_{-28}$ & 0.25 \\
J1545+0858 & 4.37 & -18.40 & $7.75\pm0.03$ & $7.50^{+0.43}_{-0.26}$ & $0.11\pm0.036$ & $8.6\pm0.3$ & $113^{+33}_{-33}$ & 0.33 \\
J1612+0817 & 2.70 & -21.12 & $8.18\pm0.19$ & $9.78^{+0.28}_{-0.26}$ & $0.29\pm0.008$ & $0.7\pm0.1$ & $459^{+63}_{-63}$ & 0.20 \\
\enddata
    \tablecomments{(1) object name; (2) UV flux at 1500 \AA\ from \citet{Berg2022}; (3) UV absolute magnitude at 1500 \AA; (4) metallicity from \citet{Berg2022}; (5) stellar mass; (6) dust extinction from \citet{Berg2022}; (7) O32 ratio; (8) velocity of Si \textsc{ii} absorption line; (9) half light radius from \citet{Xu2022}.}
\end{deluxetable*}

\begin{deluxetable*}{c c c c c c c c}
    \tablecaption{Best-fit parameters of the second attempt \label{tab:tlac2}}
        \tablehead{
        object & $z_{tlac}$ & $v_\mathrm{exp}$ & log $N_\mathrm{HI}$ & log T  & log $\tau$ & $\sigma_i$ & EW$_i$ \\
         & & km s$^{-1}$ & & K & & km s $^{-1}$ & \AA \\
         (1) & (2) & (3) & (4) & (5) & (6) & (7) & (8)}
        \startdata
J0021+0052 & $0.098902$ & $214^{+1}_{-2}$ & $18.79^{+0.11}_{-0.08}$ & $3.8^{+0.2}_{-0.1}$ & $-2.05^{+1.01}_{-0.11}$ & $117^{+1}_{-1}$ & $16.8^{+0.8}_{-0.7}$ \\ 
J0036-3333 & $0.020939$ & $207^{+2}_{-1}$ & $18.68^{+0.09}_{-0.06}$ & $3.5^{+0.2}_{-0.1}$ & $-2.10^{+1.13}_{-0.06}$ & $93^{+1}_{-1}$ & $6.6^{+0.5}_{-0.1}$ \\ 
J0808+3948 & $0.091384$ & $365^{+1}_{-1}$ & $16.76^{+0.08}_{-0.04}$ & $3.4^{+0.1}_{-0.1}$ & $-1.57^{+0.22}_{-0.11}$ & $103^{+1}_{-1}$ & $8.7^{+0.1}_{-0.1}$ \\ 
J0926+4427 & $0.180817$ & $131^{+2}_{-3}$ & $19.23^{+0.05}_{-0.08}$ & $3.7^{+0.1}_{-0.1}$ & $-1.74^{+0.11}_{-0.21}$ & $248^{+2}_{-3}$ & $31.8^{+1.0}_{-1.0}$ \\ 
J0938+5428 & $0.102513$ & $21^{+4}_{-2}$ & $19.79^{+0.08}_{-0.08}$ & $4.2^{+0.2}_{-0.1}$ & $-0.68^{+0.74}_{-0.87}$ & $308^{+5}_{-4}$ & $74.5^{+5.3}_{-4.4}$ \\ 
J0942+3547 & $0.015121$ & $86^{+1}_{-1}$ & $18.20^{+0.06}_{-0.06}$ & $3.1^{+0.1}_{-0.1}$ & $-3.31^{+0.38}_{-0.09}$ & $167^{+1}_{-1}$ & $18.3^{+0.1}_{-0.1}$ \\ 
J0944-0038 & $0.005187$ & $119^{+4}_{-4}$ & $18.60^{+0.09}_{-0.07}$ & $3.4^{+0.2}_{-0.2}$ & $-1.35^{+0.20}_{-0.12}$ & $218^{+3}_{-4}$ & $2130.2^{+1851.2}_{-1159.2}$ \\ 
J0944+3442 & $0.020226$ & $72^{+18}_{-17}$ & $19.44^{+0.21}_{-0.23}$ & $4.1^{+0.7}_{-0.8}$ & $0.44^{+0.46}_{-0.43}$ & $182^{+12}_{-11}$ & $39.6^{+21.4}_{-14.0}$ \\ 
J1016+3754 & $0.004131$ & $96^{+2}_{-1}$ & $18.85^{+0.07}_{-0.11}$ & $4.3^{+0.1}_{-0.2}$ & $0.10^{+0.80}_{-0.75}$ & $142^{+1}_{-2}$ & $26.2^{+1.9}_{-1.9}$ \\ 
J1024+0524 & $0.033425$ & $176^{+2}_{-1}$ & $19.01^{+0.08}_{-0.09}$ & $5.1^{+0.1}_{-0.2}$ & $-1.89^{+0.92}_{-0.16}$ & $82^{+1}_{-1}$ & $16.2^{+0.6}_{-0.4}$ \\ 
J1025+3622 & $0.126786$ & $167^{+3}_{-2}$ & $18.99^{+0.09}_{-0.07}$ & $4.2^{+0.2}_{-0.1}$ & $-0.88^{+0.40}_{-0.55}$ & $245^{+3}_{-3}$ & $41.3^{+2.1}_{-1.9}$ \\ 
J1044+0353 & $0.013070$ & $175^{+10}_{-15}$ & $18.18^{+0.20}_{-0.19}$ & $3.5^{+0.3}_{-0.4}$ & $-1.14^{+0.25}_{-0.10}$ & $248^{+14}_{-9}$ & $9.9^{+0.9}_{-0.7}$ \\ 
J1112+5503 & $0.131707$ & $210^{+3}_{-3}$ & $19.55^{+0.10}_{-0.05}$ & $4.6^{+0.3}_{-0.1}$ & $0.55^{+1.10}_{-1.17}$ & $270^{+3}_{-4}$ & $20.9^{+1.5}_{-1.6}$ \\ 
J1119+5130 & $0.004536$ & $0^{+2}_{-1}$ & $20.42^{+0.24}_{-0.65}$ & $4.0^{+0.4}_{-0.3}$ & $-1.80^{+1.21}_{-0.03}$ & $316^{+10}_{-8}$ & $6.5^{+6.6}_{-2.4}$ \\ 
J1144+4012 & $0.126832$ & $108^{+11}_{-8}$ & $20.20^{+0.07}_{-0.08}$ & $4.9^{+0.3}_{-0.7}$ & $0.40^{+0.46}_{-0.52}$ & $259^{+9}_{-11}$ & $281.6^{+50.7}_{-41.7}$ \\ 
J1148+2546 & $0.045710$ & $279^{+5}_{-4}$ & $19.19^{+0.09}_{-0.07}$ & $3.8^{+0.2}_{-0.2}$ & $0.53^{+0.67}_{-0.56}$ & $281^{+4}_{-4}$ & $46.3^{+3.7}_{-3.6}$ \\ 
J1157+3220 & $0.010230$ & $163^{+1}_{-2}$ & $20.06^{+0.04}_{-0.09}$ & $5.0^{+0.1}_{-0.2}$ & $0.69^{+1.98}_{-1.60}$ & $191^{+1}_{-2}$ & $200.1^{+2.7}_{-3.5}$ \\ 
J1200+1343 & $0.066942$ & $173^{+1}_{-2}$ & $16.56^{+0.09}_{-0.05}$ & $5.4^{+0.1}_{-0.1}$ & $-2.40^{+0.65}_{-0.12}$ & $266^{+1}_{-1}$ & $53.6^{+1.7}_{-1.8}$ \\ 
J1253-0312 & $0.023087$ & $184^{+0}_{-1}$ & $18.67^{+0.03}_{-0.06}$ & $4.3^{+0.1}_{-0.1}$ & $-2.70^{+0.48}_{-0.00}$ & $256^{+1}_{-1}$ & $31.2^{+0.5}_{-0.4}$ \\ 
J1323-0132 & $0.022534$ & $37^{+1}_{-1}$ & $17.93^{+0.05}_{-0.02}$ & $3.1^{+0.0}_{-0.1}$ & $-2.40^{+0.24}_{-0.12}$ & $121^{+0}_{-0}$ & $75.7^{+1.1}_{-1.1}$ \\ 
J1359+5726 & $0.034107$ & $205^{+2}_{-1}$ & $19.03^{+0.13}_{-0.09}$ & $4.7^{+0.1}_{-0.2}$ & $-1.11^{+0.10}_{-0.47}$ & $128^{+2}_{-2}$ & $51.8^{+2.5}_{-2.2}$ \\ 
J1416+1223 & $0.123181$ & $0^{+2}_{-2}$ & $19.59^{+0.09}_{-0.08}$ & $4.6^{+0.2}_{-0.3}$ & $-0.22^{+0.70}_{-0.73}$ & $289^{+6}_{-6}$ & $49.5^{+4.1}_{-3.9}$ \\ 
J1418+2102 & $0.009016$ & $98^{+2}_{-2}$ & $18.18^{+0.09}_{-0.06}$ & $3.0^{+0.1}_{-0.1}$ & $0.68^{+1.65}_{-1.61}$ & $230^{+1}_{-1}$ & $101.8^{+4.8}_{-4.9}$ \\ 
J1428+1653 & $0.181780$ & $121^{+3}_{-4}$ & $18.80^{+0.10}_{-0.07}$ & $3.5^{+0.1}_{-0.2}$ & $-1.12^{+0.08}_{-0.40}$ & $52^{+1}_{-1}$ & $17.8^{+1.9}_{-1.0}$ \\ 
J1429+0643 & $0.173984$ & $112^{+2}_{-3}$ & $19.21^{+0.08}_{-0.09}$ & $3.1^{+0.2}_{-0.2}$ & $-0.38^{+0.67}_{-0.82}$ & $323^{+5}_{-5}$ & $46.8^{+3.6}_{-3.1}$ \\ 
J1521+0759 & $0.094771$ & $215^{+2}_{-2}$ & $19.08^{+0.12}_{-0.11}$ & $3.0^{+0.2}_{-0.1}$ & $-1.04^{+0.16}_{-0.28}$ & $93^{+2}_{-2}$ & $9.2^{+1.8}_{-0.6}$ \\ 
J1525+0757 & $0.075913$ & $136^{+2}_{-1}$ & $18.58^{+0.09}_{-0.06}$ & $4.5^{+0.2}_{-0.1}$ & $-1.39^{+0.00}_{-0.44}$ & $52^{+1}_{-1}$ & $19.2^{+1.2}_{-1.0}$ \\ 
J1545+0858 & $0.038336$ & $194^{+1}_{-2}$ & $18.41^{+0.07}_{-0.10}$ & $3.0^{+0.2}_{-0.1}$ & $0.69^{+1.91}_{-1.58}$ & $156^{+1}_{-1}$ & $55.3^{+1.5}_{-1.8}$ \\ 
J1612+0817 & $0.149267$ & $246^{+3}_{-2}$ & $19.61^{+0.08}_{-0.10}$ & $5.3^{+0.2}_{-0.2}$ & $0.09^{+1.03}_{-1.11}$ & $112^{+2}_{-2}$ & $35.6^{+2.3}_{-2.0}$ \\ 
    \enddata
    \tablecomments{(1) object name; (2) redshift estimated by the shell model; (3) outflow velocity of expanding shell; (4) H \textsc{i} column density; (5)temperature; (6) dust extinction; (7) intrinsic line width; (8) equivalent width.}
\end{deluxetable*}

\begin{deluxetable*}{c c c c c c c c}
    \tablecaption{Best-fit parameters of the third attempt \label{tab:tlac}}
        \tablehead{
        object & $z_{tlac}$ & $v_\mathrm{exp}$ & log $N_\mathrm{HI}$ & log T  & log $\tau$ & $\sigma_i$ & EW$_i$ \\
         & & km s$^{-1}$ & & K & & km s $^{-1}$ & \AA \\
         (1) & (2) & (3) & (4) & (5) & (6) & (7) & (8)}
        \startdata
J0021+0052 & $0.098535$ & $163^{+1}_{-2}$ & $19.17^{+0.16}_{-0.05}$ & $4.5^{+0.2}_{-0.1}$ & $-0.67^{+0.19}_{-0.07}$ & $225^{+1}_{-1}$ & $26.7^{+4.0}_{-1.1}$ \\ 
J0036-3333 & $0.020553$ & $120^{+2}_{-2}$ & $19.17^{+0.08}_{-0.05}$ & $3.0^{+0.1}_{-0.1}$ & $-0.63^{+0.04}_{-0.04}$ & $142^{+2}_{-2}$ & $9.0^{+0.3}_{-0.3}$ \\ 
J0808+3948 & $0.091375$ & $348^{+2}_{-2}$ & $16.17^{+0.08}_{-0.05}$ & $3.7^{+0.1}_{-0.1}$ & $-1.67^{+0.46}_{-0.73}$ & $102^{+1}_{-1}$ & $8.7^{+0.1}_{-0.1}$ \\ 
J0926+4427 & $0.180816$ & $133^{+1}_{-1}$ & $19.19^{+0.07}_{-0.06}$ & $3.8^{+0.1}_{-0.2}$ & $-1.73^{+0.24}_{-0.38}$ & $244^{+3}_{-3}$ & $29.9^{+1.3}_{-1.2}$ \\ 
J0938+5428 & $0.102247$ & $11^{+1}_{-2}$ & $20.63^{+0.05}_{-0.11}$ & $4.3^{+0.1}_{-0.2}$ & $-2.44^{+0.45}_{-0.56}$ & $291^{+5}_{-4}$ & $31.8^{+3.9}_{-1.7}$ \\ 
J0942+3547 & $0.015010$ & $69^{+0}_{-0}$ & $18.58^{+0.03}_{-0.03}$ & $3.3^{+0.0}_{-0.0}$ & $-3.67^{+0.58}_{-0.73}$ & $170^{+1}_{-1}$ & $18.4^{+0.2}_{-0.2}$ \\ 
J0944-0038 & $0.004912$ & $70^{+3}_{-3}$ & $19.42^{+0.06}_{-0.08}$ & $3.9^{+0.3}_{-0.2}$ & $-2.21^{+0.54}_{-0.71}$ & $217^{+3}_{-4}$ & $302.0^{+23.0}_{-24.3}$ \\ 
J0944+3442 & $0.020138$ & $59^{+9}_{-10}$ & $19.61^{+0.11}_{-0.10}$ & $3.3^{+0.9}_{-0.4}$ & $0.34^{+0.23}_{-0.35}$ & $179^{+11}_{-11}$ & $55.8^{+18.3}_{-18.1}$ \\ 
J1016+3754 & $0.004024$ & $81^{+2}_{-2}$ & $19.04^{+0.13}_{-0.11}$ & $3.8^{+0.3}_{-0.1}$ & $-0.58^{+0.09}_{-0.08}$ & $148^{+2}_{-1}$ & $21.6^{+2.4}_{-1.3}$ \\ 
J1024+0524 & $0.033304$ & $139^{+3}_{-3}$ & $19.05^{+0.12}_{-0.11}$ & $4.6^{+0.1}_{-0.1}$ & $-0.69^{+0.14}_{-0.08}$ & $192^{+2}_{-2}$ & $20.0^{+2.5}_{-0.8}$ \\ 
J1025+3622 & $0.126552$ & $131^{+2}_{-4}$ & $19.43^{+0.07}_{-0.08}$ & $3.9^{+0.3}_{-0.2}$ & $-1.31^{+0.34}_{-0.35}$ & $243^{+3}_{-3}$ & $38.2^{+3.6}_{-2.3}$ \\ 
J1044+0353 & $0.012998$ & $158^{+9}_{-9}$ & $18.40^{+0.11}_{-0.13}$ & $3.4^{+0.3}_{-0.3}$ & $-1.16^{+0.44}_{-0.63}$ & $253^{+12}_{-8}$ & $9.9^{+0.9}_{-0.8}$ \\ 
J1112+5503 & $0.131497$ & $153^{+3}_{-4}$ & $19.83^{+0.06}_{-0.09}$ & $3.8^{+0.2}_{-0.2}$ & $0.25^{+0.08}_{-0.07}$ & $257^{+4}_{-4}$ & $26.8^{+2.7}_{-2.4}$ \\ 
J1119+5130 & $0.004532$ & $1^{+1}_{-2}$ & $20.18^{+0.10}_{-0.12}$ & $4.6^{+0.1}_{-0.2}$ & $-0.91^{+0.31}_{-0.26}$ & $225^{+9}_{-9}$ & $13.2^{+6.2}_{-4.6}$ \\ 
J1144+4012 & $0.126862$ & $111^{+12}_{-7}$ & $20.19^{+0.08}_{-0.07}$ & $4.8^{+0.3}_{-1.0}$ & $0.48^{+0.13}_{-0.19}$ & $252^{+8}_{-10}$ & $324.8^{+52.3}_{-63.3}$ \\ 
J1148+2546 & $0.045233$ & $187^{+6}_{-5}$ & $19.66^{+0.15}_{-0.11}$ & $5.0^{+0.1}_{-0.2}$ & $-1.50^{+0.58}_{-0.60}$ & $265^{+7}_{-8}$ & $28.3^{+4.7}_{-2.5}$ \\ 
J1157+3220 & $0.011074$ & $392^{+2}_{-2}$ & $19.46^{+0.12}_{-0.11}$ & $5.0^{+0.1}_{-0.2}$ & $0.56^{+0.03}_{-0.02}$ & $118^{+1}_{-1}$ & $33.6^{+4.8}_{-0.7}$ \\ 
J1200+1343 & $0.066828$ & $131^{+3}_{-3}$ & $19.23^{+0.17}_{-0.08}$ & $4.5^{+0.2}_{-0.1}$ & $-0.04^{+0.04}_{-0.04}$ & $219^{+1}_{-1}$ & $86.3^{+4.6}_{-3.1}$ \\ 
J1253-0312 & $0.022846$ & $131^{+1}_{-1}$ & $19.28^{+0.02}_{-0.04}$ & $4.2^{+0.1}_{-0.1}$ & $-3.19^{+0.70}_{-0.69}$ & $246^{+1}_{-1}$ & $30.2^{+0.4}_{-0.4}$ \\ 
J1323-0132 & $0.022511$ & $26^{+1}_{-1}$ & $18.22^{+0.06}_{-0.08}$ & $3.2^{+0.0}_{-0.1}$ & $-0.95^{+0.04}_{-0.05}$ & $112^{+1}_{-1}$ & $87.5^{+1.3}_{-1.2}$ \\ 
J1359+5726 & $0.033938$ & $175^{+3}_{-2}$ & $19.30^{+0.13}_{-0.08}$ & $4.6^{+0.1}_{-0.2}$ & $-0.82^{+0.23}_{-0.18}$ & $125^{+2}_{-2}$ & $55.7^{+5.5}_{-2.9}$ \\ 
J1416+1223 & $0.123174$ & $-5^{+4}_{-4}$ & $19.20^{+0.08}_{-0.08}$ & $3.6^{+0.6}_{-0.3}$ & $0.38^{+0.10}_{-0.10}$ & $306^{+7}_{-8}$ & $41.7^{+2.9}_{-2.6}$ \\ 
J1418+2102 & $0.008699$ & $21^{+1}_{-1}$ & $19.73^{+0.05}_{-0.02}$ & $4.1^{+0.1}_{-0.1}$ & $-3.38^{+0.58}_{-0.76}$ & $246^{+1}_{-1}$ & $67.4^{+2.9}_{-2.9}$ \\ 
J1428+1653 & $0.181544$ & $122^{+2}_{-3}$ & $19.23^{+0.05}_{-0.09}$ & $3.0^{+0.2}_{-0.1}$ & $0.21^{+0.07}_{-0.08}$ & $168^{+2}_{-3}$ & $46.5^{+3.3}_{-3.4}$ \\ 
J1429+0643 & $0.173649$ & $21^{+2}_{-2}$ & $19.95^{+0.09}_{-0.04}$ & $3.4^{+0.2}_{-0.1}$ & $-1.30^{+0.06}_{-0.05}$ & $321^{+5}_{-6}$ & $41.6^{+2.1}_{-2.1}$ \\ 
J1521+0759 & $0.094335$ & $141^{+3}_{-4}$ & $19.40^{+0.08}_{-0.08}$ & $5.0^{+0.1}_{-0.1}$ & $-1.21^{+0.22}_{-0.23}$ & $97^{+4}_{-4}$ & $9.5^{+1.1}_{-0.7}$ \\ 
J1525+0757 & $0.075916$ & $133^{+1}_{-2}$ & $18.58^{+0.09}_{-0.07}$ & $4.7^{+0.1}_{-0.2}$ & $-1.09^{+0.12}_{-0.11}$ & $44^{+2}_{-2}$ & $20.1^{+1.0}_{-1.0}$ \\ 
J1545+0858 & $0.037834$ & $92^{+2}_{-3}$ & $19.43^{+0.06}_{-0.06}$ & $4.2^{+0.1}_{-0.2}$ & $-1.87^{+0.34}_{-0.40}$ & $72^{+3}_{-2}$ & $37.0^{+1.8}_{-1.1}$ \\ 
J1612+0817 & $0.149198$ & $187^{+4}_{-2}$ & $19.79^{+0.08}_{-0.07}$ & $5.0^{+0.2}_{-0.2}$ & $0.42^{+0.03}_{-0.02}$ & $84^{+2}_{-2}$ & $94.9^{+5.1}_{-4.8}$ \\ 
    \enddata
    \tablecomments{(1) object name; (2) redshift estimated by the shell model; (3) outflow velocity of expanding shell; (4) H \textsc{i} column density; (5)temperature; (6) dust extinction; (7) intrinsic line width; (8) equivalent width.}
\end{deluxetable*}

\begin{figure*}[ht]
    \addtocounter{figure}{-18}
    \centering
    \includegraphics[width=3.3in]{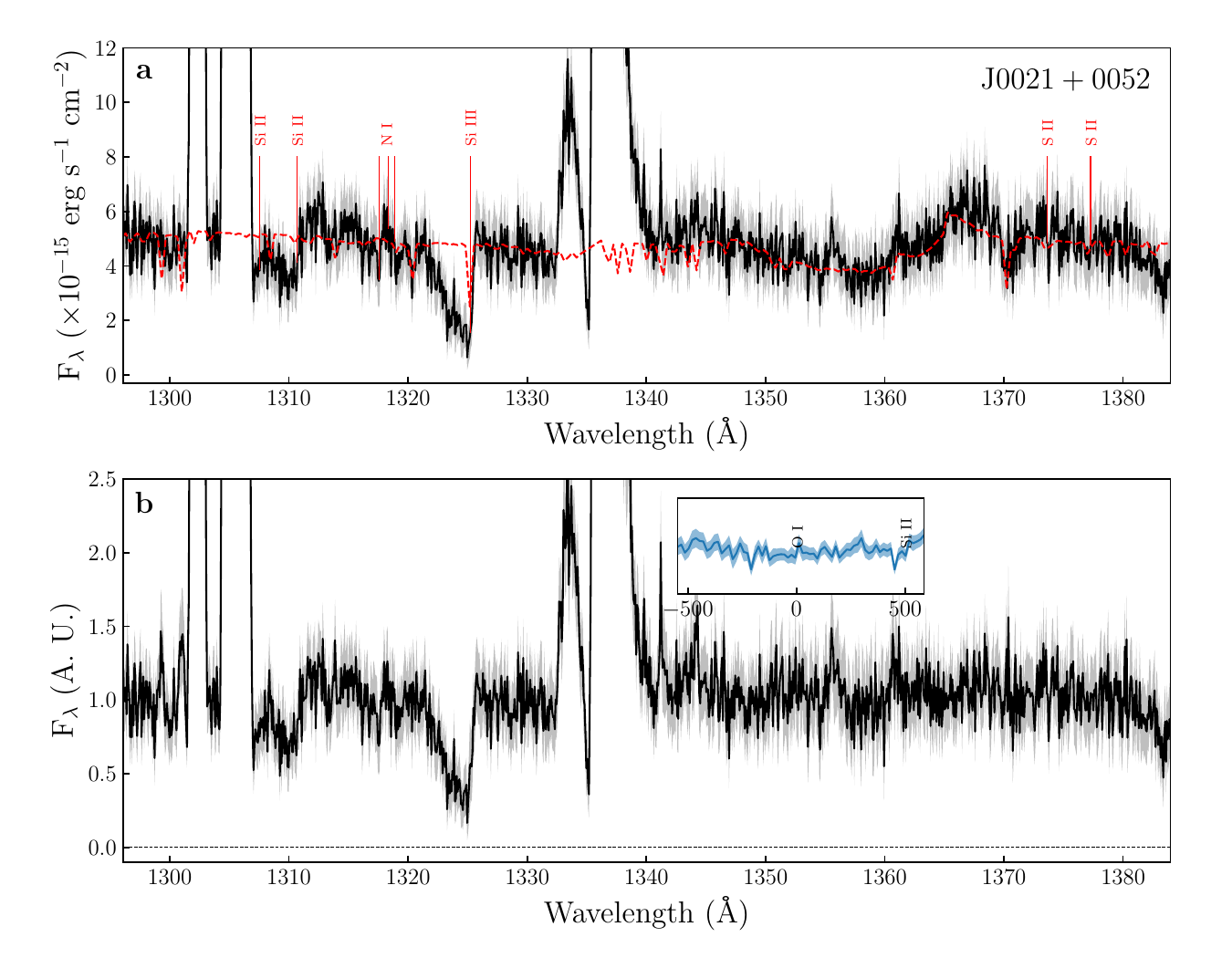}
    \includegraphics[width=3.3in]{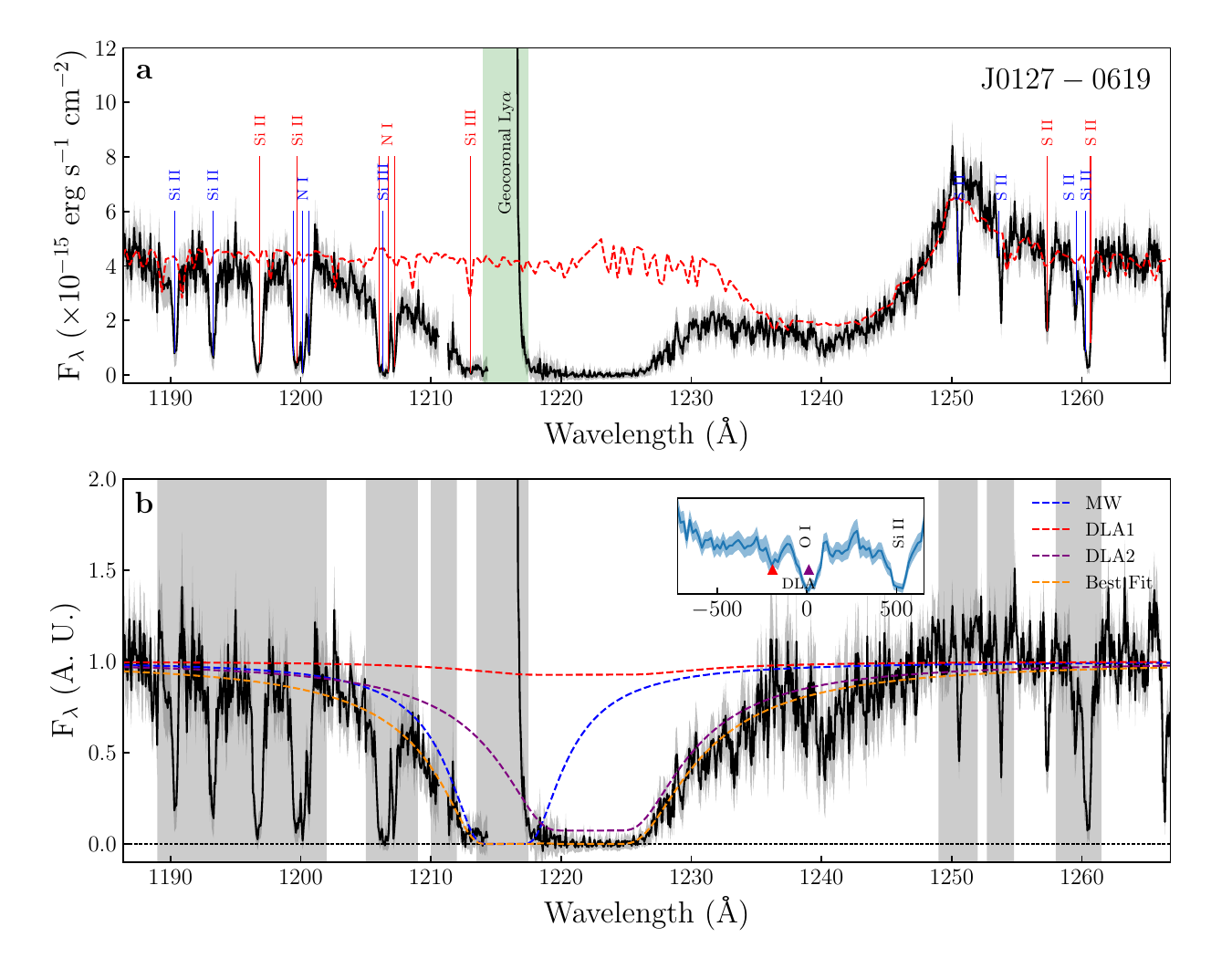}
    \includegraphics[width=3.3in]{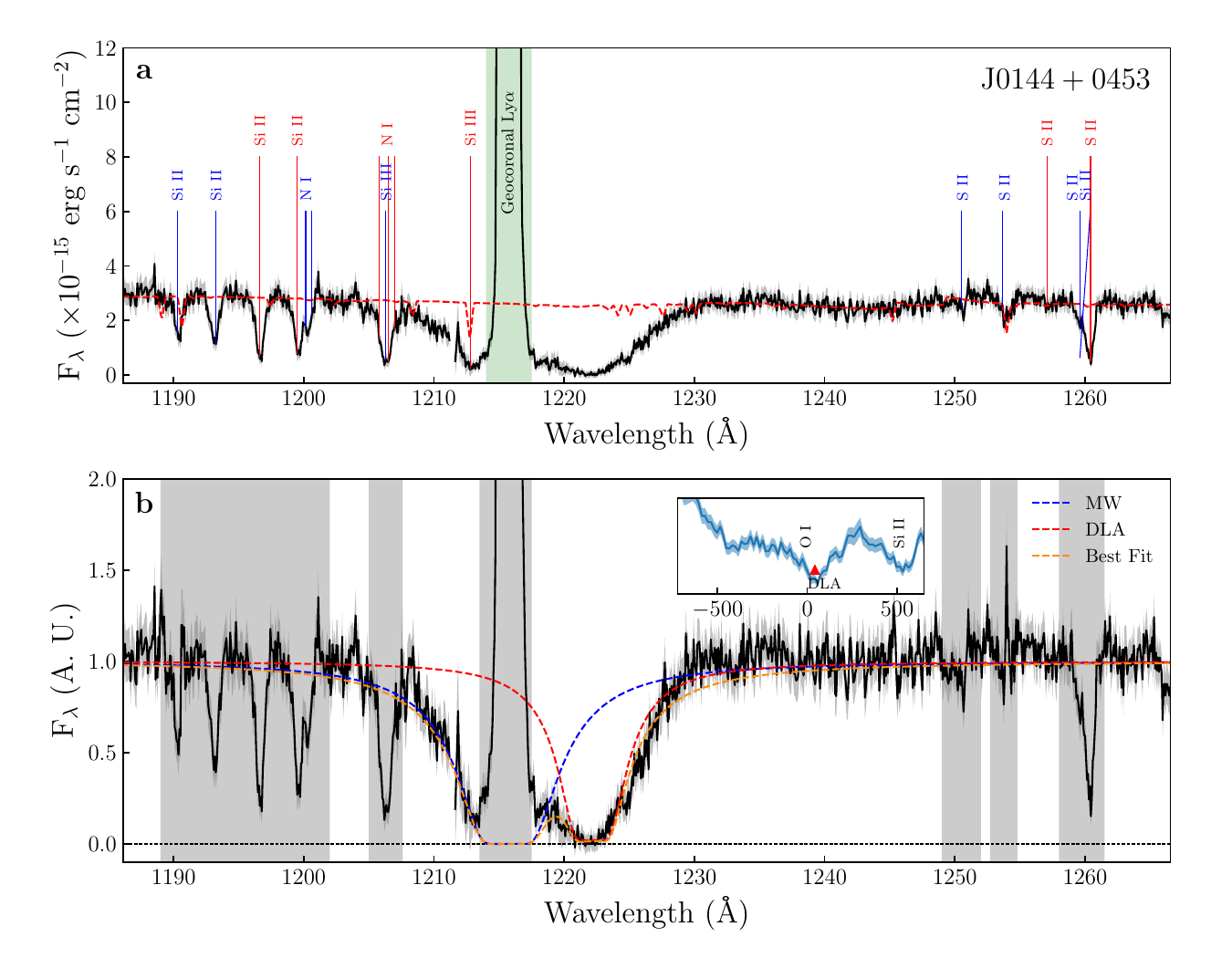}
    \includegraphics[width=3.3in]{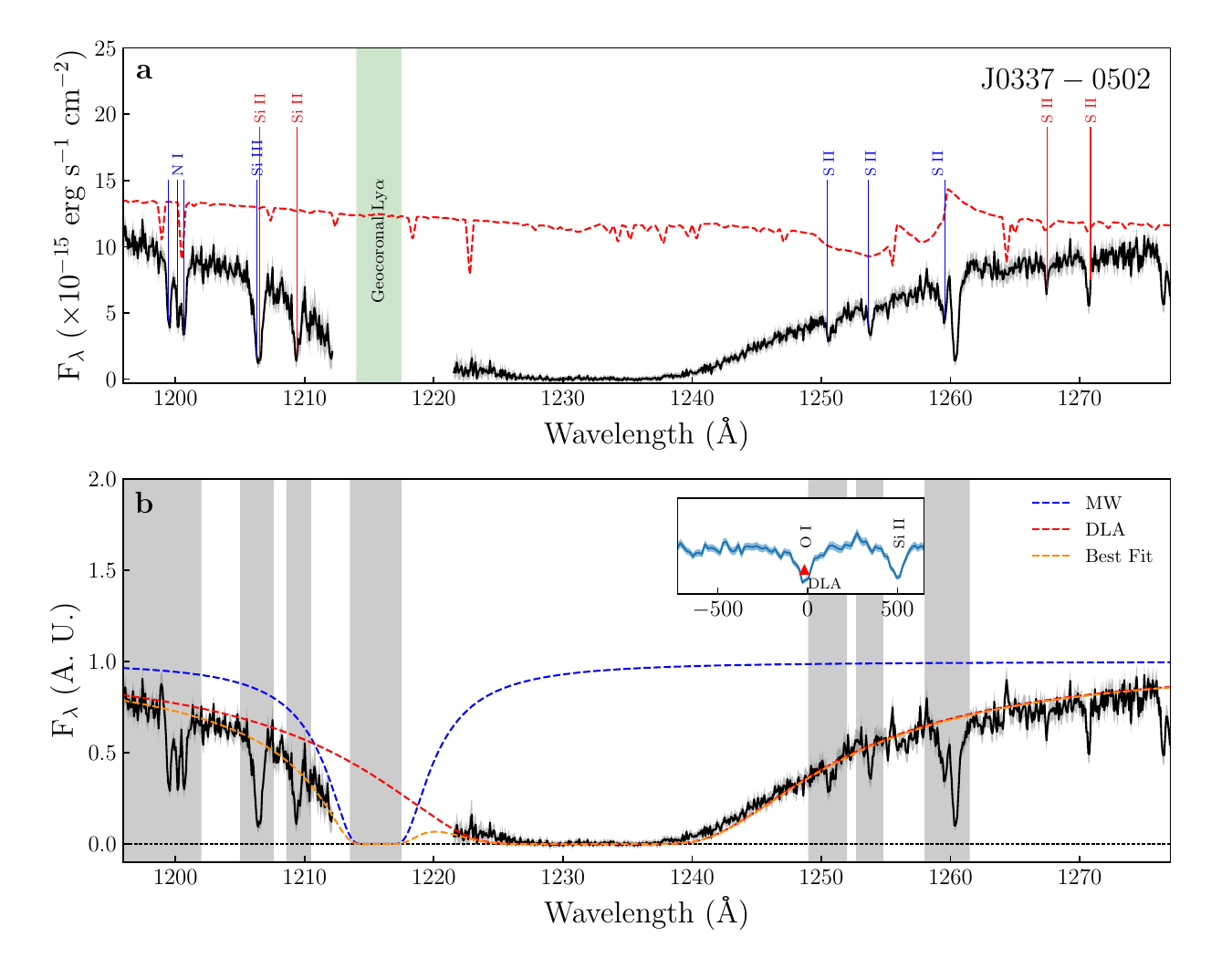}
    \includegraphics[width=3.3in]{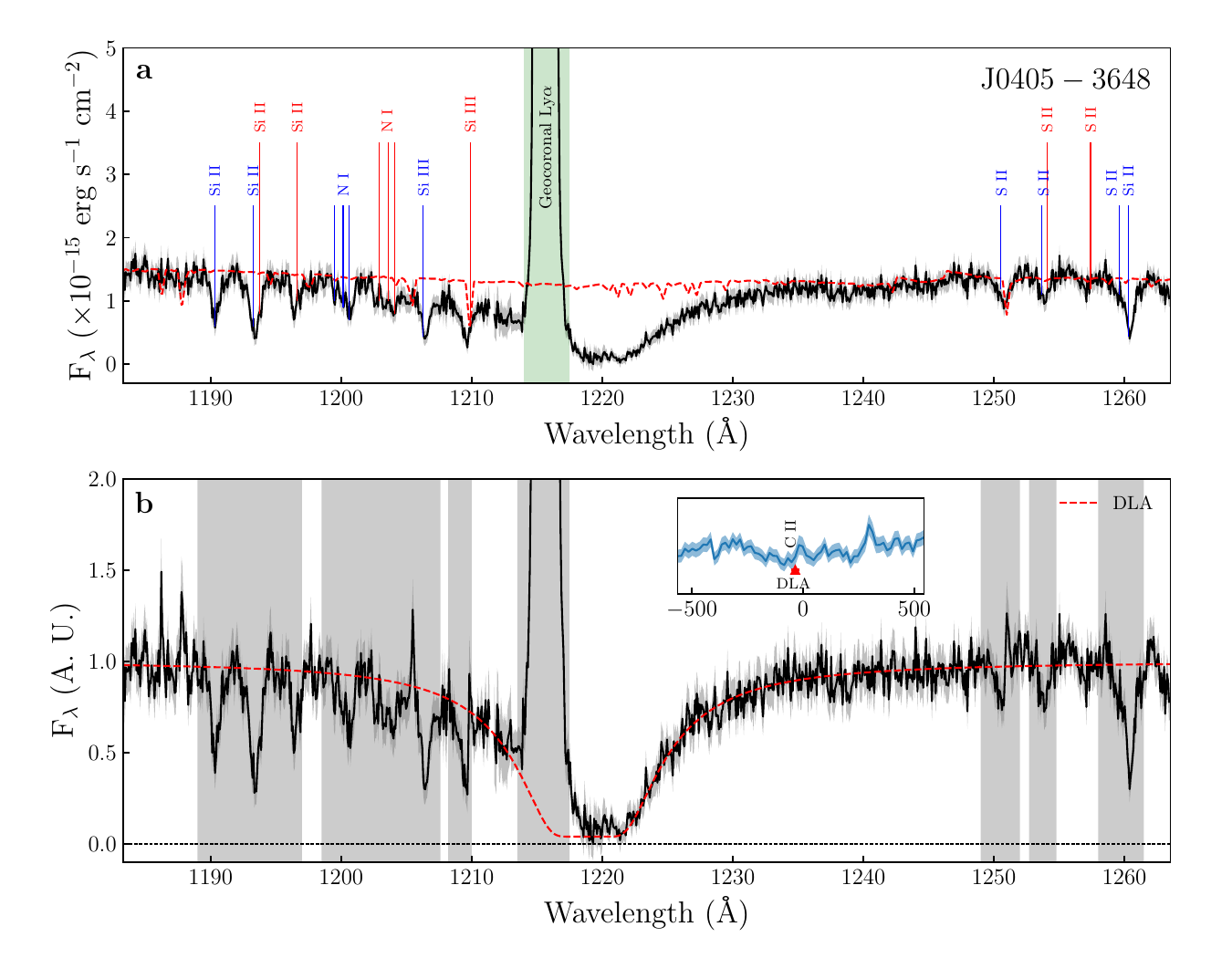}
    \includegraphics[width=3.3in]{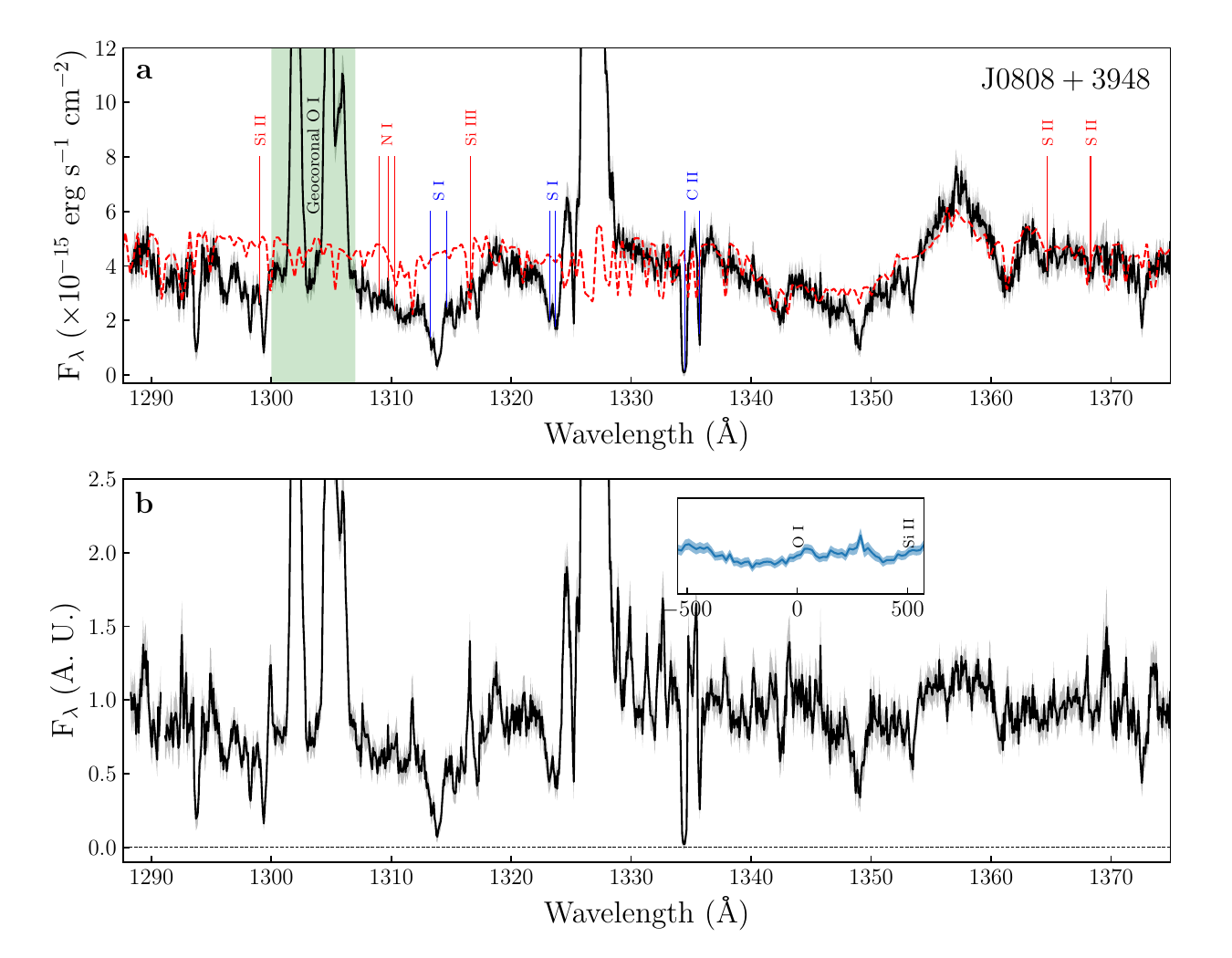}
    \caption{\label{fig:allspec} Continued.}
\end{figure*}

\begin{figure*}[ht]
    \addtocounter{figure}{-1}
    \centering
    \includegraphics[width=3.3in]{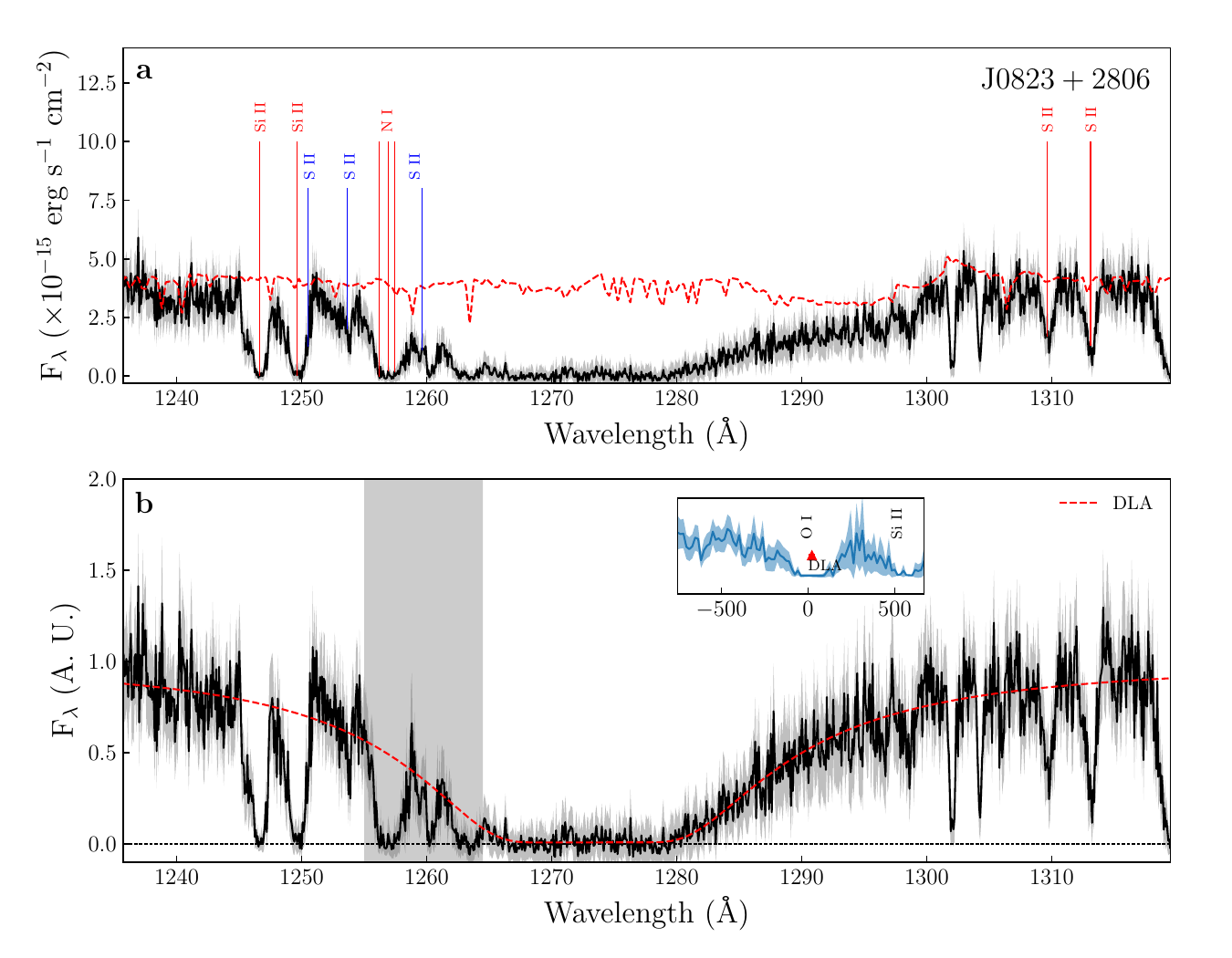}
    \includegraphics[width=3.3in]{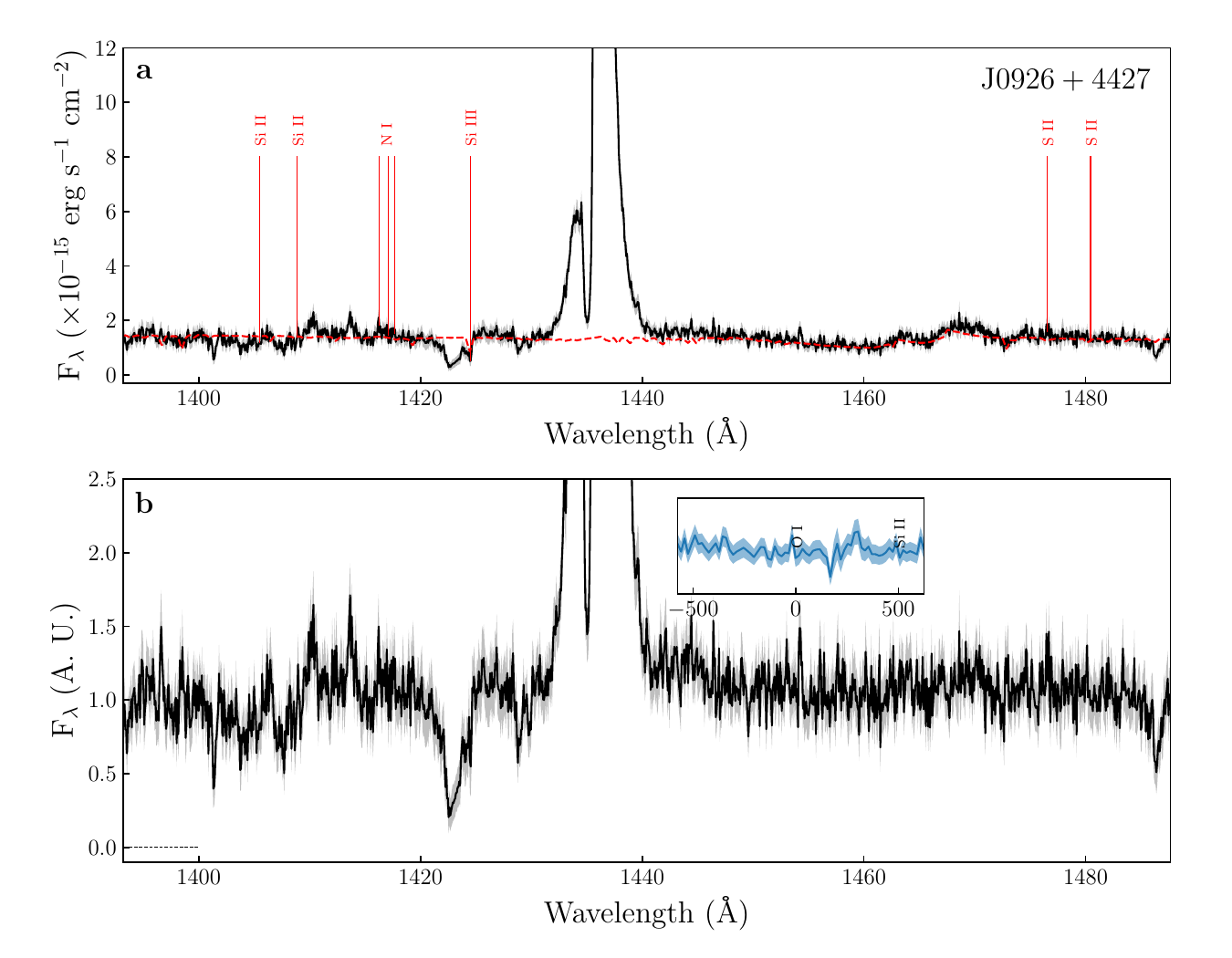}
    \includegraphics[width=3.3in]{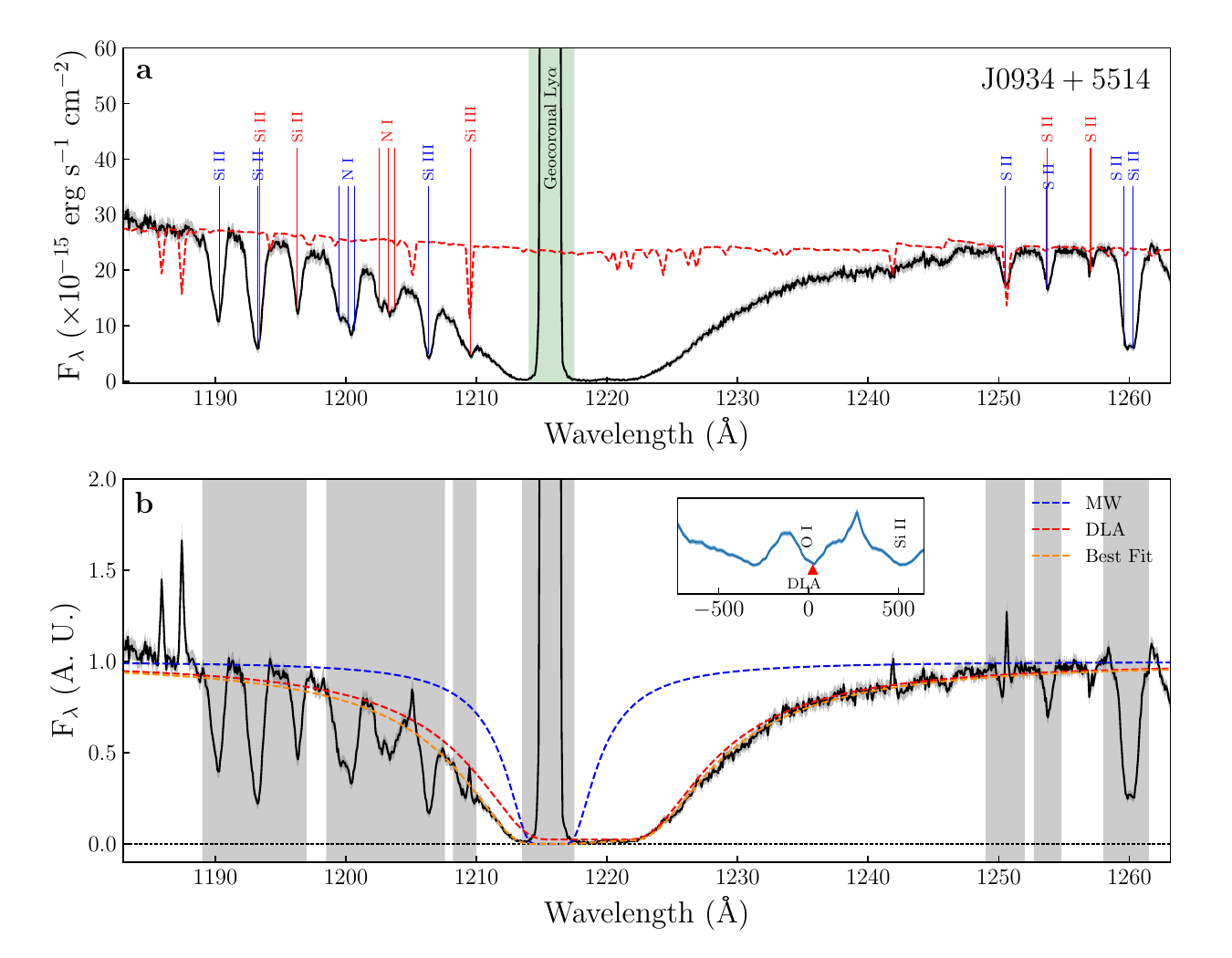}
    \includegraphics[width=3.3in]{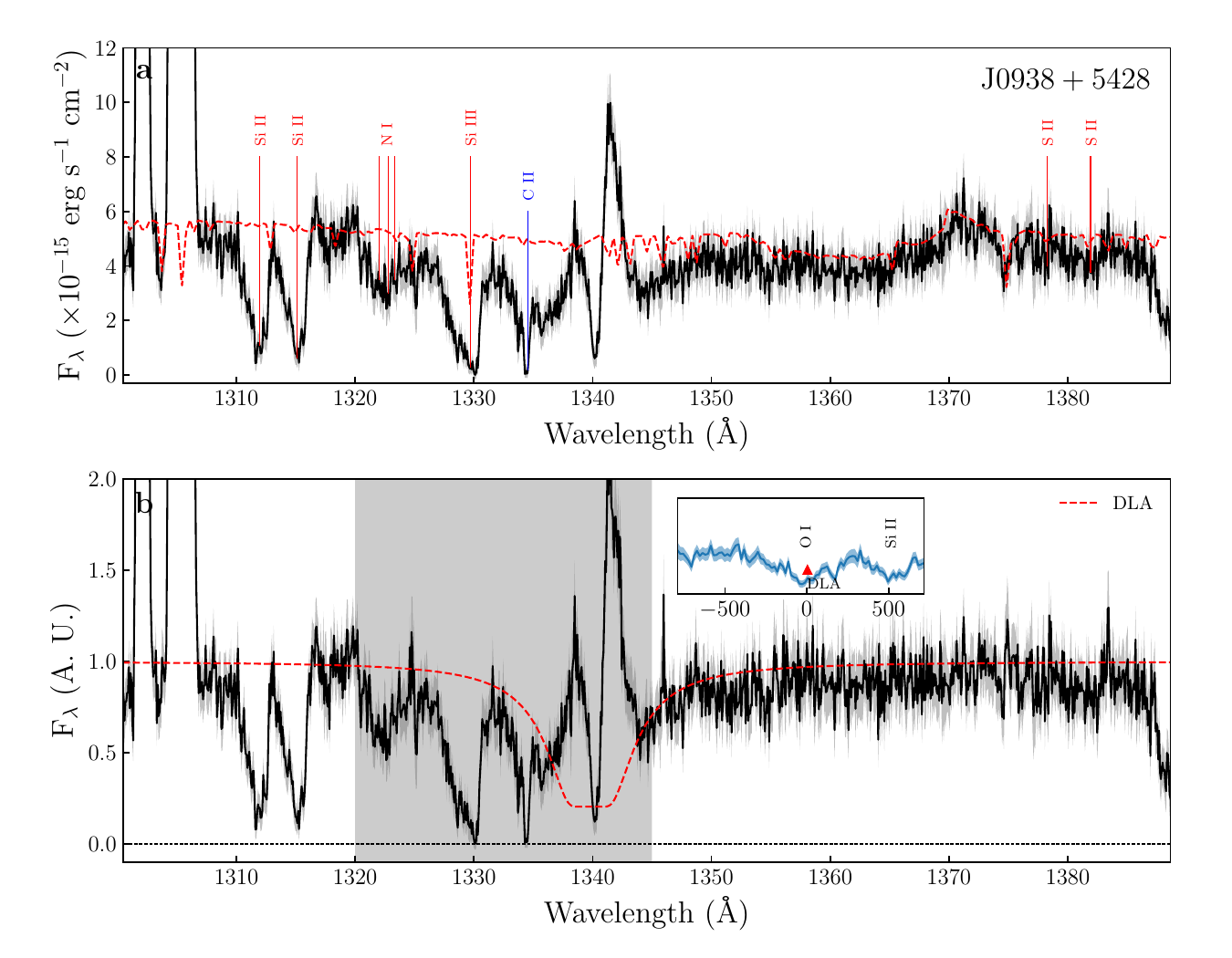}
    \includegraphics[width=3.3in]{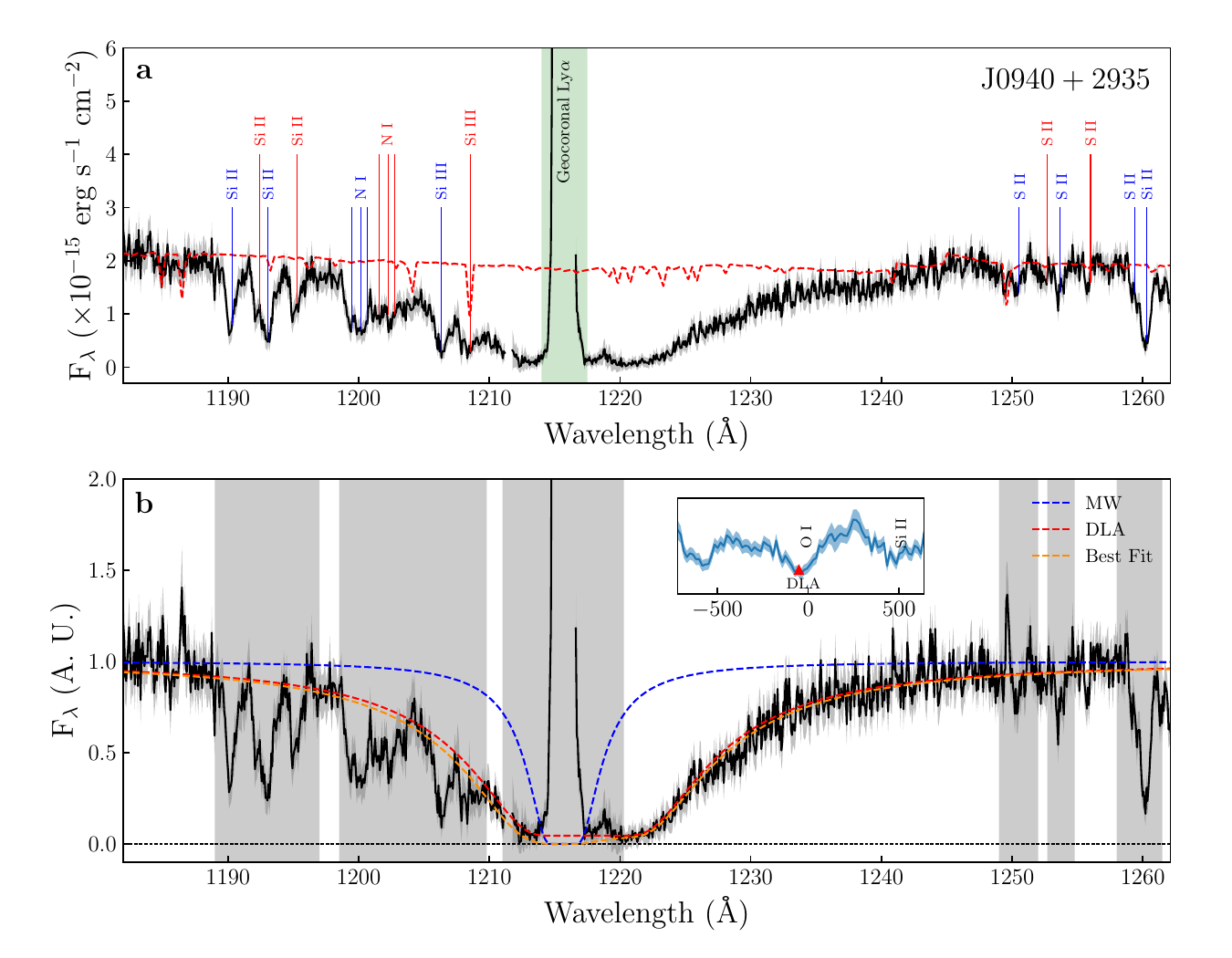}
    \includegraphics[width=3.3in]{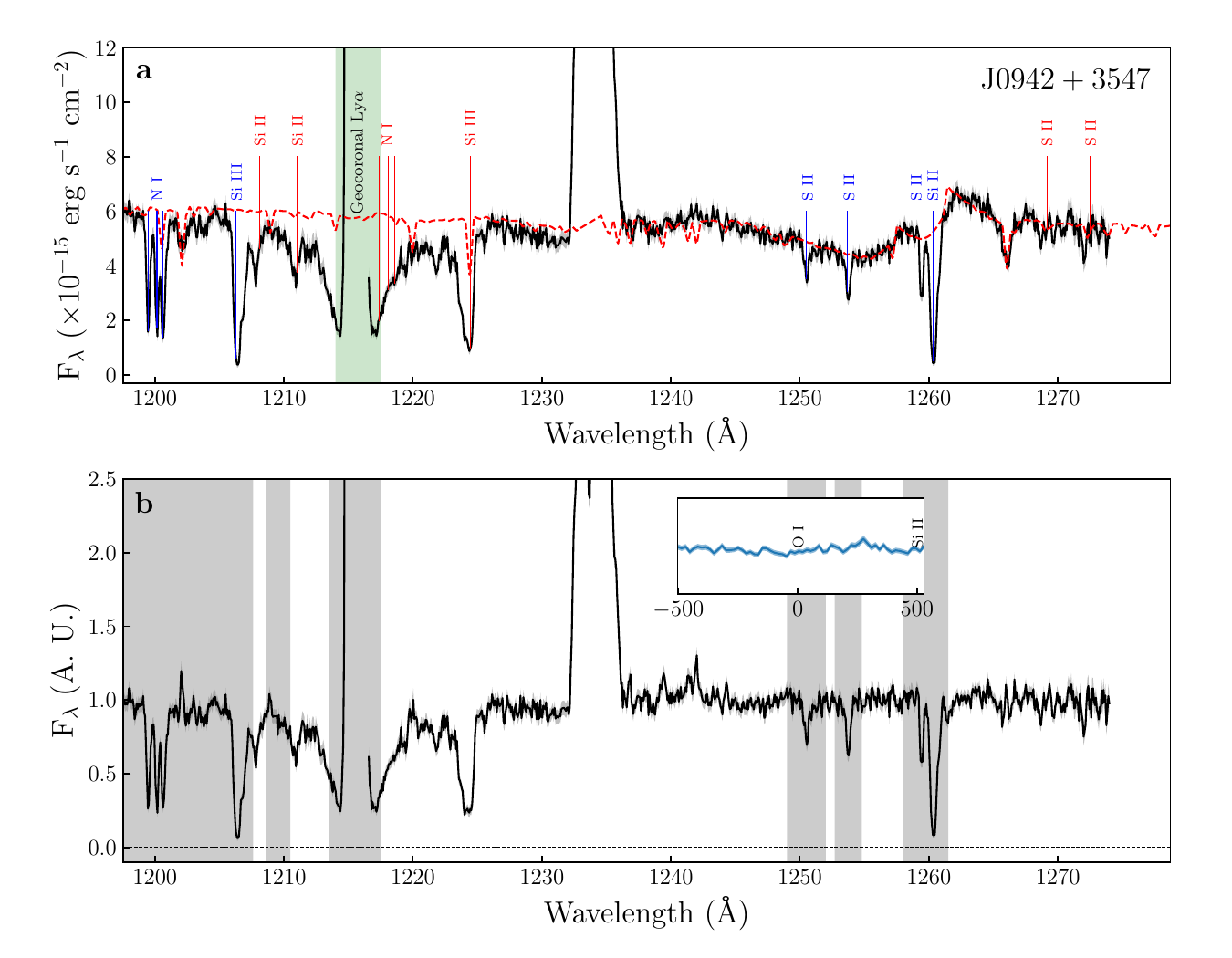}
    \caption{Continued.}
\end{figure*}

\begin{figure*}[ht]
    \addtocounter{figure}{-1}
    \centering
    \includegraphics[width=3.3in]{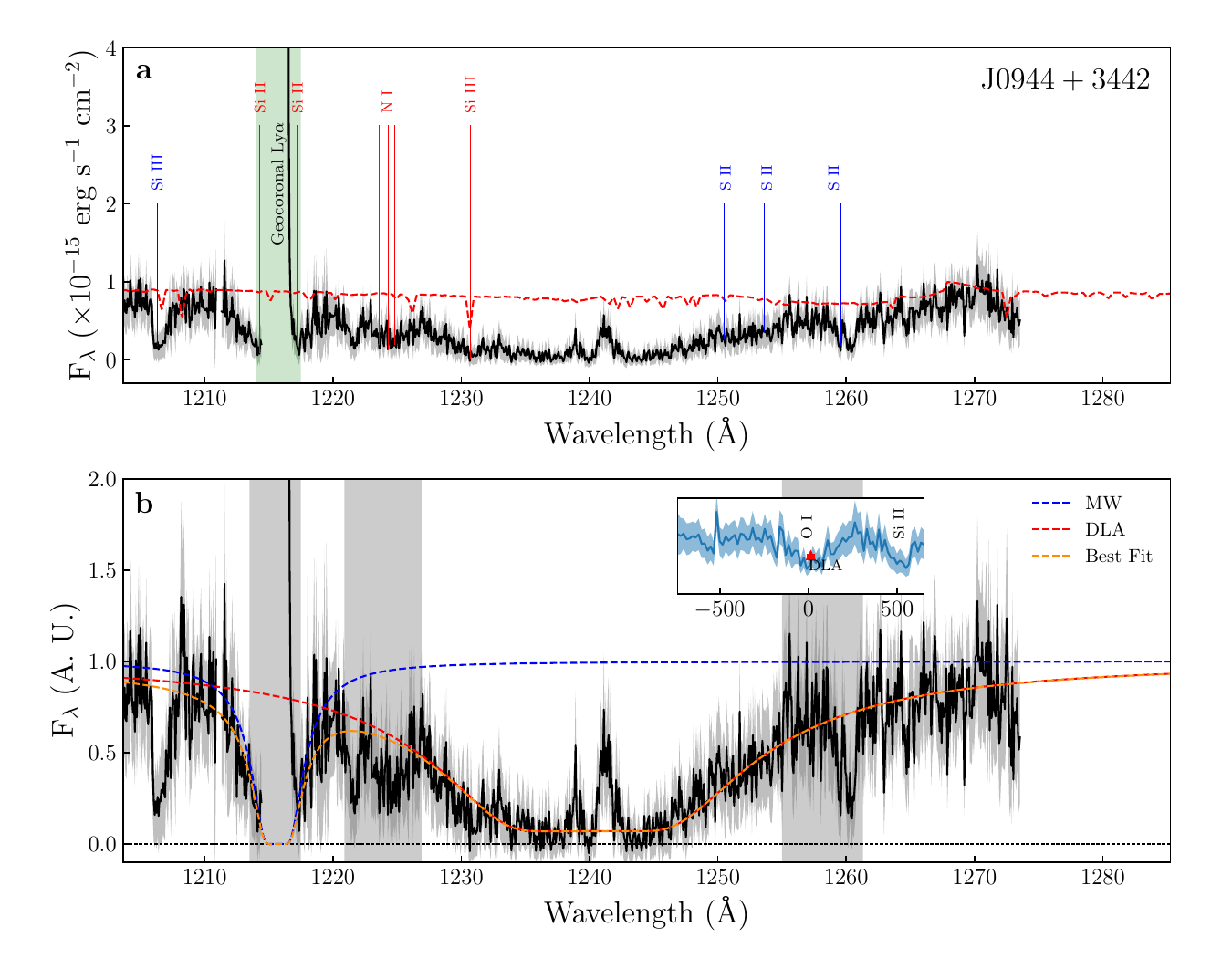}
    \includegraphics[width=3.3in]{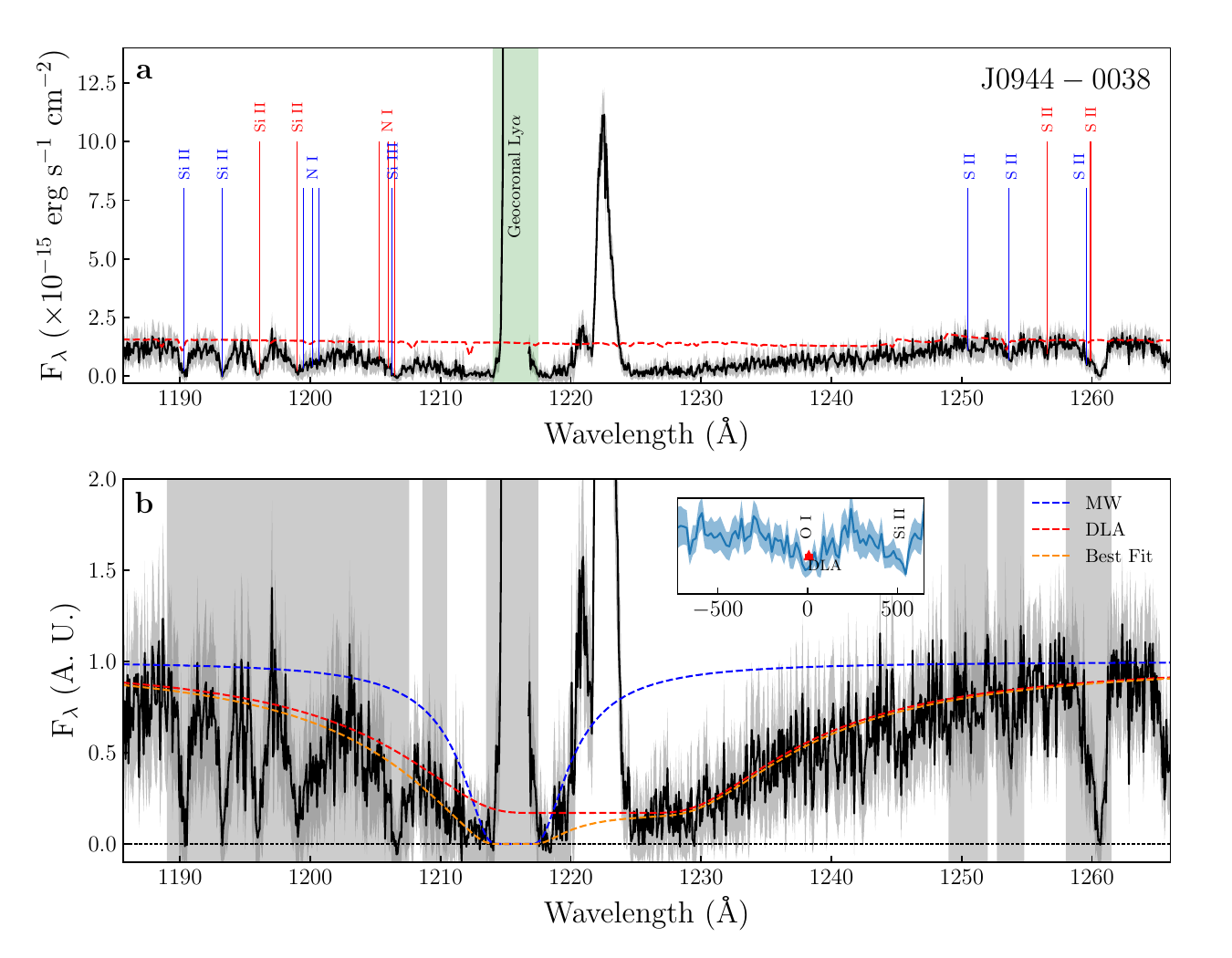}
    \includegraphics[width=3.3in]{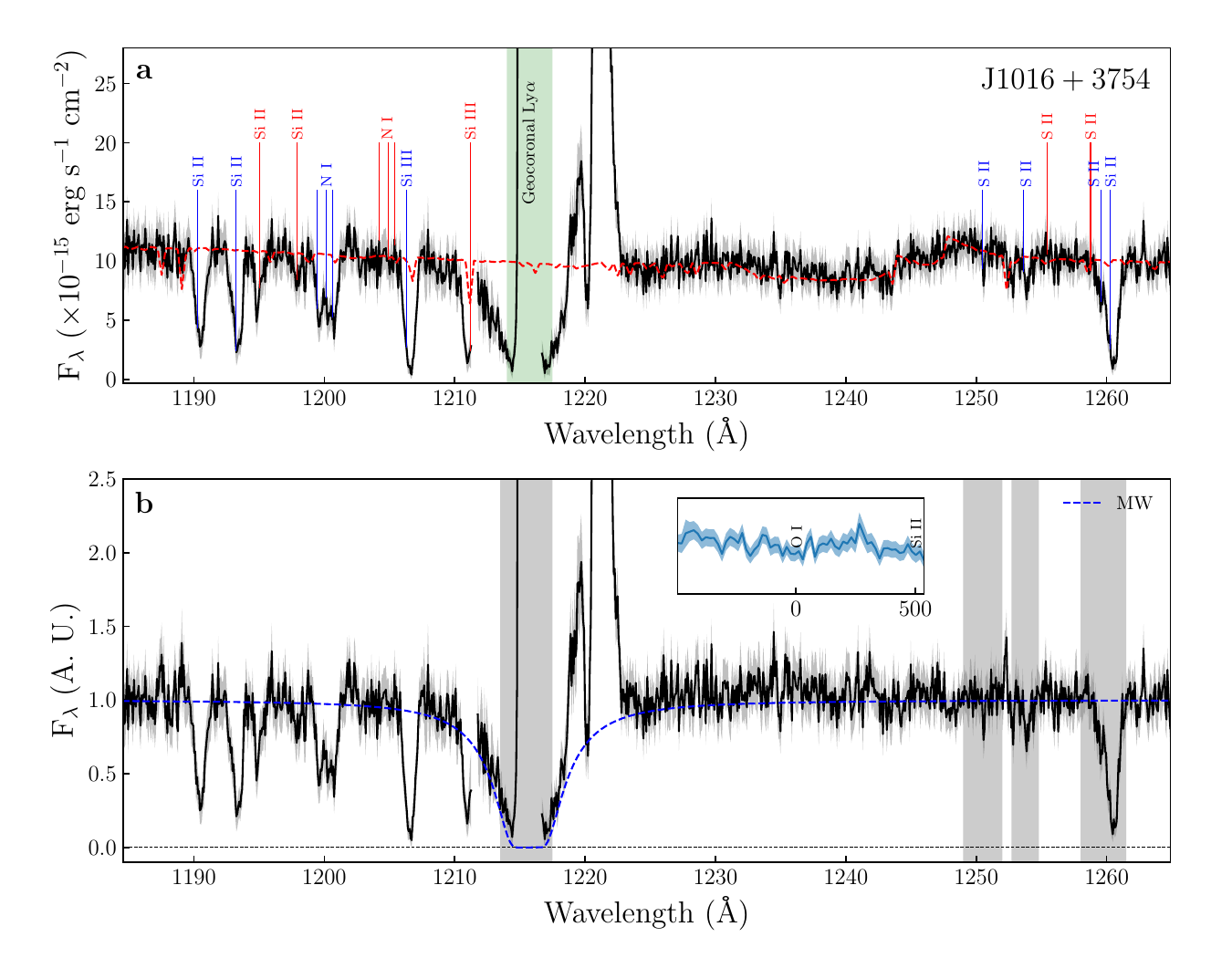}
    \includegraphics[width=3.3in]{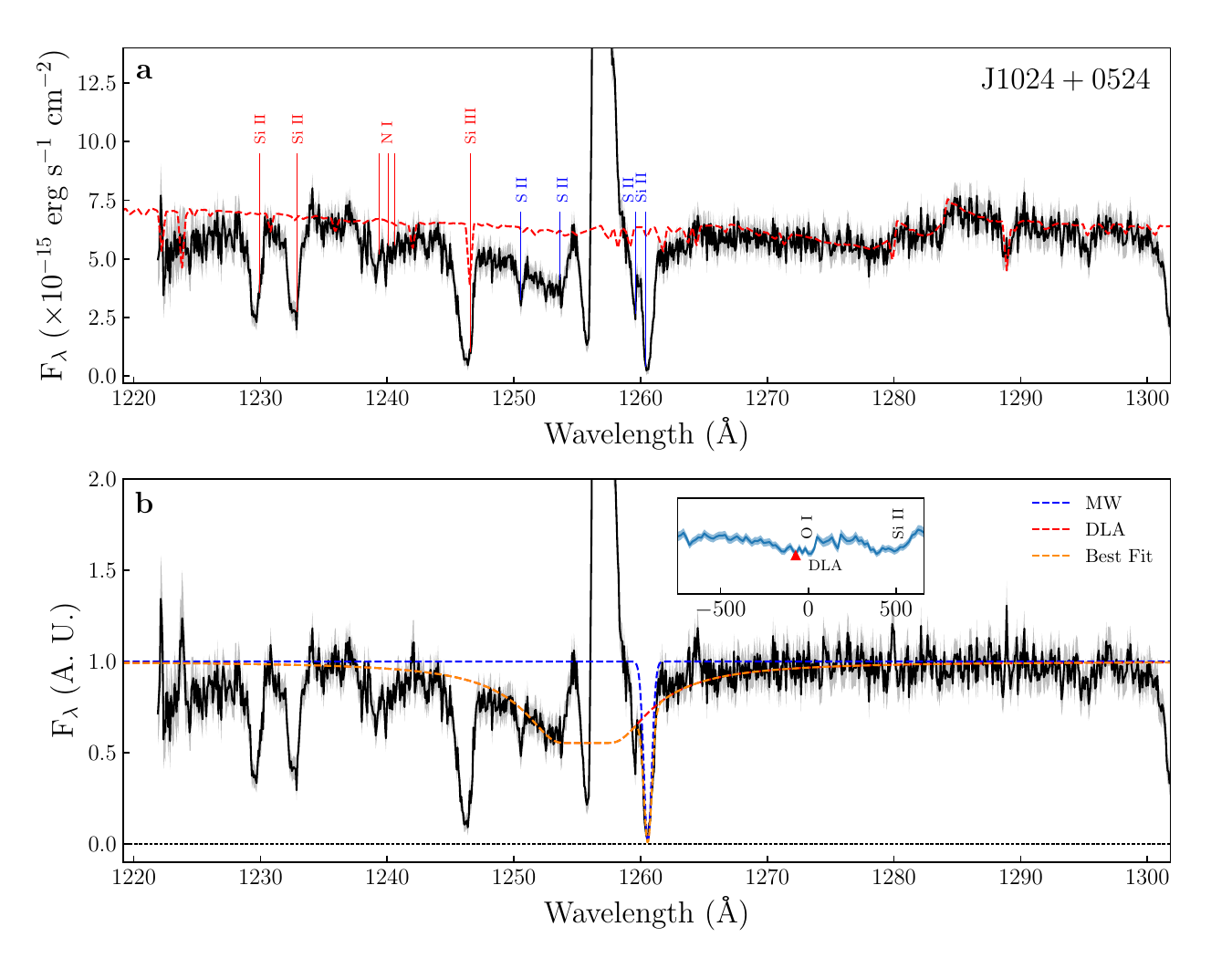}
    \includegraphics[width=3.3in]{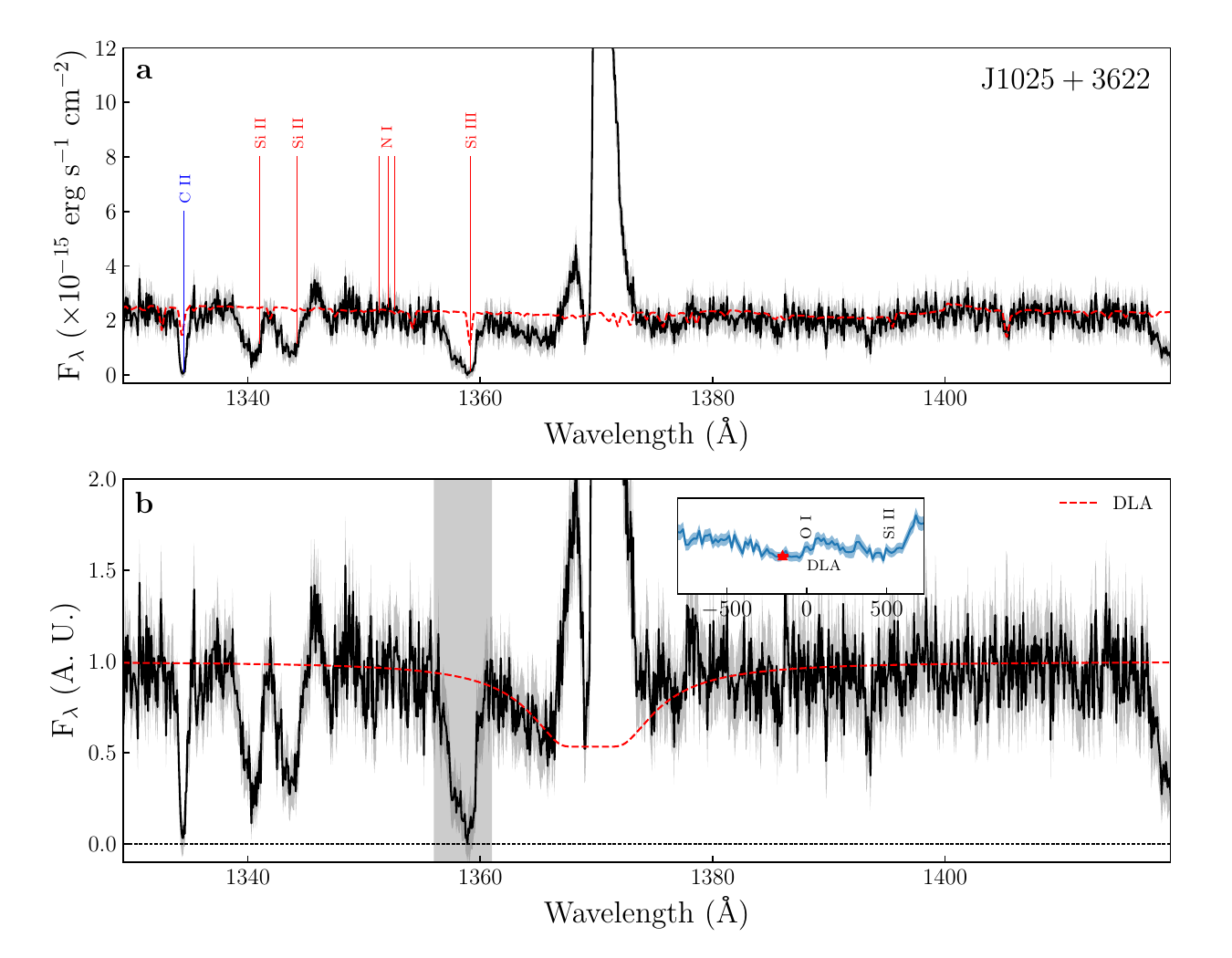}
    \includegraphics[width=3.3in]{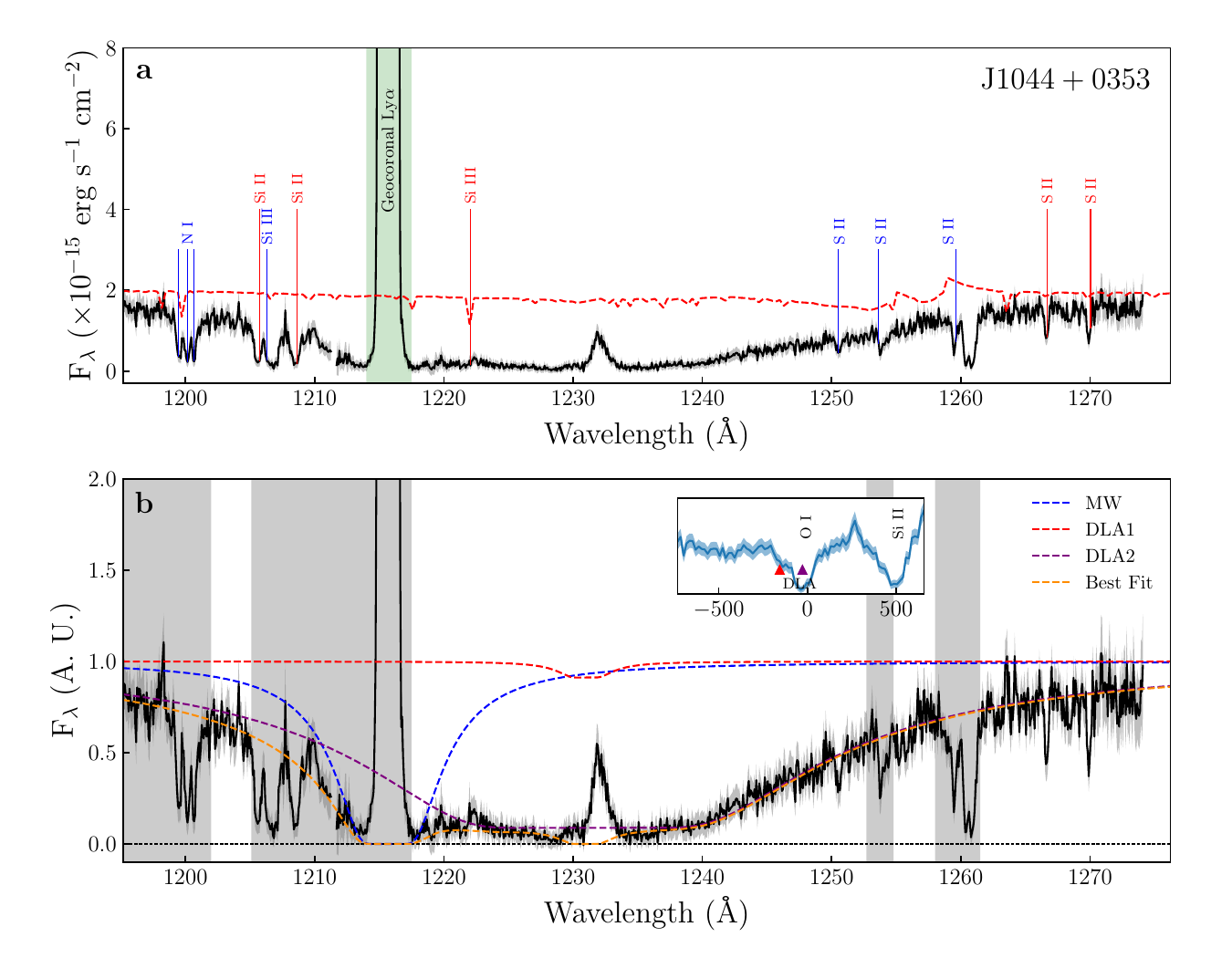}
    \caption{Continued.}
\end{figure*}

\begin{figure*}[ht]
    \addtocounter{figure}{-1}
    \centering
    \includegraphics[width=3.3in]{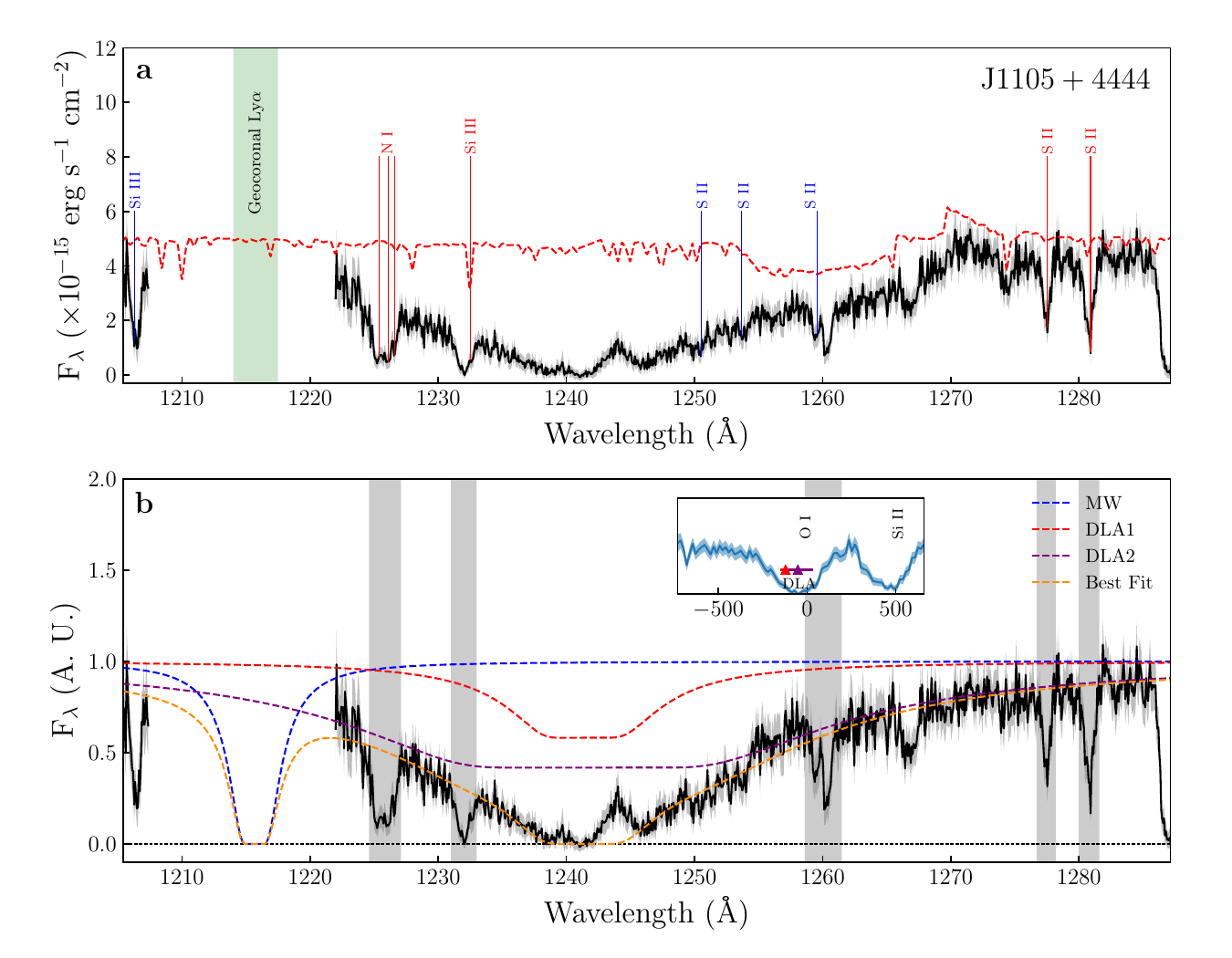}
    \includegraphics[width=3.3in]{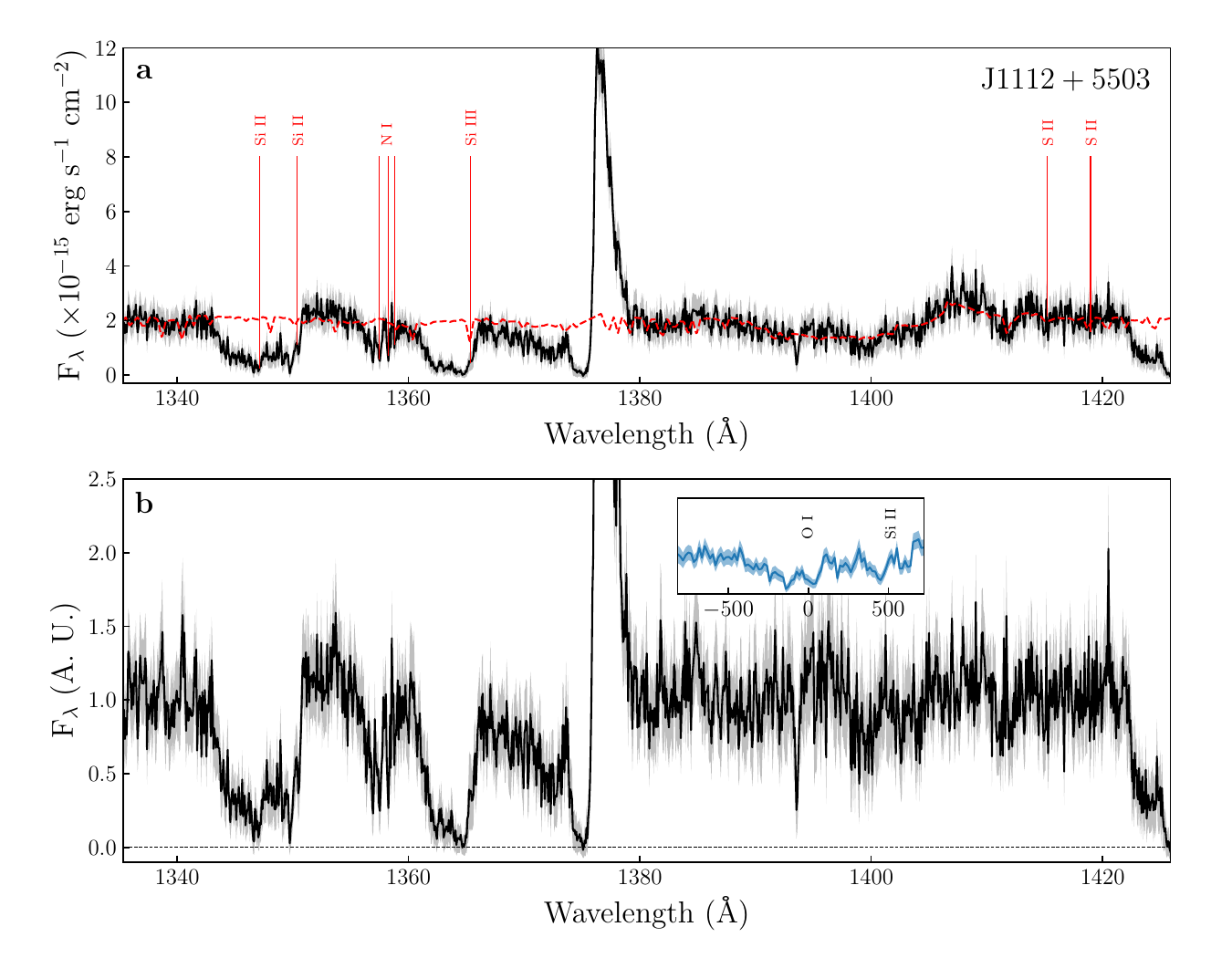}
    \includegraphics[width=3.3in]{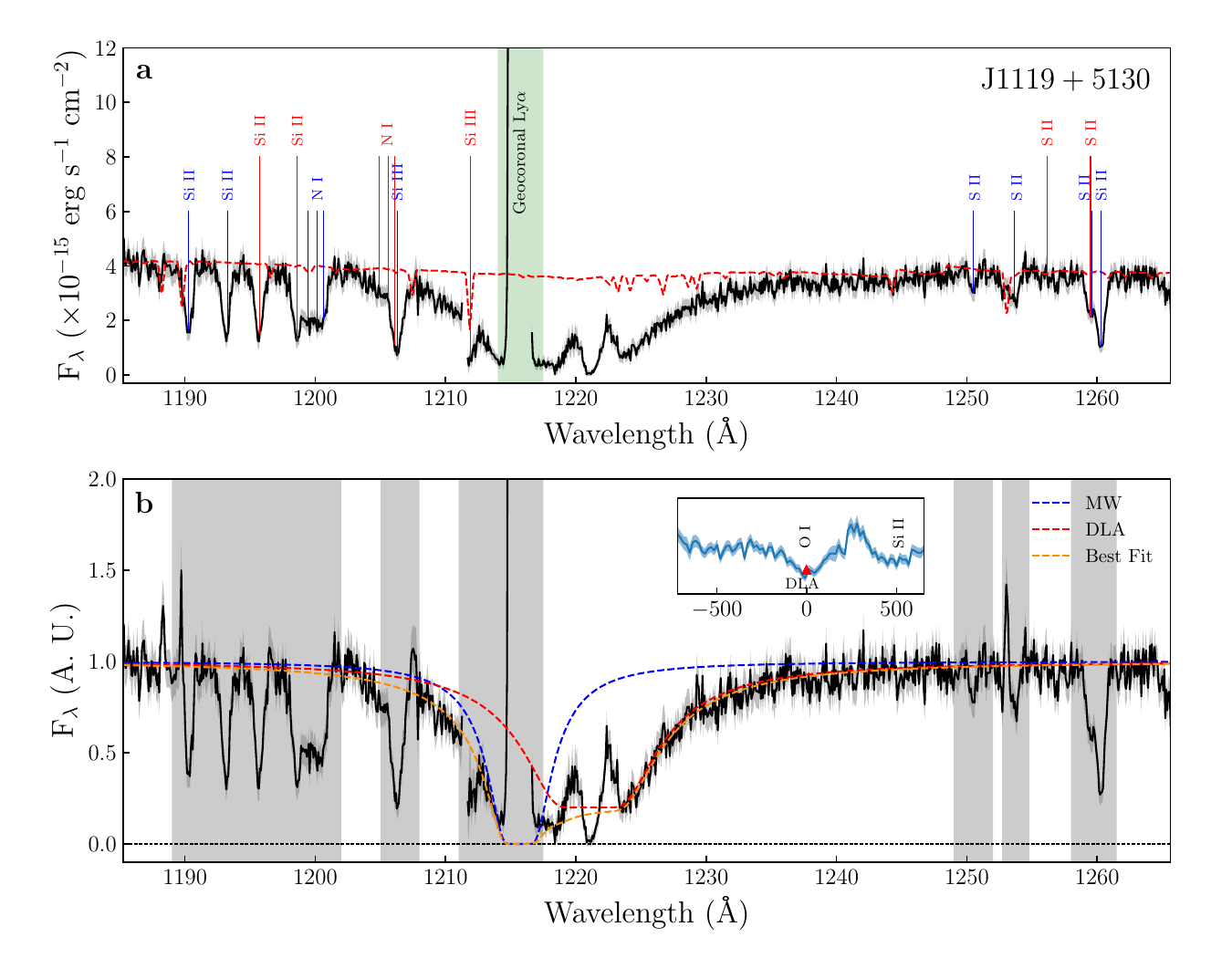}
    \includegraphics[width=3.3in]{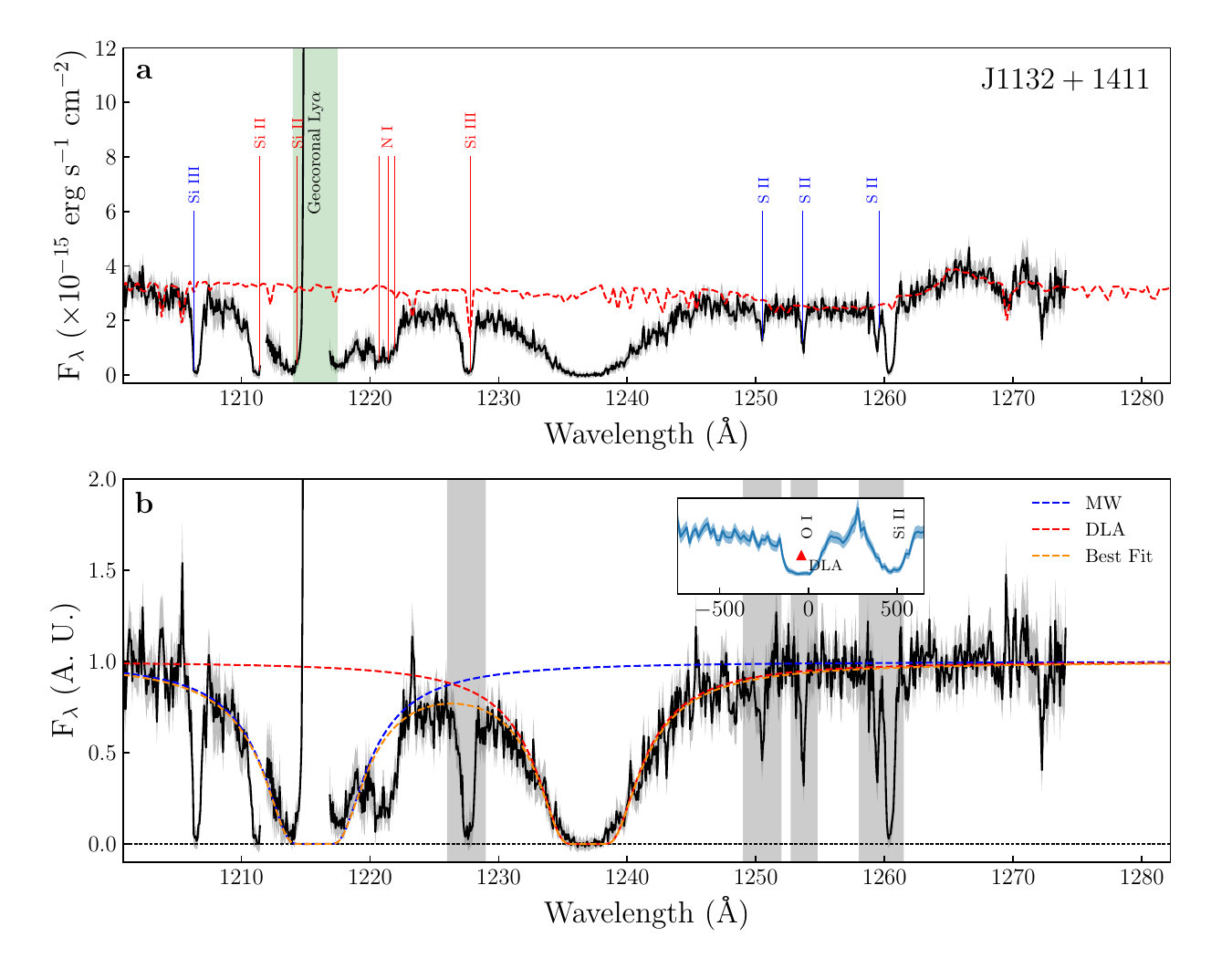}
    \includegraphics[width=3.3in]{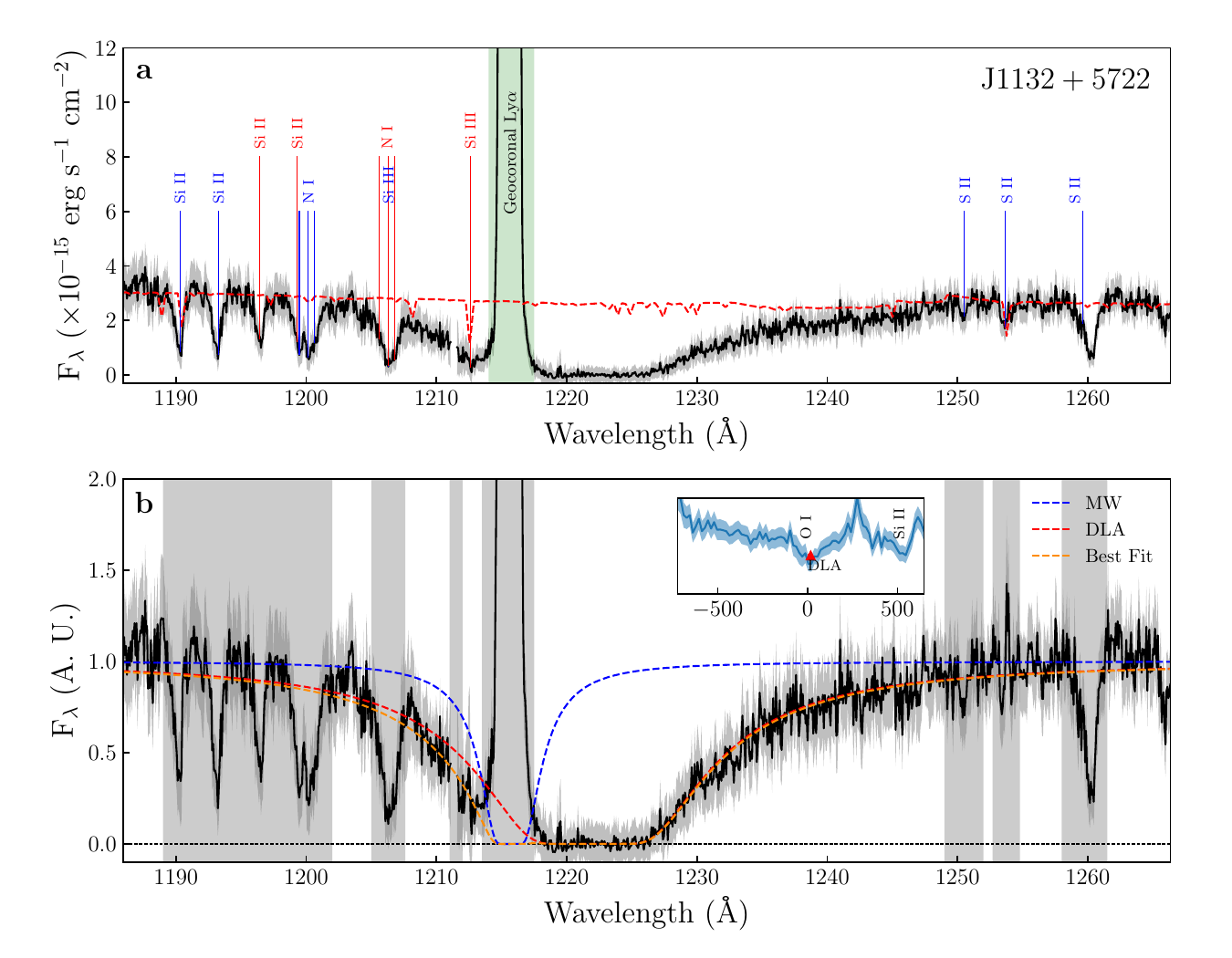}
    \includegraphics[width=3.3in]{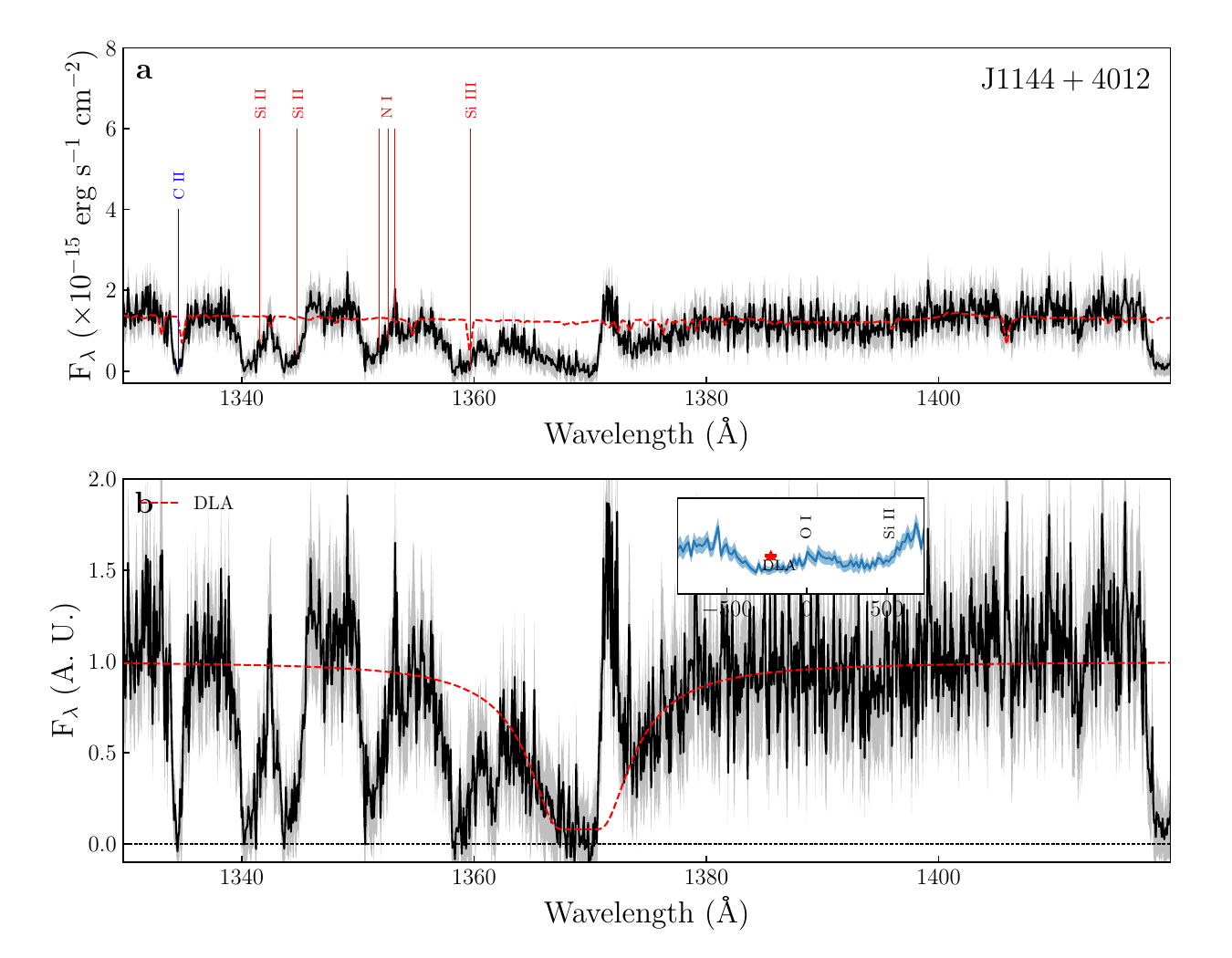}
    \caption{Continued.}
\end{figure*}

\begin{figure*}[ht]
    \addtocounter{figure}{-1}
    \centering
    \includegraphics[width=3.3in]{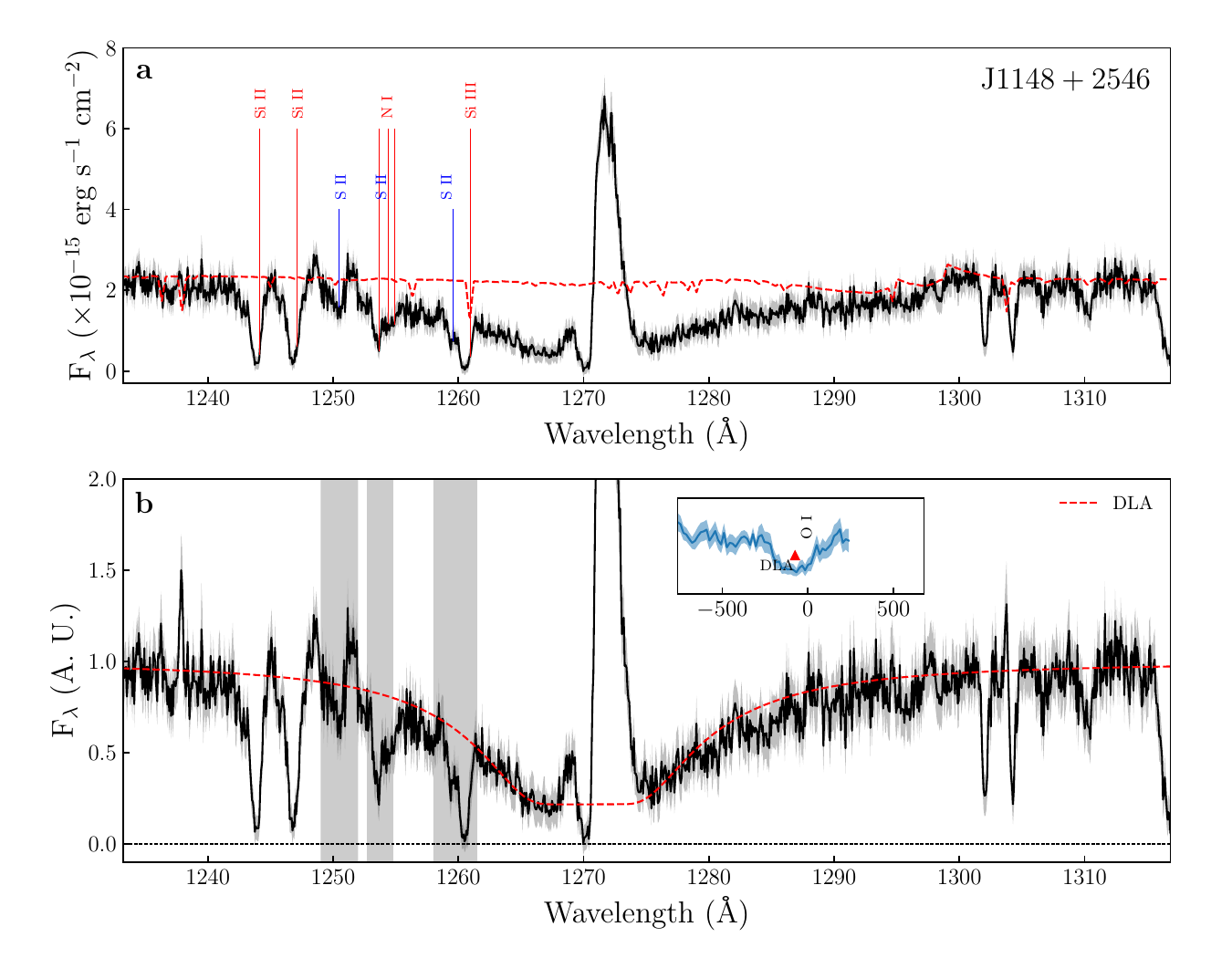}
    \includegraphics[width=3.3in]{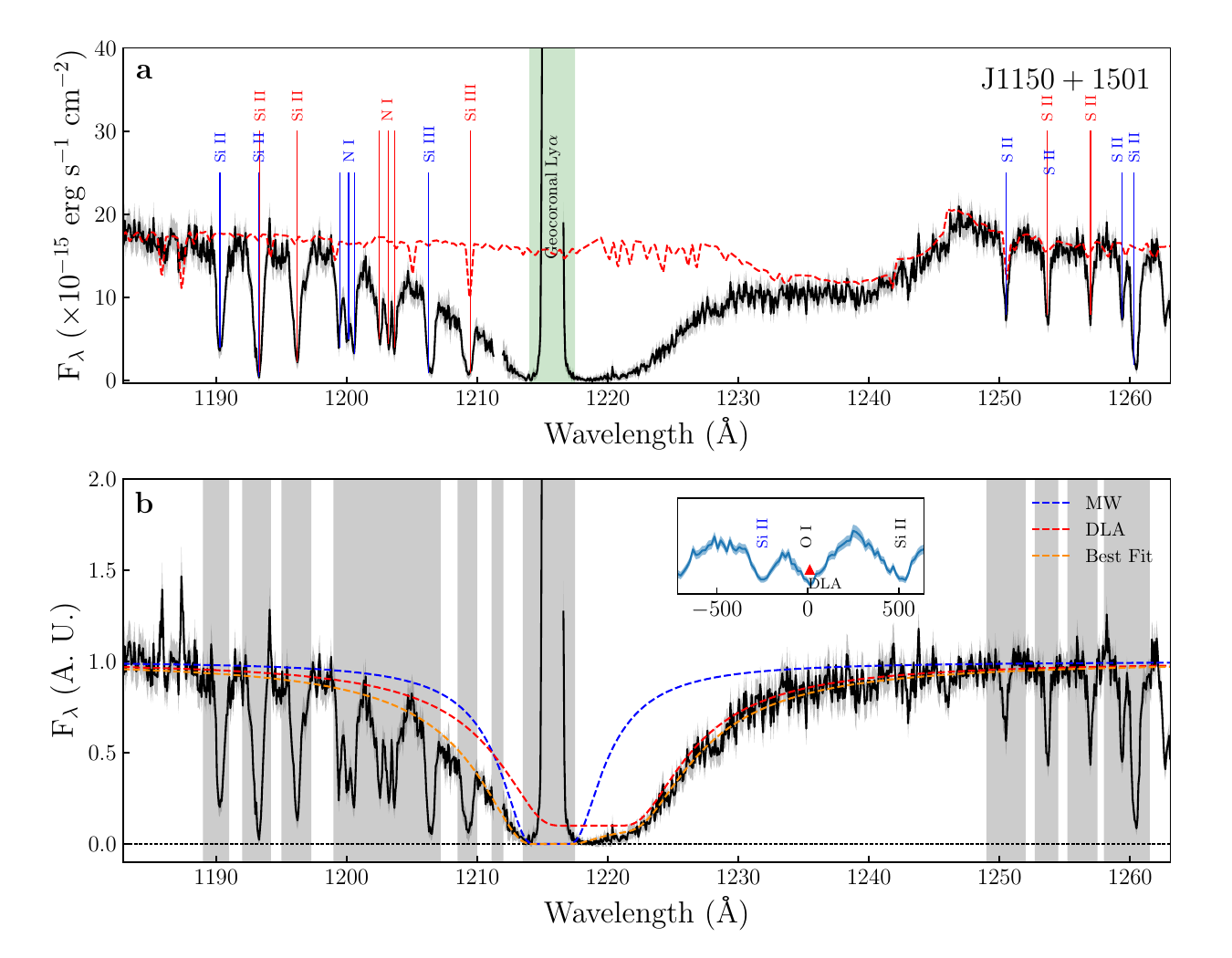}
    \includegraphics[width=3.3in]{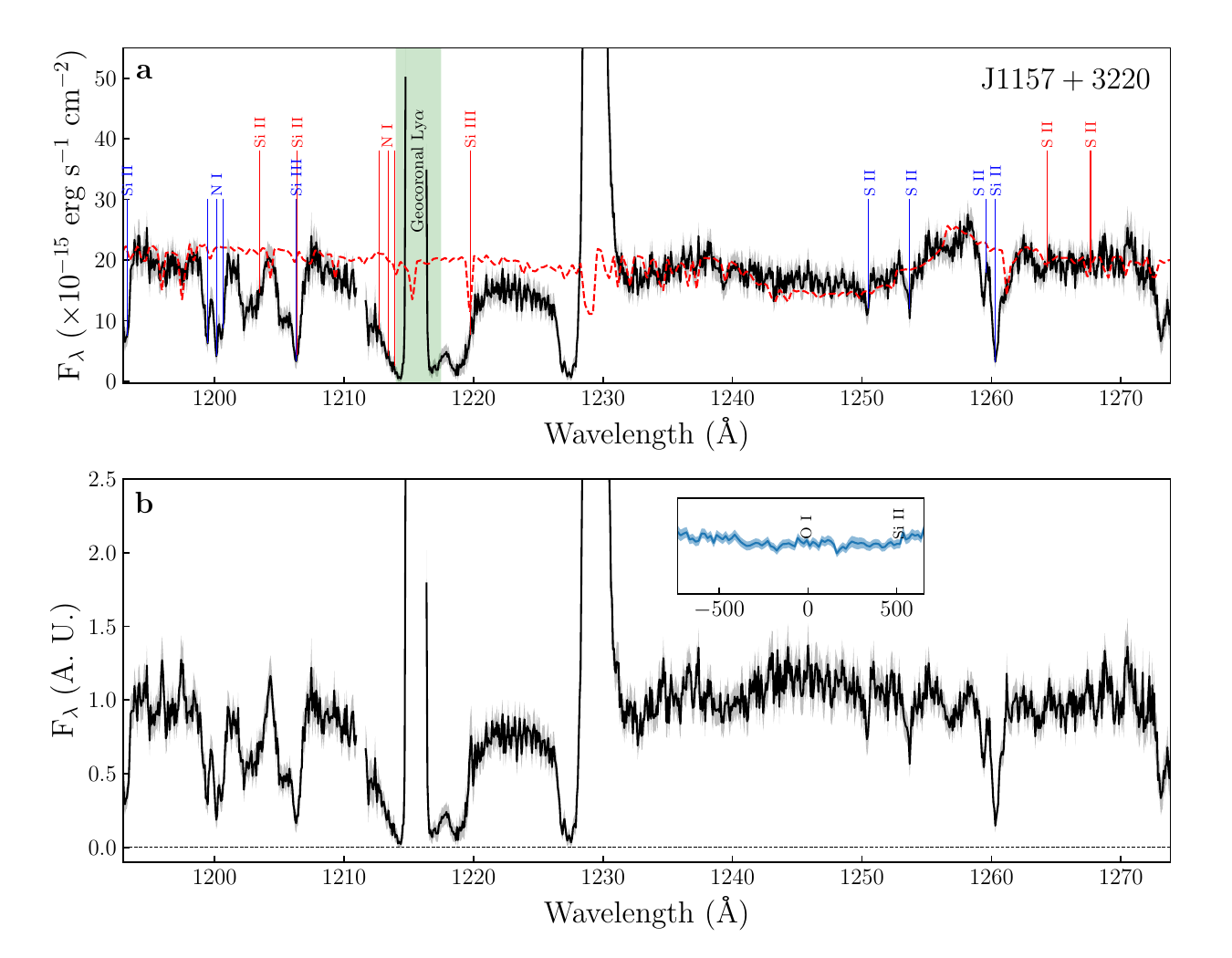}
    \includegraphics[width=3.3in]{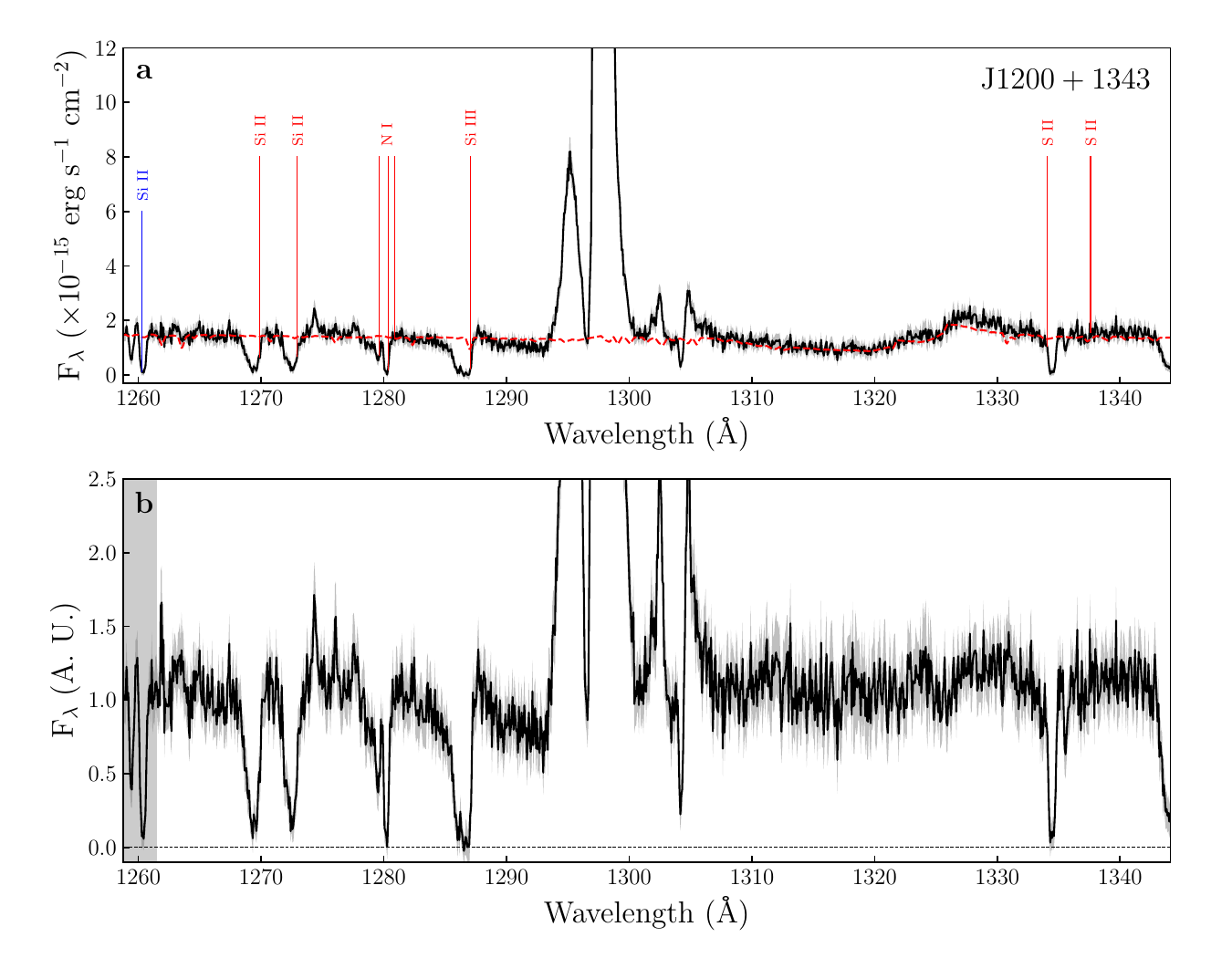}
    \includegraphics[width=3.3in]{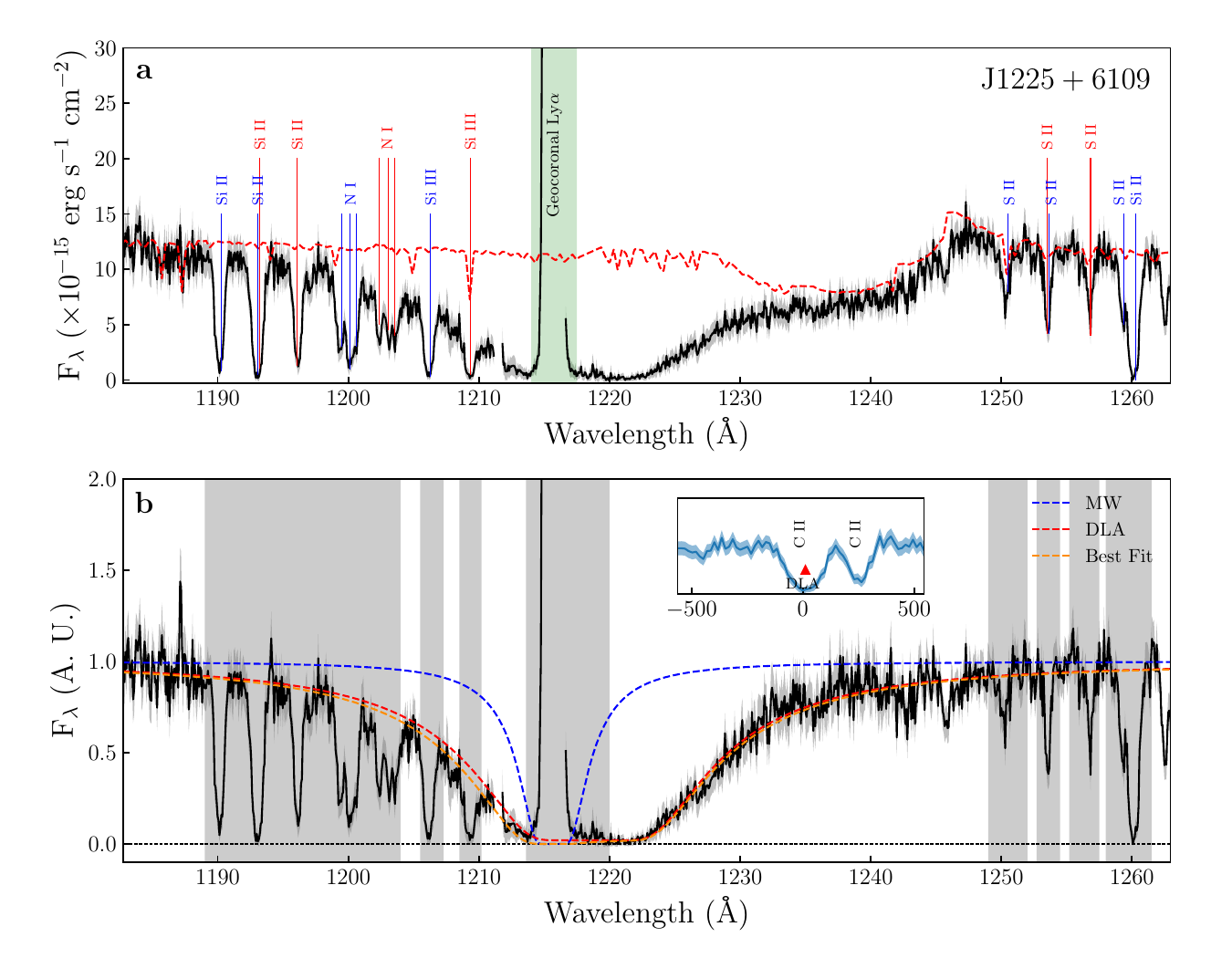}
    \includegraphics[width=3.3in]{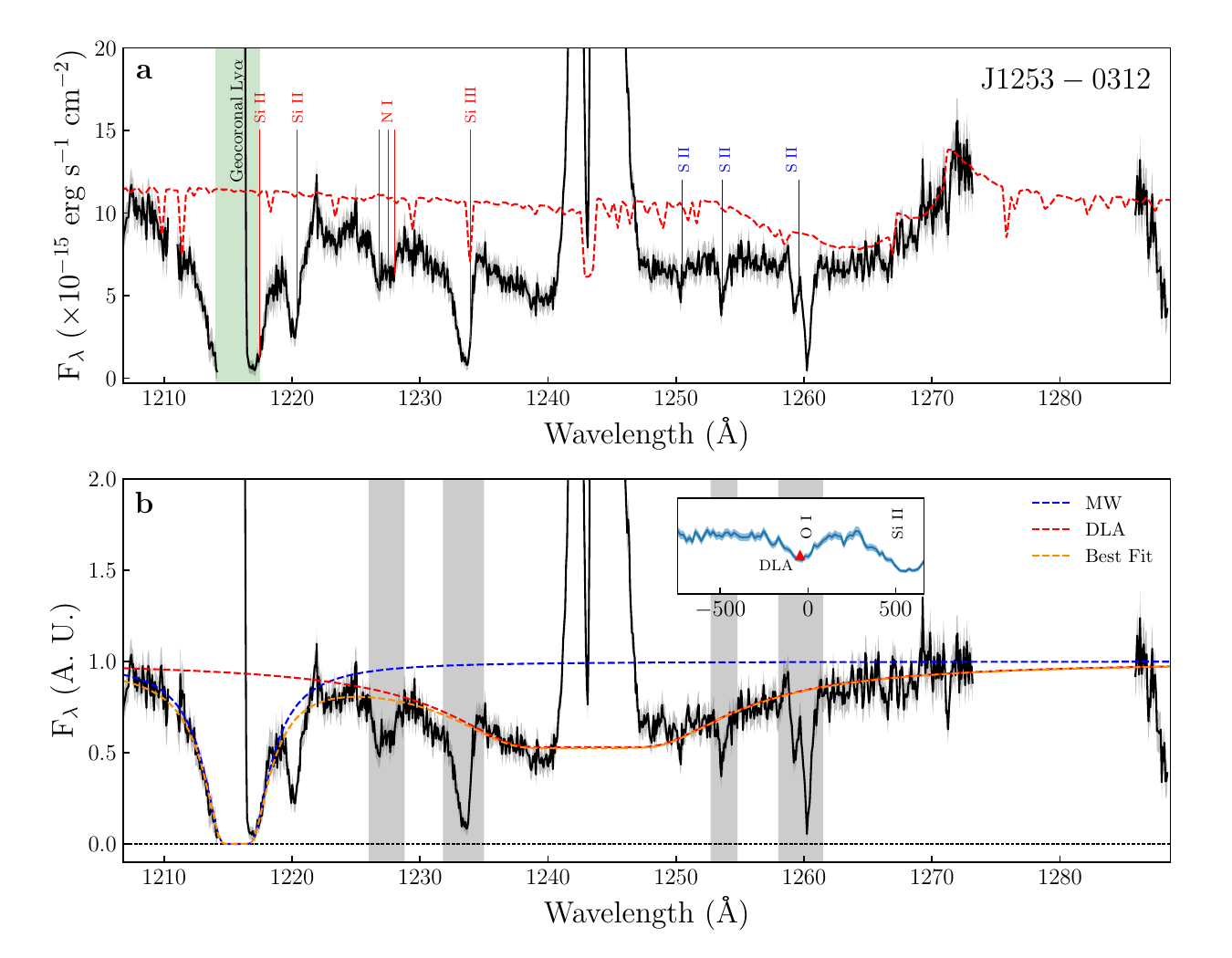}
    \caption{Continued.}
\end{figure*}

\begin{figure*}[ht]
    \addtocounter{figure}{-1}
    \centering
    \includegraphics[width=3.3in]{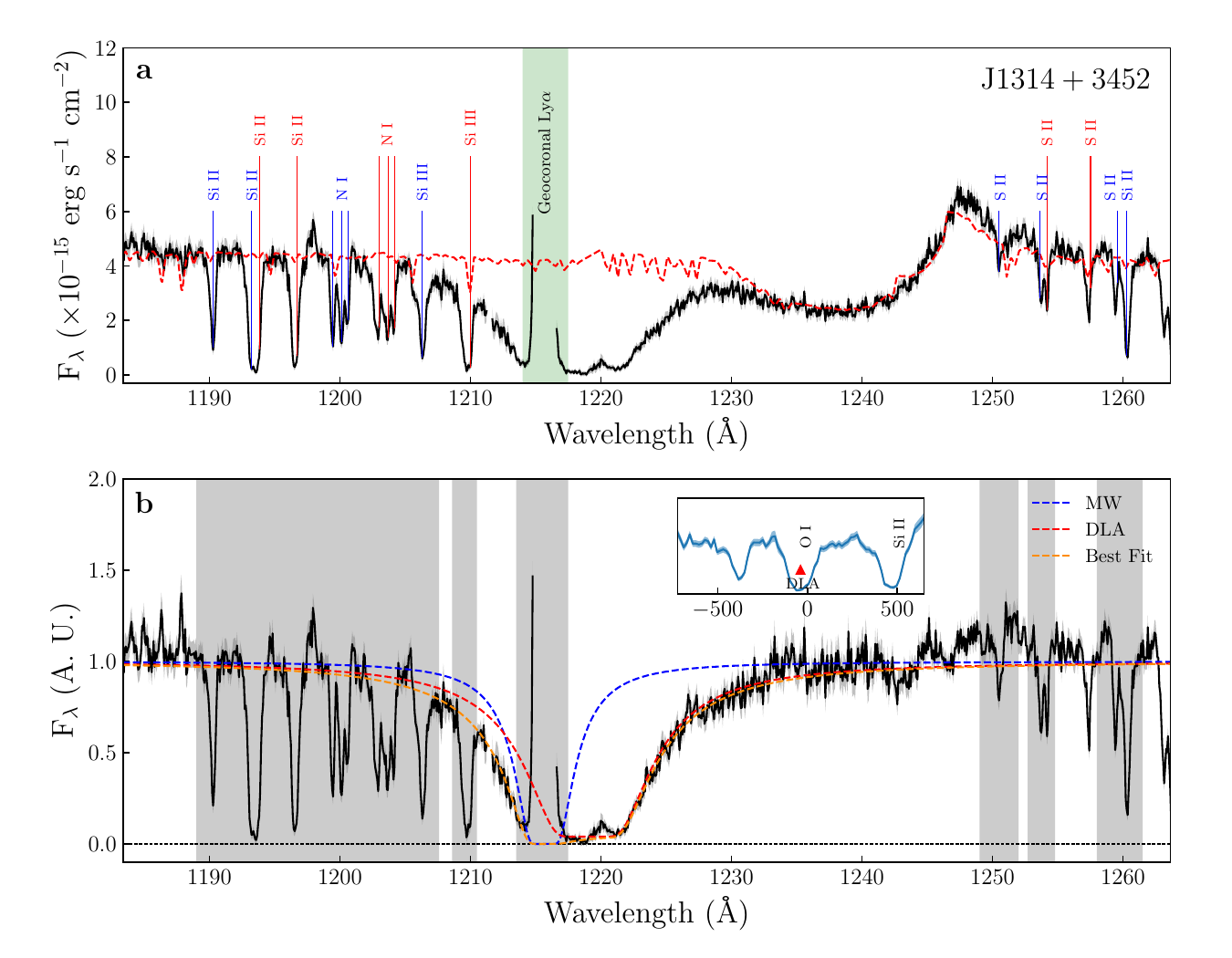}
    \includegraphics[width=3.3in]{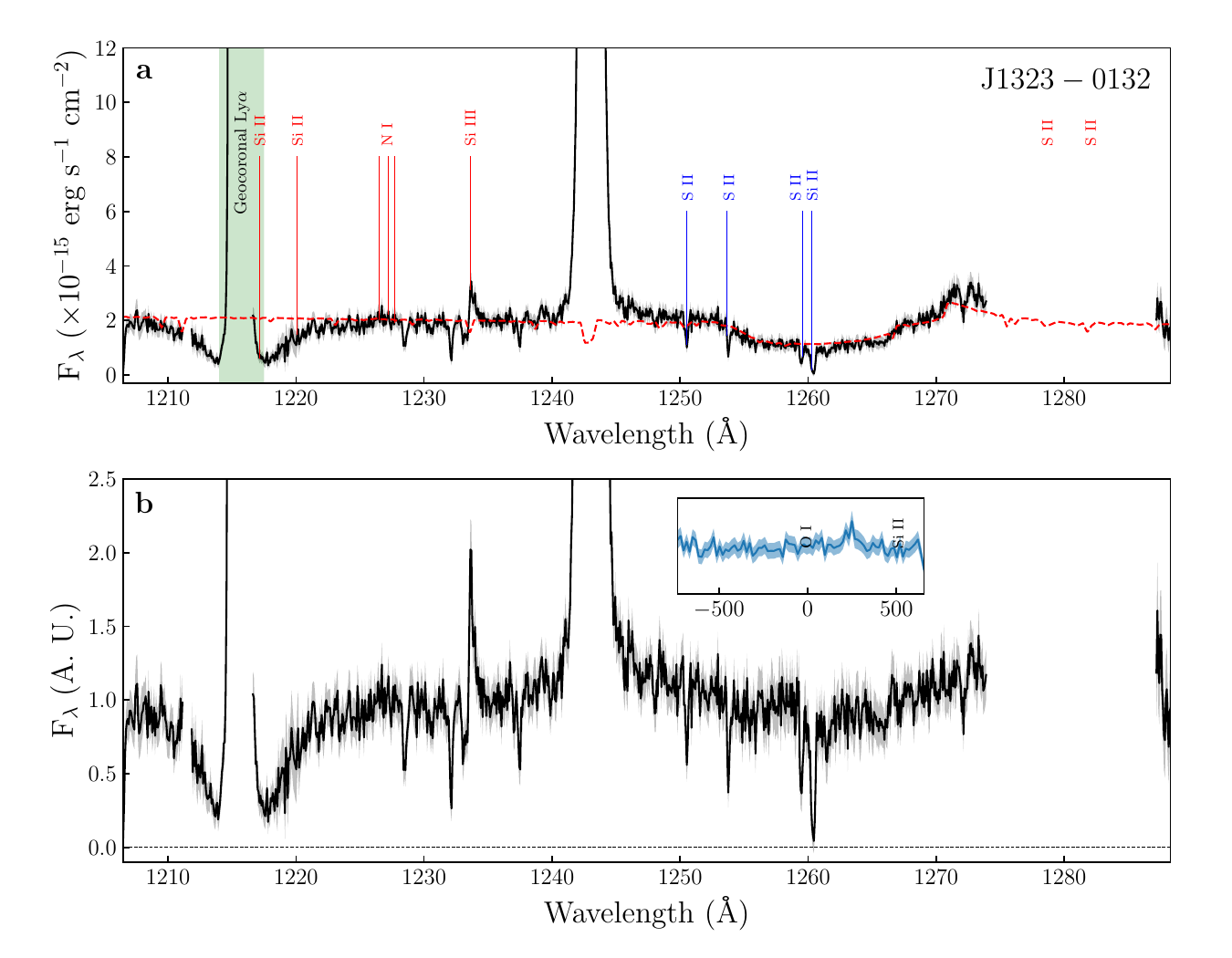}
    \includegraphics[width=3.3in]{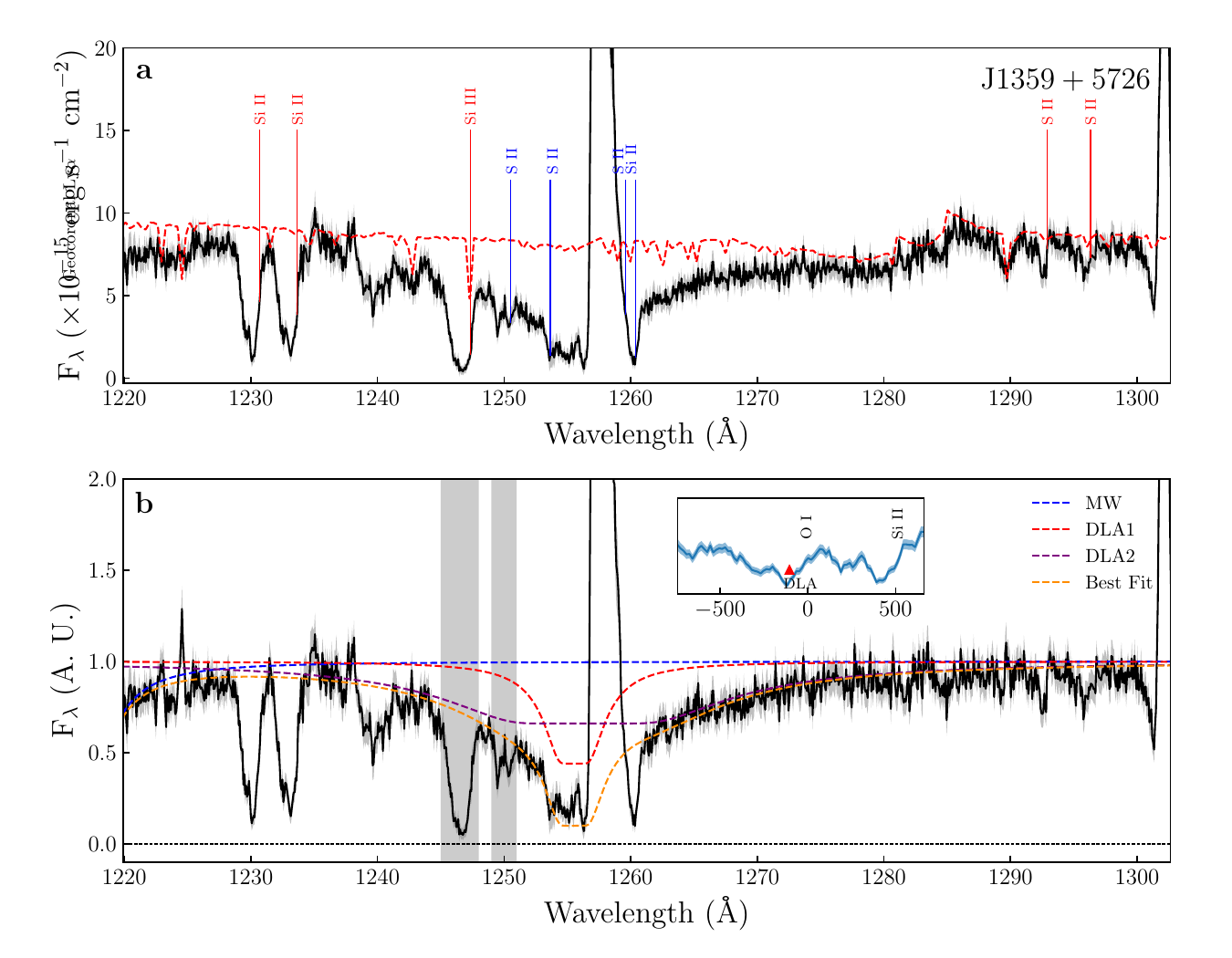}
    \includegraphics[width=3.3in]{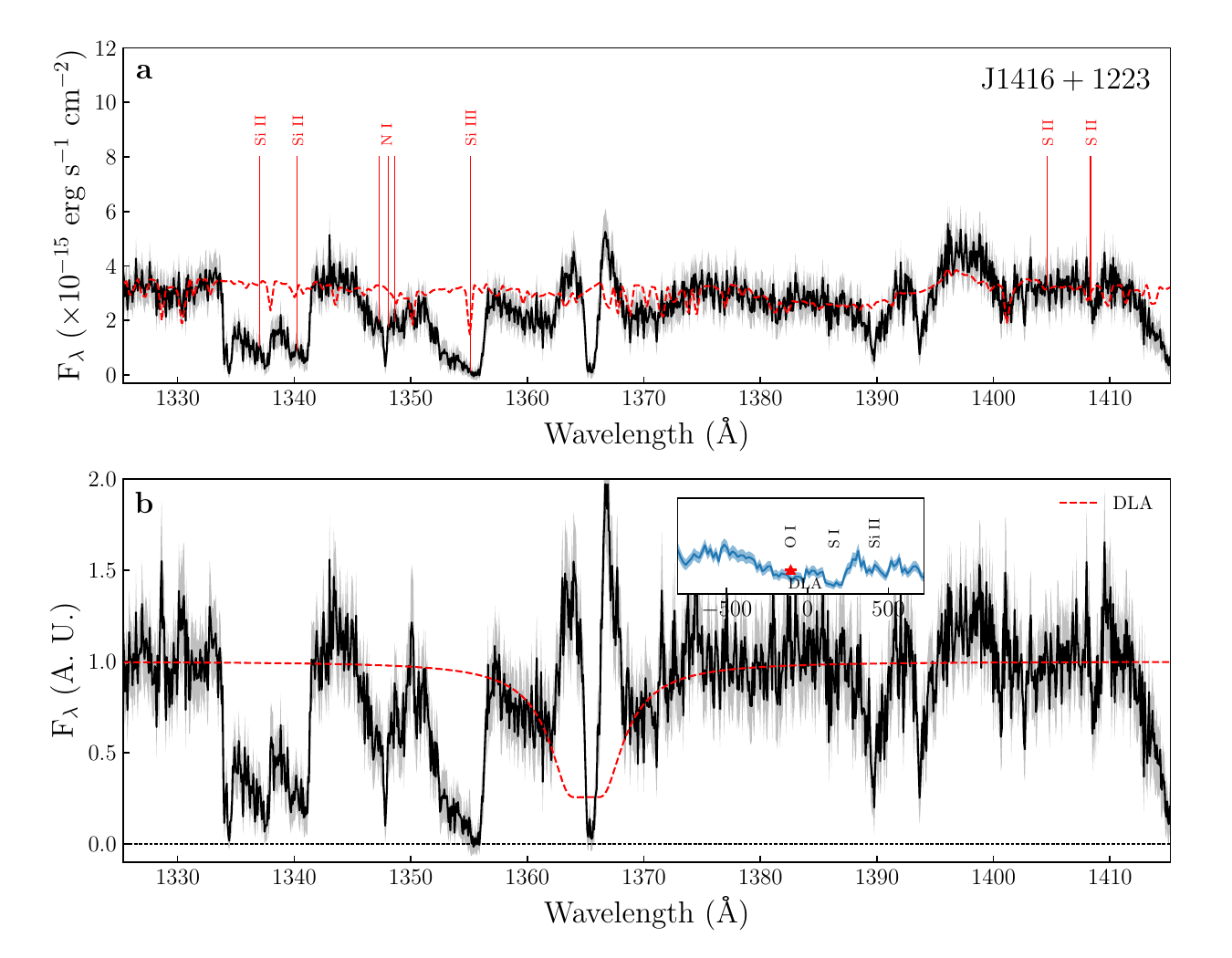}
    \includegraphics[width=3.3in]{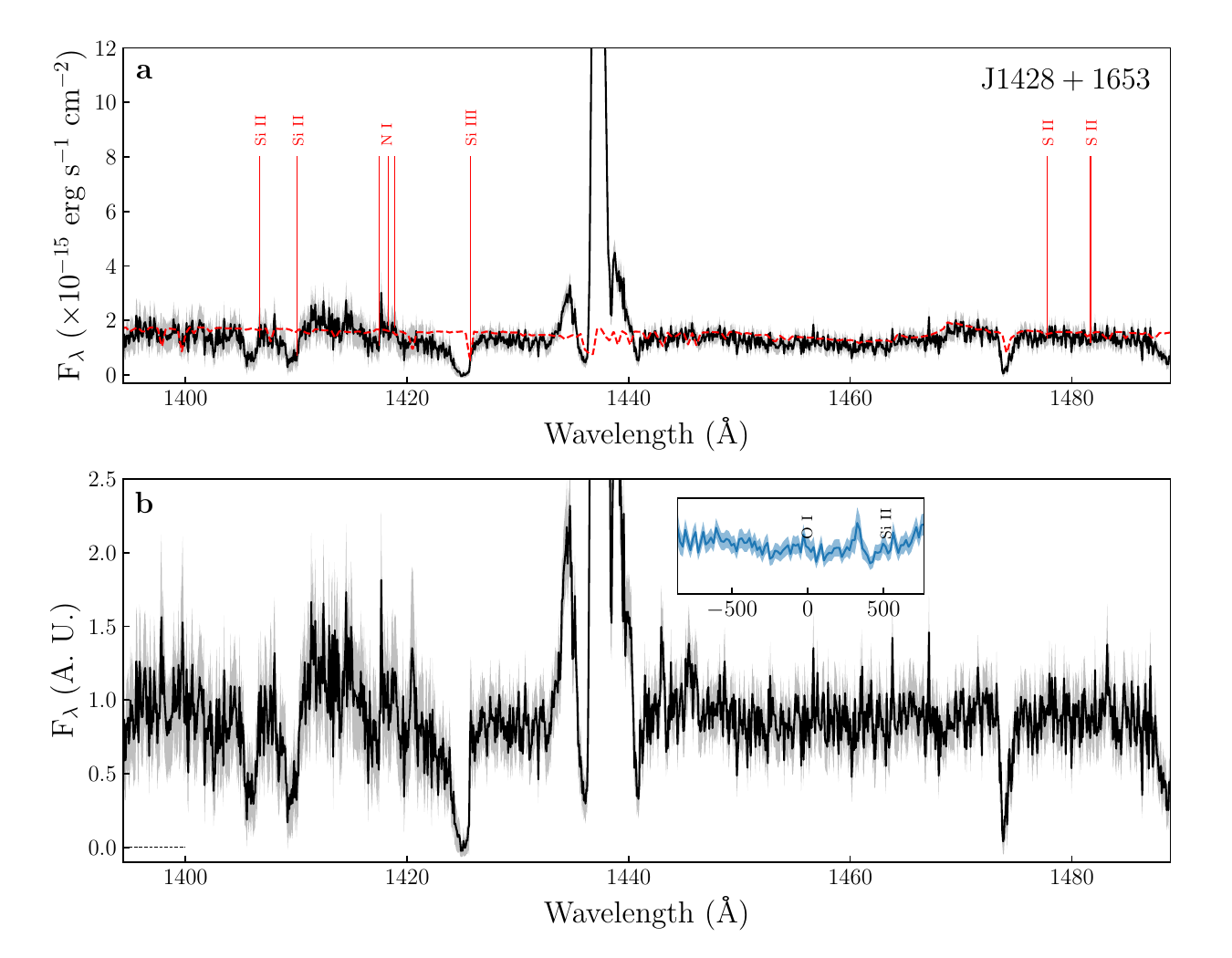}
    \includegraphics[width=3.3in]{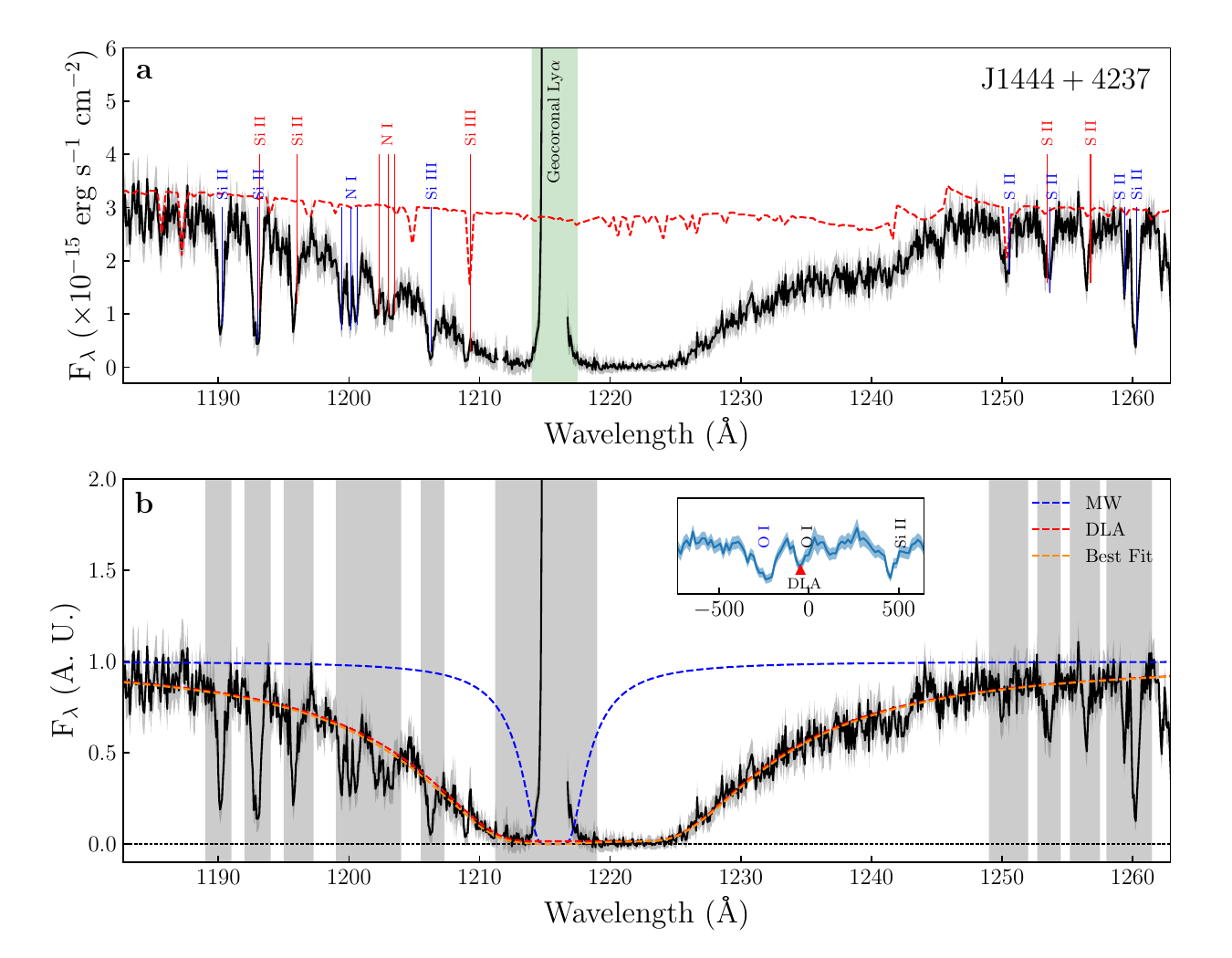}
    \caption{Continued.}
\end{figure*}

\begin{figure*}[ht]
    \addtocounter{figure}{-1}
    \centering
    \includegraphics[width=3.3in]{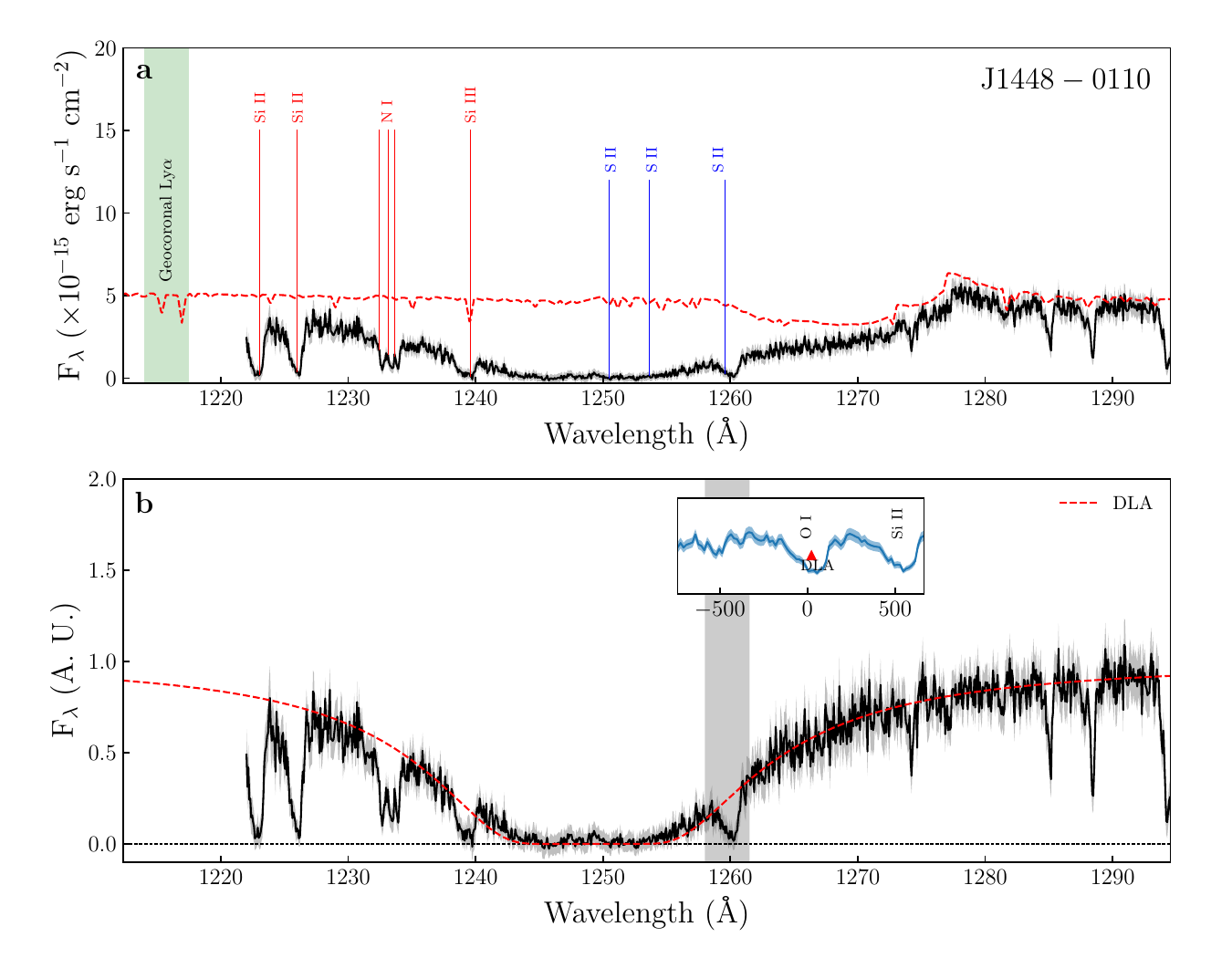}
    \includegraphics[width=3.3in]{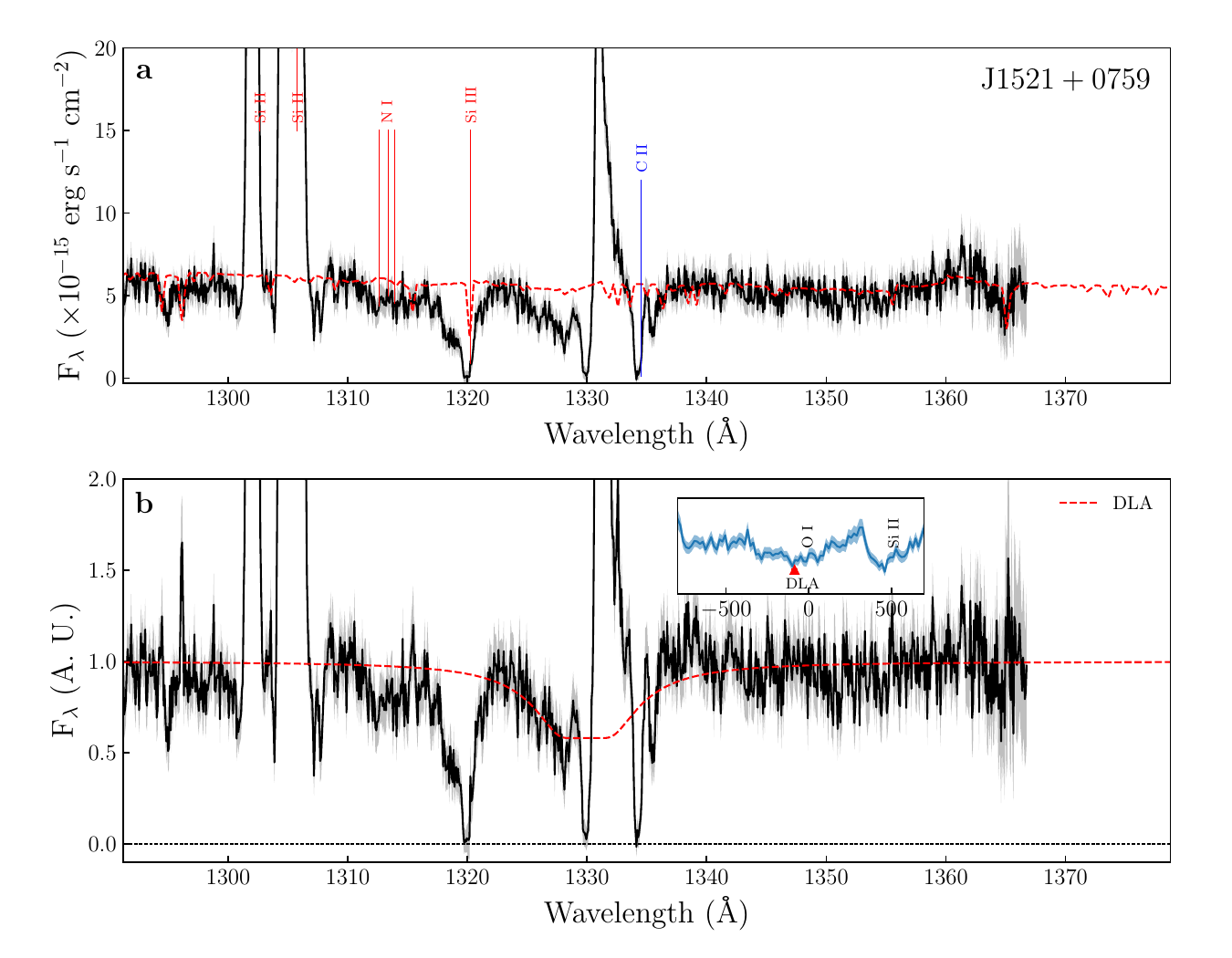}
    \includegraphics[width=3.3in]{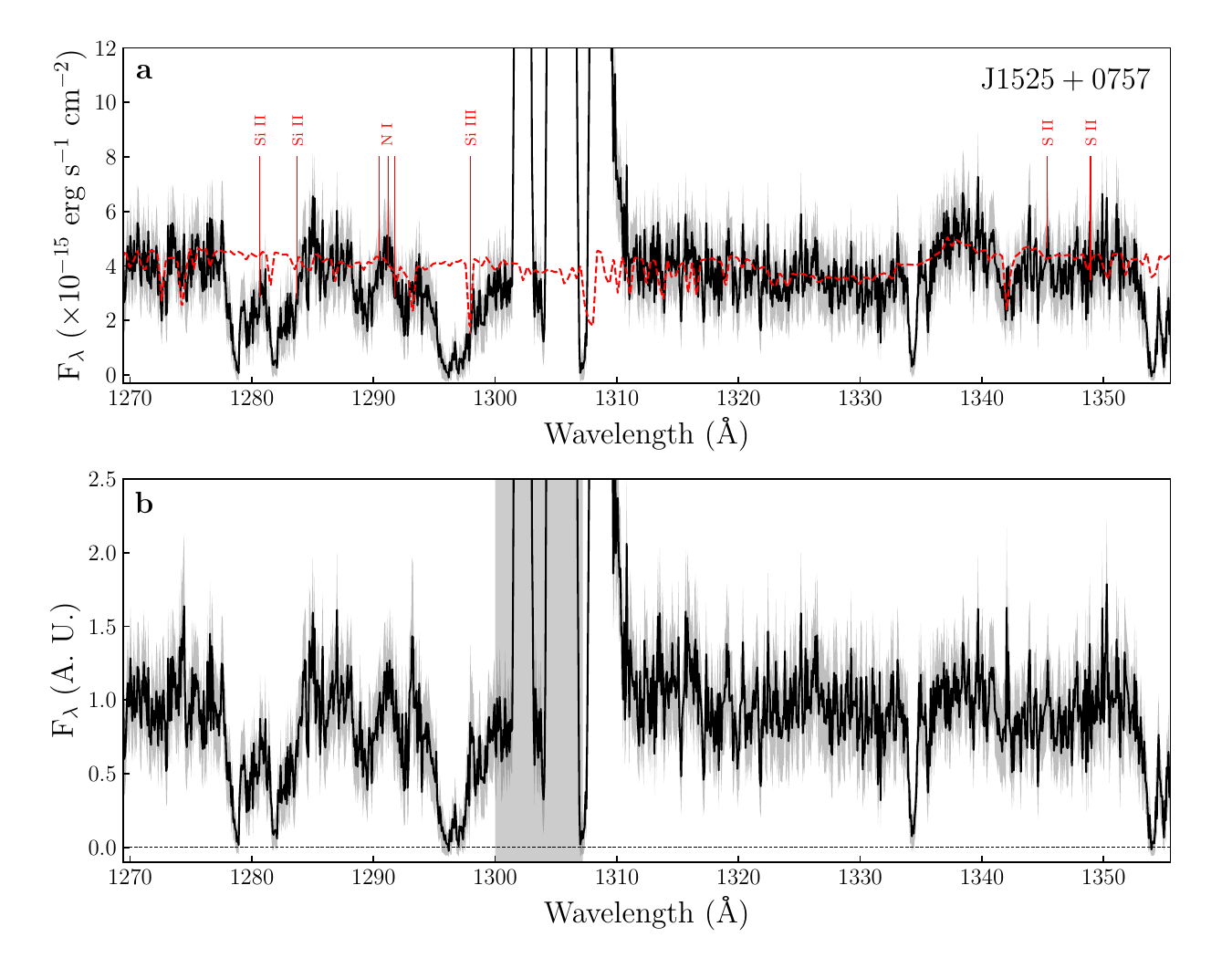}
    \includegraphics[width=3.3in]{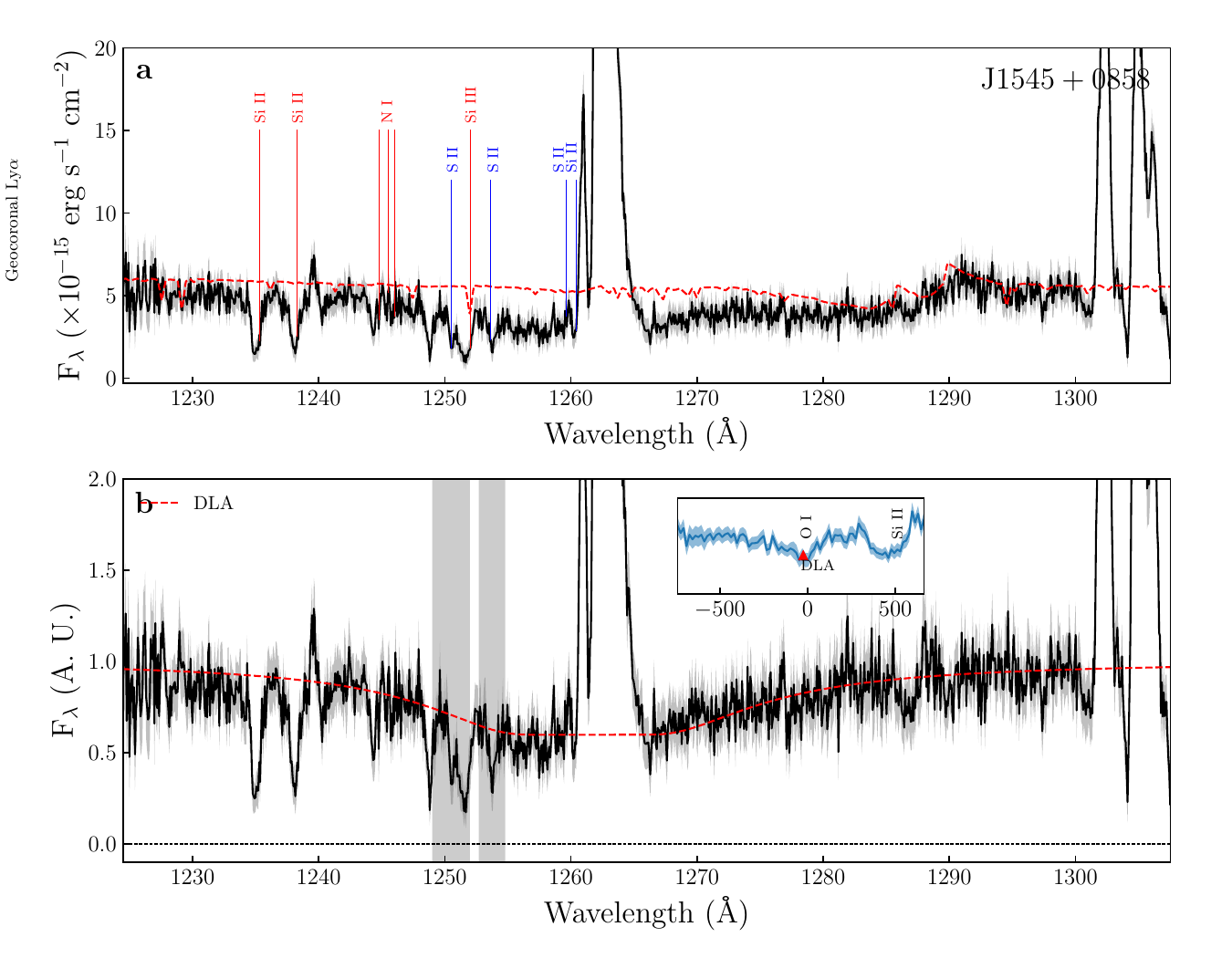}
    \includegraphics[width=3.3in]{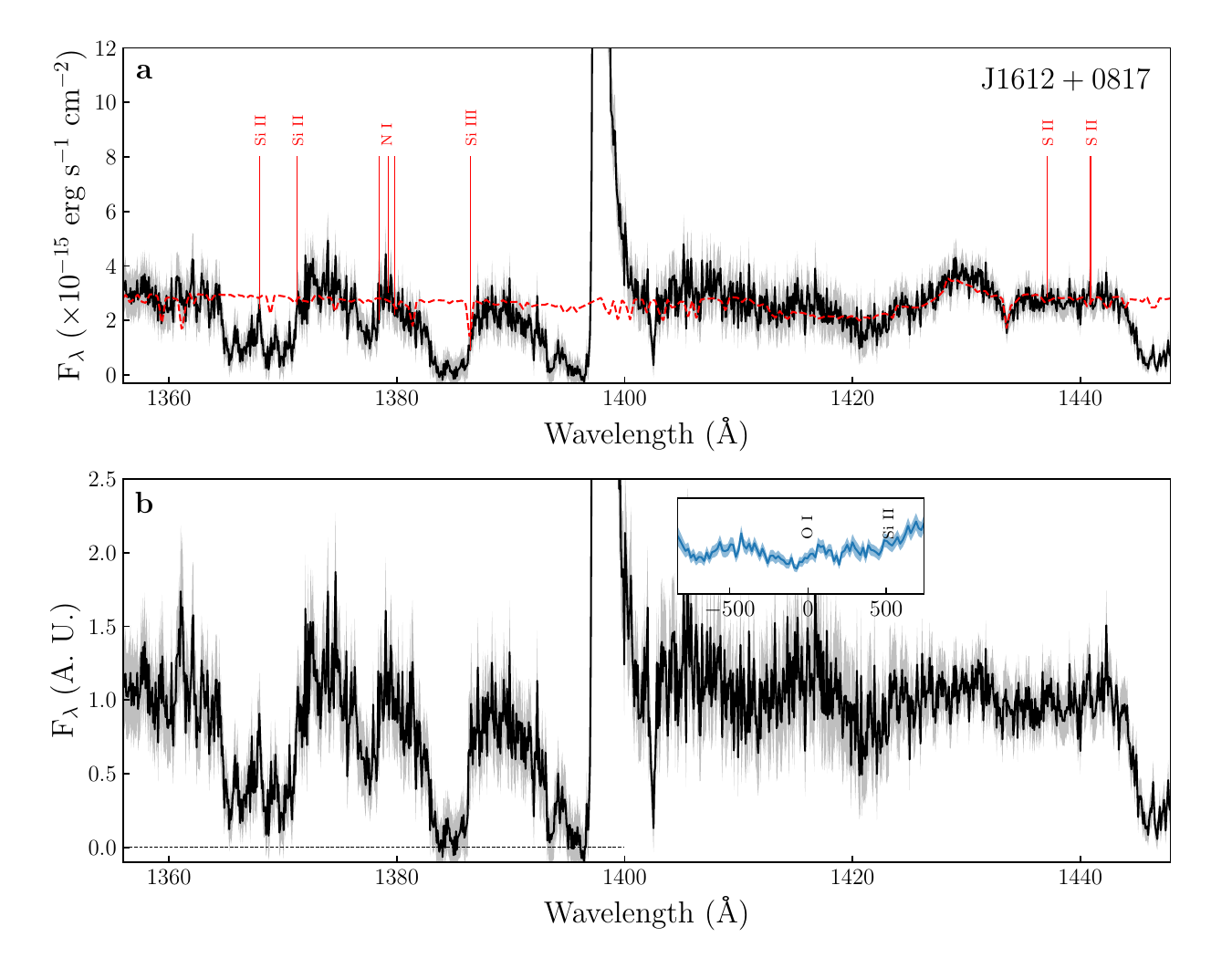}  
    \caption{Continued.}
\end{figure*}

\end{document}